\documentclass[preprint,superscriptaddress,tightenlines]{revtex4}
\usepackage{times}
\usepackage{amsmath}
\usepackage{graphicx}
\usepackage{epsfig}
\usepackage{dcolumn}
\usepackage{bm}
\usepackage{color}
\usepackage[utf8]{inputenc}

\newcommand{\ECM}{{\rm E_{\rm CM}}}
\newcommand{\ECMp}{{\rm E^{\prime}_{\rm CM}}}
\newcommand{\ECMsqr}{{\rm E_{\rm CM}^2}}
\newcommand{\ACP}{A_{\rm CP}}
\newcommand{\BR}{{\cal B}}
\newcommand{\pp}{\pi^+\pi^-}
\newcommand{\ks}{K_S^0}
\newcommand{\pip}{\pi^+}
\newcommand{\pim}{\pi^-}
\newcommand{\kk}{K^+K^-}

\newcommand{\EE}{e^+e^-}

\newcommand{\etap}{\eta^\prime}
\newcommand{\psp}{\psi(3686)}
\newcommand{\psip}{\psi(3686)}

\newcommand{\jpsi}{J/\psi}
\newcommand{\piz}{\pi^0}

\newcommand{\XYZ}{\rm XYZ}
\newcommand{\xyz}{\rm XYZ}
\newcommand{\x}{X(3872)}
\newcommand{\y}{Y(4260)}

\newcommand{\zc}{Z_c(3900)}
\newcommand{\zcp}{Z_c(4020)}

\newcommand{\ppjpsi}{\pi^+\pi^-J/\psi}
\newcommand{\hc}{h_c}
\newcommand{\pphc}{\pi^+\pi^-\hc}

\newcommand{\BBb}{B\bar{B}}
\newcommand{\LLb}{\Lambda\bar{\Lambda}}
\newcommand{\ccb}{c\bar{c}}
\newcommand{\ppb}{p\bar{p}}
\newcommand{\rt}{\rightarrow}
\newcommand{\lrarrow}{\leftrightarrow}

\begin{document}

\title{\boldmath The BESIII Physics Programme}

%\noaffiliation

\affiliation{Institute of High Energy Physics, Chinese Academy of Sciences, Beijing 100049, China}
\affiliation{University of Chinese Academy of Sciences, Beijing 100049, China}

\author{Chang-Zheng Yuan}\affiliation{Institute of High Energy Physics, Chinese Academy of Sciences, Beijing 100049, China}
\affiliation{University of Chinese Academy of Sciences, Beijing 100049, China}
\author{Stephen Lars Olsen}\affiliation{University of Chinese Academy of Sciences, Beijing 100049, China}

\date{July 29, 2019}

\begin{abstract}

The standard model of particle physics is a well-tested
theoretical framework, but there are still a number of issues that
deserve further experimental and theoretical investigation. For
quark physics, such questions include: the nature of quark
confinement, the mechanism that connects the quarks and gluons of
the standard model theory to the strongly interacting particles;
and the weak decays of quarks, which may provide insights into new
physics mechanisms responsible for the matter-antimatter asymmetry
of the Universe.  These issues are addressed by the Beijing
Spectrometer III (BESIII) experiment at the  Beijing
Electron-Positron Collider II (BEPCII) storage ring, which for the
past decade has been studying particles produced in
electron-positron collisions in the tau-charm energy-threshold
region, and has by now accumulated the world's largest datasets
that enables searches for nonstandard hadrons, weak decays of the
charmed particles, and new physics phenomena beyond the standard
model. Here, we review the contributions of BESIII to such studies
and discuss future prospects for BESIII and other experiments.

\end{abstract}

%%\pacs{14.40.Rt, 14.40.Pq, 13.66.Bc, ...}

\maketitle

\section{Introduction}

The two main questions addressed by particle physics are `what are
the most elementary building blocks of matter?' and `what are the
forces between them?'. In the standard model (SM) of particle
physics, the well-established theoretical description of the
fundamental particles and their mutual interactions, the building
blocks are quarks and leptons, and the forces between them are
mediated by the exchange of gauge bosons associated with the weak,
electromagnetic, and strong interactions~\cite{Weinberg:2018apv}.
Although the SM has been successfully tested by numerous
experiments, many of its detailed predictions remain to be
confirmed (for example, the existence of nonstandard hadrons and
the quark mixing properties in weak interaction), and substantial
theoretical and experimental efforts are currently being devoted
to this end.

Particle accelerators and experiments specifically aimed at
studies of these issues include: the BaBar~\cite{Aubert:2001tu}
and Belle~\cite{Abashian:2000cg} experiments at the PEPII and KEKB
$B$-factory colliders operating during 1999-2008 and 1999-2010,
respectively; the currently operating BESIII~\cite{Ablikim:2009aa}
experiment at the BEPCII electron-positron collider and the
LHCb~\cite{Alves:2008zz} experiment at the high energy Large
Hadron Collider (LHC) proton-proton collider; the
Belle~II/SuperKEKB project~\cite{Kou:2018nap} that is now starting
to take data; and proposed electron-positron `super tau-charm'
factories: STCF in Hefei, China~\cite{Luo:2018njj} and SCTF in
Novosibirsk, Russia~\cite{Levichev:2018}.

In the early 2000s, the BaBar and Belle experiments established
that the experimentally observed $CP$
violations~\cite{Bevan:2014iga} in $B\bar{B}$ system can be
explained as consequences of an irreducible complex phase in the
Cabibbo-Kobayashi-Maskawa (CKM) six-quark flavour-mixing matrix,
as first suggested in Ref.~\cite{Kobayashi:1973fv}. In addition,
these experiments discovered a number of nonstandard hadronic
states~\cite{Brambilla:2010cs,Chen:2016qju,Olsen:2017bmm,Guo:2017jvc,Brambilla:2019esw}
with properties that indicate their substructures are more complex
than the quark-antiquark mesons and three-quark baryons of the
conventional quark model~\cite{GellMann:1964nj,Zweig:1981pd}.

The BESIII experiment~\cite{Ablikim:2009aa} (see
Appendix~\ref{appA} for a short description), the third (and current) phase of a
thirty-years-old research program based at the Institute of High
Energy Physics in Beijing, investigates both hadronic states and
weak decay of charmed quark with electron-positron ($\EE$)
collision data in the center-of-mass energy ($\ECM$) range between
2.0 and 4.6~GeV produced by the BEPC collider, with a luminosity
that started at $10^{29}$~cm$^{-2}$s$^{-1}$ at the Beijing
Spectrometer in 1988~\cite{Bai:1994zm}, improved to
$10^{31}$~cm$^{-2}$s$^{-1}$ with BESII in 2002~\cite{Bai:2001dw},
and, with BESIII and the BEPCII collider, reached
$10^{33}$~cm$^{-2}$s$^{-1}$ in 2016. In 2008, the single-ring BEPC
collider, in which the counter-rotating $e^+$ and $e^-$ beams
shared the same magnets and vacuum chamber, was replaced by
BEPCII~\cite{Qin:2010zzd}, a high performance, two-ring $\EE$
collider, in which each beam has its own magnet and vacuum
systems. At the same time, BESII was replaced by BESIII, a
state-of-the-art detector with substantially improved
capabilities.

The $2\sim 4.6$~GeV center-of-mass (CM) energy region covers the
thresholds for the production of pairs of particles that contain
charm-quarks ($c$) and charm-antiquarks ($\bar{c}$) and includes
the narrow charmonium ($\ccb$) resonances $\jpsi$ and $\psip$,
which are prolific sources of hadrons comprising of the up ($u$),
down ($d$) and strange ($s$) light quarks. The cross section for
$D^0$ ($c\bar{u}$) and $D^+$ ($c\bar{d}$) mesons and their
antiparticles has a maximum at 3.77~GeV, the peak of the
$\psi(3770)$ resonance, and large samples of $D_s$ ($c\bar{s}$)
mesons are produced at the nearby $\psi(4160)$ resonance. The
energy threshold for pair production of the $\Lambda_c$ ($cud$),
the lightest baryon that contains a charm quark, is $\ECM=
4.573$~GeV, and also accessible to BEPCII. In addition, this
energy region includes the thresholds for $\tau$-lepton and all of
the stable hyperons (baryons that contain one or more $s$-quarks).
Studies at the near-threshold energy for the production and decays
of these particles have a number of unique advantages (large
production rate, clean environment, very low background, high
detection efficiency, and so on) over measurements done at higher
energies. As a result the BES program addresses a diverse range of
interesting physics with unprecedented sensitivity.

Notable achievements of the BES and BESII phases of this program
include: BES's first result, which was a five-fold improvement in
the precision of the $\tau$-lepton's measured mass that was two
standard deviations lower than the previous world average value at
that time and cleared up a discrepancy with SM
expectations~\cite{Bai:1992bu}; the first observation of the
purely leptonic $D^+_s\rt\mu^+\nu$ decay
process~\cite{Bai:1994qz}; precision measurements of the
annihilation cross section for $\EE\to$~{\it hadrons} that
provided essential inputs to the SM predictions for the Higgs
boson mass and the anomalous magnetic moment of the muon,
$(g-2)_{\mu}$~\cite{Bai:2001ct}; the discoveries of the $\sigma
(500)$~\cite{Ablikim:2004qna} and $\kappa
(700)$~\cite{Ablikim:2005ni} scalar mesons; and a comprehensive
study of the so-called $\rho\pi$ puzzle in vector charmonium
decays~\cite{Mo:2006cy}.

During its first ten years of operation, BESIII accumulated the
world's largest data sets of $D$ and $D_s$ meson decays, 10
billion $\jpsi$ and 450 million $\psip$ events, and about 100
million events with $\ECM$ between 4~and~4.6~GeV for studies of
nonstandard hadrons and the $\Lambda_c$ baryon. BESIII data
provide stringent constraints on the CKM quark-flavour mixing
scheme with precision measurements of the $|V_{cs}|$ and
$|V_{cd}|$ CKM matrix elements that modify the strengths of weak
interaction $c\rt s$ and $c\rt d$ quark transitions, and
strong-interaction phases in $D$-meson decays. These strong phases
are basic quantities that are essential inputs to other
experiments that determine $\gamma$, the $CP$-violating complex
phase angle of the $V_{ub}$ CKM matrix element responsible for
$B$-meson decays. The CKM matrix elements are fundamental
constants of the SM that have to be measured in experiment. Their
values are strictly constrained by unitarity; any deviation from
unitarity would be an unambiguous signal for new, non-SM physics.

In the conventional quark model, mesons comprise of one quark and
one antiquark, whereas baryons comprise of three quarks. This
simple picture successfully describes almost all of the hadrons
that were observed prior to the operation of the BaBar and Belle
$B$-factory experiments. However, other, nonstandard
configurations have been proposed since the very inception of the
quark model~\cite{GellMann:1964nj,Zweig:1981pd}. Although these
were the subject of a long series of experimental
searches~\cite{Jaffe:2004ph}, the results were inconclusive.
However, starting in 2003, Belle and BaBar discovered a number of
meson states that decay to final states that contain both a $c$-
and a $\bar{c}$-quark~\cite{Bevan:2014iga}. Whereas some of these
states have properties that fit well the expectations for the
conventional $\ccb$ mesons of the charmonium
model~\cite{Deng:2016stx}, others have properties that do not
match those of any $\ccb$ meson and can only be accommodated by
nonstandard, multi-quark
configurations~\cite{Brambilla:2010cs,Brodsky:2014xia,Chen:2016qju,Olsen:2017bmm,Guo:2017jvc,Brambilla:2019esw}.
These latter charmonium-like states are collectively referred to
as the $\xyz$ mesons to indicate that their underlying structure
is still not well understood. During the same time period, the
Belle group discovered candidates for nonstandard mesons in the
bottom-quark sector~\cite{Belle:2011aa} and the LHCb group found
strong candidates for five-quark (pentaquark)
baryons~\cite{Aaij:2015tga}.

With the capability of adjusting the $\EE$ CM energy to the peaks
of resonances and to just-below and just-above the energy
thresholds for particle-antiparticle pair formation, combined with
the clean experimental environments due to near-threshold
operation (production of one additional hadron needs to increase
the CM energy by more than 100~MeV), BESIII is uniquely able to
perform a broad range of critical measurements of the weak decays
of strange and charmed particles, and the production and decays of
many of the nonstandard $\xyz$ meson states. Here, we briefly
review some highlights of the BESIII program, including: precision
measurements of CKM matrix elements; studies of charmed particle
decays; discoveries  of new $\xyz$ mesons; in-depth investigations
of light hadrons; refined measurements of the fundamental
properties of baryons; contributions to stringent tests of the SM,
such as the anomalous muon magnetic moment $(g-2)_{\mu}$, with
precision measurements of the cross section for $\EE$ annihilation
into hadrons; and precision measurements of the production
cross-sections and decay properties of hyperons. We present our
perspective on the potential future results from BESIII and other
experiments, and discuss the opportunities for studies at proposed
next-generation facilities.

\section{\boldmath Measurement improvements of $m_\tau$ and $\sigma(\EE\to {\rm hadrons})$ }

\subsection{\boldmath Three decades of $m_{\tau}$ measurements}

The $\tau$-lepton mass ($m_{\tau}$) is a fundamental parameter of
the SM and a precise knowledge of its value is essential for tests
of the lepton flavour universality (LFU). The current experimental
precision is primarily due to a 2014 BESIII  measurement
$m_{\tau}=1776.91\pm 0.18$~MeV/$c^2$~\cite{Ablikim:2014uzh} that
is in good agreement with the original 1992 BES value: $1776.9\pm
0.5$~MeV/$c^2$~\cite{Bai:1992bu}, but with significantly better
precision. The BESIII measurement benefited from the
implementation of a new laser backscattering beam energy
measurement system, BEMS~\cite{Abakumova:2011rp}.

This level of precision is still three orders of magnitude poorer
than that for the muon-lepton mass $m_{\mu}$ and continued
improvements are needed.  With refinements of the BEMS and more
data near the energy threshold for $\tau^+\tau^-$ production,
BESIII will further improve the precision of this measurement to
$\pm 0.10$~MeV/$c^2$, and maintain the program's leading role in
this direction~\cite{Zhang:2018gol}.

\subsection{Precision measurement of vacuum polarization of virtual photons}

The measured value of $(g-2)_\mu$ from Brookhaven National
Laboratory experiment E821~\cite{Bennett:2006fi} is $\sim$3.5
standard deviations higher than the SM
prediction~\cite{Davier:2017zfy,Keshavarzi:2018mgv}, a discrepancy
that has inspired elaborate follow-up experiments at
Fermilab~\cite{Grange:2015fou} and J-PARC~\cite{Otani:2015jra}.
The SM prediction for $(g-2)_{\mu}$ is very sensitive to the
effects of hadronic vacuum polarization (HVP) of the virtual
photon, which are about 100 times larger than the current
experimental uncertainty and, thus, must be determined with high
precision. Vacuum polarization also has a critical influence on
precision tests of the electroweak theory, which rely on a precise
knowledge of $\alpha(s)$, the running quantum electrodynamics
(QED) coupling constant. Because of vacuum polarization,
$\alpha^{-1}(m^2_Z)=128.95\pm 0.01$~\cite{Davier:2017zfy} is about
6\% below its long-distance value of $\alpha^{-1}(0)=137.04$.
About half of this difference is due to HVP.

Since HVP effects are non-perturbative, they cannot be directly
computed from first principles. Recent computer-based lattice
quantum chromodynamics (LQCD) calculations have made significant
progress, but the uncertainties are still
large~\cite{Miura:2019xtd,Davies:2019efs}. Instead, the most
reliable determinations of the HVP contributions to $(g-2)_\mu$
and $\alpha(m^2_Z)$ use dispersion relations with input from
experimental measurements of cross-sections for the $\EE$
annihilation into
hadrons~\cite{Davier:2017zfy,Keshavarzi:2018mgv}. The data used
for the most recent determinations are mostly from the
SND~\cite{Achasov:2006vp}, BaBar~\cite{Lees:2012cj},
BESIII~\cite{Ablikim:2015orh},
CMD-2~\cite{Akhmetshin:2003zn,Akhmetshin:2006bx}, and
KLOE~\cite{Anastasi:2017eio} experiments. BaBar and KLOE
operations have been terminated (although the data analysis
continues), leaving SND, CMD-3~\cite{Akhmetshin:2016dtr}, and
BESIII as the only running facilities with the capability to
provide the improvements in precision that will be essential for
the evaluation of $(g-2)_\mu$ with a precision that will match
that of the new experimental measurements at
Fermilab~\cite{Grange:2015fou} and J-PARC~\cite{Otani:2015jra}.

\begin{figure*}[htbp]
  \centering
  \includegraphics[height=8cm]{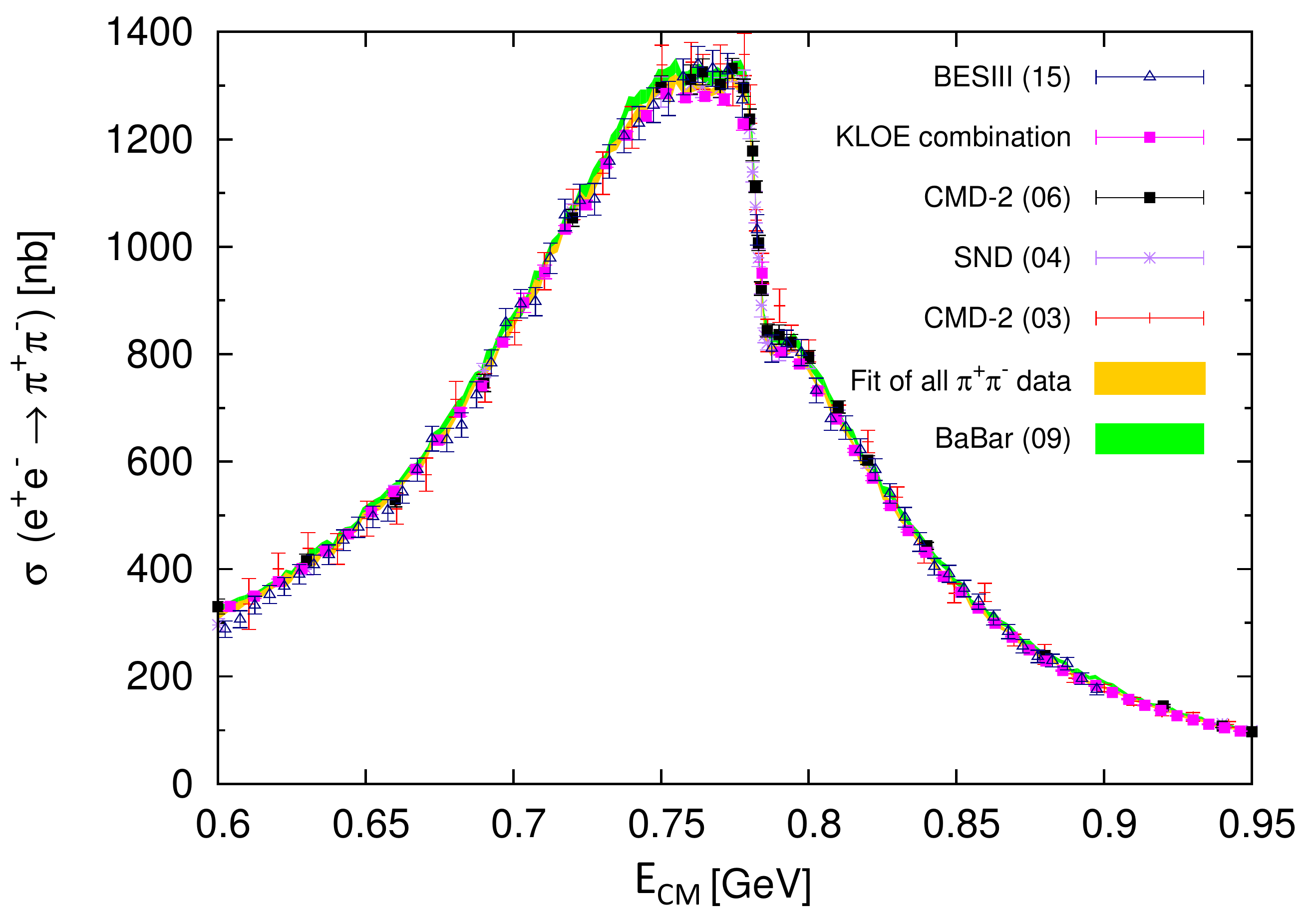}
\caption{Measurements of cross-section $\sigma(e^+e^-\to \pp)$
from SND~\cite{Achasov:2006vp},
CMD-2~\cite{Akhmetshin:2003zn,Akhmetshin:2006bx},
BaBar~\cite{Lees:2012cj}, KLOE~\cite{Anastasi:2017eio}, and
BESIII~\cite{Ablikim:2015orh}. The structure near $\ECM=0.78$~GeV
is due to interference between $\rho\to \pp$ and $\omega\to \pp$;
the orange band is the result of a fit from
Ref.~\cite{Keshavarzi:2018mgv}. The numbers in parentheses in the
figure legend denote the year when the respective reference was
published. Reproduced from
Ref.~\cite{Keshavarzi:2018mgv}.}
  \label{fpi}
\end{figure*}

With data taken at $\ECM=3.773$~GeV (primarily for studies of
$D$-meson decays) BESIII measured the cross-sections for
$\EE\to\pp$, where $\pi^+$ and $\pi^-$ denote pions, at $\ECM$
between 0.6 and 0.9~GeV~\cite{Ablikim:2015orh}, which covers the
$\rho\to\pp$ peak, where $\rho$ denotes a rho meson, the major
contributor to the HVP dispersion relation integral. These
measurements used initial state radiation (ISR) events in which
one of the incoming beam particles radiates a $\gamma$-ray with
energy $E_{\rm ISR}=x\ECM /2$ before annihilating at a reduced CM
energy of $\ECMp=\sqrt{1-x}\ECM$. The relative uncertainty of the
BESIII measurements is 0.9\%, which is similar to the precision of
the BaBar~\cite{Lees:2012cj} and KLOE~\cite{Anastasi:2017eio}
results. The BESIII measured values agree well with the KLOE
results for energies below 0.8~GeV, but are systematically higher
at higher energies. In contrast, the BESIII results agree with
BaBar at higher energies, but are lower at lower energies. The
detailed comparisons are shown in Fig.~\ref{fpi}. Nevertheless,
the contributions of $\EE\to \pp$ to the $(g-2)_{\mu}$ HVP
calculation from these experiments are overall in agreement within
two standard deviations, and the observed $\sim$3.5 standard
deviation difference between the calculated muon magnetic moment
value and the E821 experimental measurement persists.

In the ISR process, the direction of the radiated $\gamma$-ray is
strongly peaked in the beam direction and does not usually
register in the detector. In the BESIII measurements reported in
Ref.~\cite{Ablikim:2015orh}, the $\gamma$-ray had to be detected,
which limited the usable data sample to only a small fraction of
the produced events. With larger data sets, not only will these
measurements have improved statistical precision, but, by
including events in which the $\gamma$-ray is emitted along or
very close to the beam direction and undetected, but with its
presence inferred from energy-momentum constraints, cross sections
at even lower $\ECMp$ values will be measured. In addition, BESIII
is currently measuring cross sections for other modes, such as
$\EE\to \pp\piz$~\cite{Ablikim:2019sjw}, $2(\pp)$, and $\pp 2\piz$, that also contribute
to the HVP integral, albeit at a lower level~\cite{Ripka:2018jsj}.
For $\ECM$ above 2~GeV, BESIII does not need to use the ISR method
and can make higher precision, direct cross-section
measurements~\cite{Ablikim:2017wlt}. These efforts will continue
to improve our understanding of the vacuum polarization of the
photon and provide essential input to other experimental programs.

\section{Precision measurements of charmed particle decays}

The charm quantum number is conserved by the strong and
electromagnetic forces. As a result the lightest charmed hadrons,
the $D^0$, $D^+$, and $D^+_s$ mesons, and the $\Lambda^+_c$
baryon, decay via weak interactions. The rates and dynamics of
these decays are sensitive to the strong interactions between the
parent charm and light quarks and those between the final-state
light quarks. The strengths of the weak decays are proportional to
the Fermi constant $G_F$ multiplied by the $V_{cd}$ and $V_{cs}$
CKM matrix elements. In principle, the strong interactions between
quarks can be computed by quantum chromodynamics (QCD), but the
calculations are extremely difficult. Instead, QCD-inspired
models, or LQCD calculations are used. Rapid advances in the
latter need to be rigorously tested by measurements of purely
leptonic and semileptonic $D$ and $D_s$ meson decays with improved
precision.

\subsection{Purely leptonic decays}

The rates for purely leptonic charmed meson decays, $D^+_q\rt
\ell^+\nu$, where $q=d$ or $s$ and $\ell =e$, $\mu$, or $\tau$,
electron, muon, or tau, respectively, and $\nu$ stands for
neutrino, are proportional to the product of $|V_{cq}|$, the
relevant CKM matrix element, and $f_{D^+_q}$, the $D^+_q$-meson
decay constant. In terms of these parameters, the SM decay width
is given by
\begin{equation}
\Gamma(D_q^+ \rightarrow \ell^+\nu)=
     \frac{G^2_F f^2_{D_q^+}} {8\pi}
     |V_{cq}|^2
      m^2_{\ell} m_{D_q^+}
    \left (1- \frac{m^2_{\ell} } {m^2_{D_q^+}}\right )^2,
\label{fdq}
\end{equation}
where $m_{\ell}$ is the lepton mass, and $m_{D_q^+}$ is the
$D_q^+$-meson mass, which are both well measured. Thus, the
determination of $\Gamma(D^+_q \rightarrow \ell^+\nu)$ directly
measures the product $|V_{cq}| f_{D^+_q}$.

In principle, the $f_{D^+_q}$ values can be computed using LQCD.
At present, the obtained precision for these calculations are at
the part~per~thousand level~\cite{Bazavov:2017lyh}, and can
translate leptonic decay-rate measurements into high-precision
determinations of the $|V_{cq}|$ CKM matrix elements.

The $\psi(3770)$ resonance is only 30~MeV/$c^2$ above the
$D\bar{D}$ mass threshold and predominantly decays to final states
with a $D\bar{D}$ meson pair and nothing else. Thus, when a
final-state $D$ (or $\bar{D}$) meson is fully reconstructed in one
of its common hadronic decay modes, the accompanying $\bar{D}$
($D$) is tagged, meaning that it must be the parent of all the
remaining particles in the event and its four-momentum is
specified. When a tagged $D^{+}$ decays to $\mu^{+}\nu$, the mass
of the (undetected) zero-mass neutrino can be inferred from
energy-momentum conservation. This tagging feature is a powerful
tool for charmed-particle decay measurements that is only possible
in near-threshold experiments.

\begin{figure*}[htbp]
\centering
 \includegraphics[height=5cm]{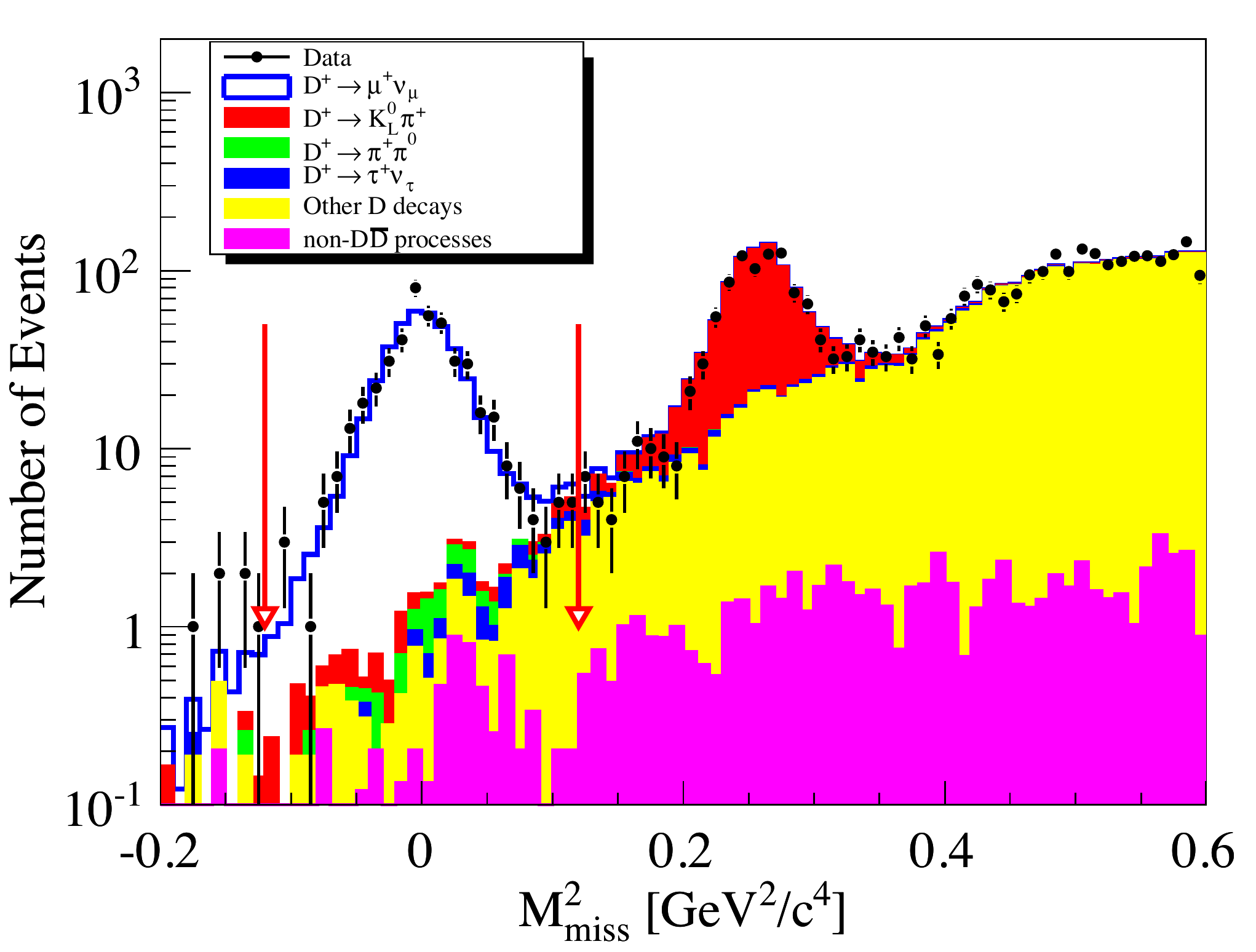}
 \includegraphics[height=5cm]{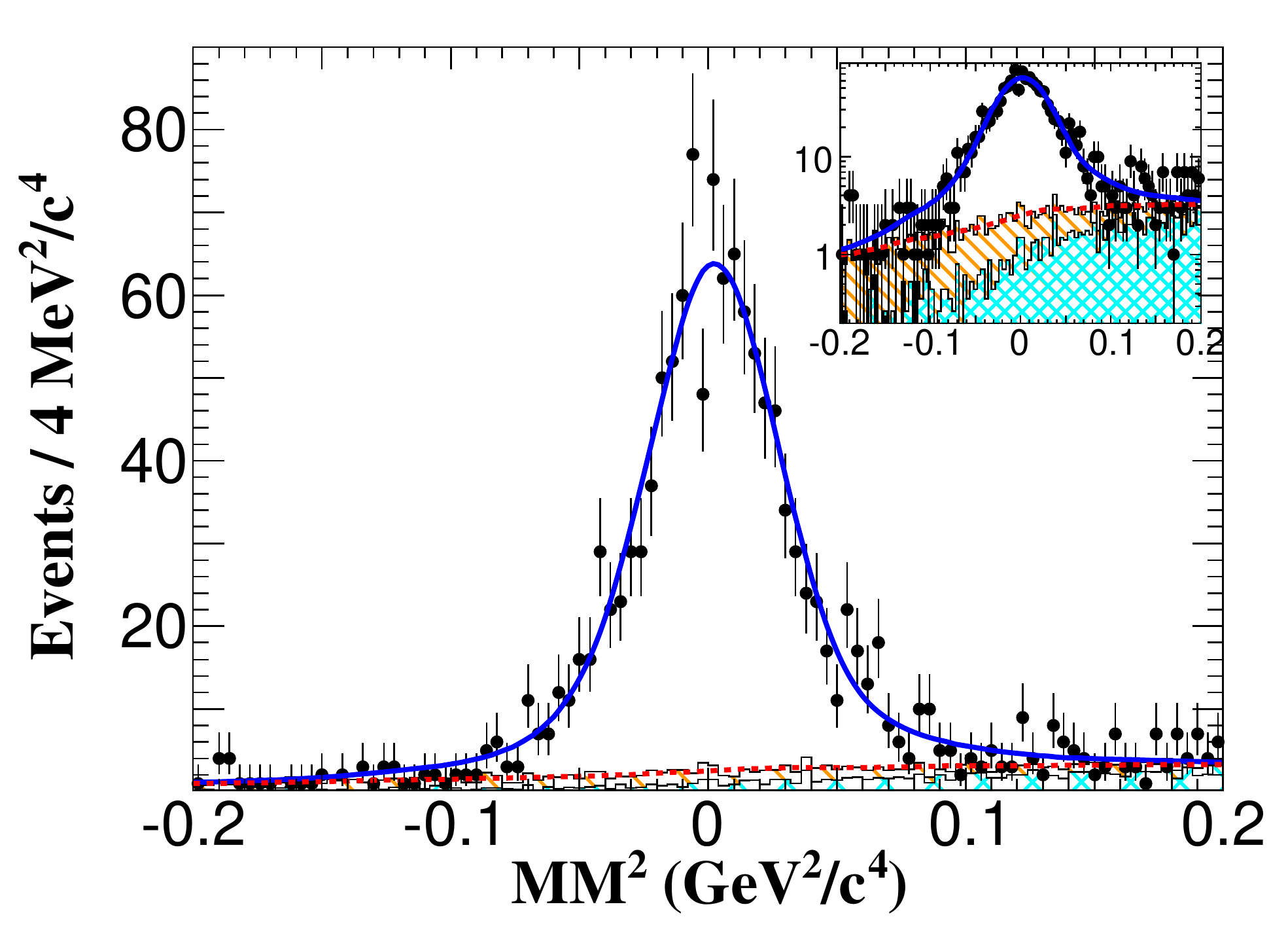}
\caption{The peaks centered at the square of the invariant mass of
the missing particles $M^2_{\rm miss}=0$ are the signals for the
purely leptonic decays $D^+\to \mu^+\nu$~\cite{Ablikim:2013uvu} (left panel) and $D^+_s\to
\mu^+\nu$~\cite{Ablikim:2018jun} (right panel) measured at BESIII. Red arrows in 
the left panel are the
boundaries of the signal region, and the inset in the right panel shows the
same distribution in log scale. } \label{fdfds}
\end{figure*}

In a sample of 1.7 million tagged $D^{\pm}$ mesons from
$\psi(3770)\rt D^+D^-$ decays, BESIII found $409\pm 21$
$D^\pm\to\mu^\pm\nu$ signal events over a small background (see
Fig.~\ref{fdfds}), corresponding to the world's best branching
fraction measurement: $\BR(D^+\to\mu^+\nu)=(3.71 \pm 0.20)\times
10^{-4}$~\cite{Ablikim:2013uvu}, which translates to
$f_{D^+}|V_{cd}|=(45.8\pm 1.3)$~MeV. This result, in conjunction
with the CKM matrix element $|V_{cd}|$ determined from a global SM
fit~\cite{Tanabashi:2018oca}, implies a value for the weak decay
constant $f_{D^+}=203.9\pm 5.6$~MeV. Alternatively, using this
result with an LQCD calculated value for $f_{D^+}$~($212.7\pm
0.6$~MeV)~\cite{Bazavov:2017lyh}, one finds $|V_{cd}|=0.2151\pm
0.0060$. In either scenario, these are the most precise results
for these quantities to date.

BESIII reported a measurement of the absolute decay branching
fractions $\BR (D_s^+\to \mu^+\nu)$ with 0.39 million tagged
$D_s^+$ mesons produced via $\EE\to D_s^+D_s^{*-}$ at $\ECM =
4.178$~GeV~\cite{Ablikim:2018jun}. In this report, the inclusion
of charged conjugate states is always implied; for example, $D_s^+
D_s^{*-}$ indicates both $D_s^+ D_s^{*-}$ and $D_s^{*+} D_s^{-}$.
The $1136\pm 33$ $D_s^\pm\to \mu^\pm\nu$ signal events (see
Fig.~\ref{fdfds}) were used to determine $\BR(D_s^+\to\mu^+\nu)
=(5.49\pm 0.22)\times 10^{-3}$, which corresponds to
$f_{D_s^+}|V_{cs}|=246.2\pm 5.0$~MeV. If $|V_{cs}|$ is fixed at
its latest global SM fit value~\cite{Tanabashi:2018oca}, the
$D_s^+$ decay constant is determined to be $f_{D_s^+}=252.9\pm
5.1$~MeV. Alternatively, fixing $f_{D_s^+}$ at its value from LQCD
calculations ($249.9\pm
0.4$~MeV)~\cite{Carrasco:2014poa,Bazavov:2017lyh}, yields
$|V_{cs}| = 0.985\pm 0.020$. Either of these measurements qualify
as the currently most precise value.

In principle, $D_q^+\to e^+\nu$ and $\tau^+\nu$ can also be
measured to obtain the same decay constant-CKM matrix element
product. According to Eq.~(\ref{fdq}), the expected relative decay
widths for the $\tau^+\nu$, $\mu^+\nu$, and $e^+\nu$ modes are
$2.67:1:2.35\times 10^{-5}$ for $D^+$ and $9.75:1:2.35\times
10^{-5}$ for $D_s^+$; the SM $D_q^+\to e^+\nu$ decay widths are
highly suppressed by helicity conservation, and $2\sim 3$~orders
of magnitude below the sensitivity of BESIII. Although the SM
$D_q^+\to \tau^+\nu$ decay widths are larger than those for
$D_q^+\to \mu^+\nu$, the presence of additional final-state
neutrinos from the $\tau$ decay voids the kinematic constraint
(the inference of the four-momentum of a single missing neutrino
from measurements of all the accompanying particles in the event)
that is available for $\mu^+\nu$ and this results in more
background. Nevertheless, $D_s^+\to\tau^+\nu$ decay is currently
being studied at BESIII with an expected result that will have a
precision comparable to that achieved for $D_s^+\to \mu^+\nu$.
This result should improve the accuracy of the $|V_{cs}|f_{D_s^+}$
measurement, and can also be used to test LFU.

\subsection{Semileptonic decays}

Semileptonic decay rates, in conjunction with form factors
determined from LQCD calculations, provide independent
measurements of the $|V_{cs}|$ and $|V_{cd}|$ CKM matrix elements.
The first high-precision, semi-leptonic decay rate measurements
were made by CLEO-c based on data sets accumulated at the
$\psi(3770)$ resonance peak~\cite{Besson:2009uv}. BESIII, with
triple the amount of $\psi(3770)$ data, significantly improved the
measurements of these quantities, thereby keeping pace with
improvements in the relevant LQCD results.

The most relevant measurements are for the $D^0\to K^- \ell^+\nu$
and $\pi^-\ell^+\nu$, and $D^+\to \bar{K}^0 \ell^+\nu$ and $\pi^0
\ell^+\nu$ decay channels, with $\ell=e$ or $\mu$ and $K$ standing
for kaons. With 2.8 million tagged $D^0$ mesons, $70727\pm 278$
$D^0\to K^- e^+\nu$ and $6297\pm 87$ $D^0\to \pi^- e^+\nu$ signal
events are observed, and absolute decay branching fractions are
determined to be $\BR(D^0 \to K^-e^+\nu)=(3.505\pm 0.035)\%$ and
$\BR(D^0 \to \pi^-e^+\nu)=(0.295\pm
0.005)\%$~\cite{Ablikim:2015ixa}. From differential decay rate
measurements, $d\Gamma/dq^2$, where $q^2=M^2(e^+\nu)$ is the
square of the four-momentum transfer between initial state $D^0$
and final state $K^-$ or $\pi^-$, the hadronic form factors at
$q^2=0$ times CKM matrix element products
$f_{+}^K(0)|V_{cs}|=0.7172\pm 0.0043$ and
$f_{+}^{\pi}(0)|V_{cd}|=0.1435\pm 0.0020$ are obtained. Since the
precision of current LQCD form factor calculations is not very
high, these products are combined with values of $|V_{cs}|$ and
$|V_{cd}|$ from a SM-constrained fit~\cite{Tanabashi:2018oca}, to
extract the hadronic form factors: $f^K_+(0) = 0.7367\pm 0.0044$
and $f^\pi_+(0) = 0.6395\pm 0.0090$. Their measured ratio,
$f_+^{\pi}(0)/f_+^{K}(0)=0.868\pm 0.013$, is in good agreement
with a light cone sum rule value $0.84\pm
0.04$~\cite{Wang:2002zba}, but with a smaller uncertainty.

The branching fractions of $D^0\to K^-\mu^+\nu$ ($47100\pm 259$
observed events) and $\pi^-\mu^+\nu$ ($2265\pm 63$ observed
events) are measured to be $(3.413\pm
0.040)\%$~\cite{Ablikim:2018evp} and $(0.272 \pm
0.010)\%$~\cite{Ablikim:2018frk}, respectively, with significantly
improved precision compared with previous measurements. With
$|V_{cs}|$ taken from a SM constrained
fit~\cite{Tanabashi:2018oca}, $f_{+}^{K}(0)=0.7327\pm 0.0049$ is
obtained, in good agreement with the form factor measured in
electronic mode with comparable precision.

For the charged $D$ meson measurements, 1.7 million $D^+D^-$ pairs
are tagged, and the absolute decay branching fractions $\BR(D^+
\to \bar{K}^0 e^+ \nu)=(8.60\pm 0.16)\%$ and $\BR(D^+ \to \pi^0
e^+ \nu)=(0.363\pm 0.009)\%$ are determined based on $26008\pm
168$ and $3402\pm 70$ observed $D^+ \to \bar{K}^0 e^+ \nu$ and
$\pi^0 e^+ \nu$ events, respectively~\cite{Ablikim:2017lks}. These
results correspond to $f^K_+(0) = 0.725\pm 0.013$ and
$f^{\pi}_+(0) = 0.622\pm 0.012$, in agreement with the
measurements using neutral $D$ decays with slightly worse
precision.

The branching fractions of $D^+\to \bar{K}^0\mu^+\nu$ ($16516\pm
130$ observed events) and $\pi^0\mu^+\nu$ ($1335\pm 42$ observed
events) are measured to be $(8.72\pm
0.19)\%$~\cite{Ablikim:2016sqt} and $(0.350 \pm
0.015)\%$~\cite{Ablikim:2018frk}, respectively. The former
measurement is ten times more precise than previous results and
the latter is a first measurement.

Since the same hadronic form factors occur in $\pi/K e^+ \nu$ and
$\pi/K \mu^+ \nu$ decays, they cancel in the ratio of branching
fractions allowing for a model-independent test of LFU.
Tantalizing indications of possible LFU violations have been
reported for semileptonic $B$ decays, and these have sparked an
interest in more stringent tests in the charm sector, which, if
nothing else, could provide useful constraints on models proposed
as explanations for the $B$ decay anomalies~\cite{Bifani:2018zmi}.

The results listed above correspond to branching fraction ratios
of
 ${\mathcal R}_\pi^0\equiv \frac{\BR(D^0\to \pi^-\mu^+\nu)}
 {\BR(D^0\to \pi^-e^+\nu)}=0.922\pm 0.037$,
 and
 ${\mathcal R}_\pi^+\equiv \frac{\BR(D^+\to \pi^0\mu^+\nu)}
 {\BR(D^+\to \pi^0e^+\nu)}=0.964\pm 0.045$~\cite{Ablikim:2018frk}.
These are compatible with LFU-based theoretical expectations:
 ${\mathcal R}_\pi=0.985\pm 0.002$~\cite{Riggio:2017zwh,Soni:2017eug},
within $1.7$ and $0.5$ standard deviations, respectively.
Likewise,
 ${\mathcal R}_K^0  \equiv \frac{\BR(D^0\to   K^-\mu^+\nu)}
 {\BR(D^0\to   K^-e^+\nu)}=0.974\pm 0.014$~\cite{Ablikim:2018evp},
 and
 ${\mathcal R}_K^+  \equiv \frac{\BR(D^+\to \bar{K}^0\mu^+\nu)}
 {\BR(D^+\to \bar{K}^0e^+\nu)}=1.014\pm 0.017$,
are in agreement with the LFU expected value of ${\mathcal
R}_K=0.975\pm 0.001$~\cite{Riggio:2017zwh,Soni:2017eug}, within
$0.1$ and $2.3$ standard deviations, respectively. These tests are
summarized in Table~\ref{tab:charmbrs}. A study of the ratios of
differential branching fractions for different four-momentum
transfer regions was also
performed~\cite{Ablikim:2018evp,Ablikim:2018frk}, and no evidence
for LFU violation was found.

\begin{table}[htbp]
\caption{Charm decay branching fractions from BESIII experiment,
where the first errors are statistical and the second ones
systematic.}
    \label{tab:charmbrs}
    \centering
    \begin{tabular}{lccc}
    \hline\hline
      Mode   &  Number of signals & Branching fraction &  Physics implication \\\hline
      $D^+\to \mu^+\nu$    &  $409\pm 21$ & $(3.71\pm 0.19\pm 0.06)\times 10^{-4}$ & $f_{D^+}|V_{cd}|=45.75\pm 1.20\pm 0.39$~MeV \\
      $D_s^+\to \mu^+\nu$  & $1136\pm 33$ & $(5.49\pm 0.16\pm 0.15)\times 10^{-3}$ & $f_{D_s^+}|V_{cs}|=246.2\pm 3.6\pm 3.6$~MeV \\\hline
      $D^0\to K^-  e^+\nu$    &  $70727\pm 278$ & $(3.505\pm 0.014\pm 0.033)\%$ &  \\
      $D^0\to K^-\mu^+\nu$    &  $47100\pm 259$ & $(3.413\pm 0.019\pm 0.035)\%$ & \raisebox{1.5ex}[0pt]{$R^0_K=0.974\pm 0.007\pm 0.012$} \\
      \hline
      $D^0\to \pi^-  e^+\nu$  &  $ 6297\pm  87$ & $(0.295\pm 0.004\pm 0.003)\%$ &  \\
      $D^0\to \pi^-\mu^+\nu$  &  $ 2265\pm  63$ & $(0.272\pm 0.008\pm 0.006)\%$ & \raisebox{1.5ex}[0pt]{$R^0_{\pi}=0.922\pm 0.030\pm 0.022$} \\
      \hline
      $D^+\to \bar{K}^0  e^+\nu$ &  $26008\pm 168$ & $(8.60\pm 0.06\pm 0.15)\%$ &  \\
      $D^+\to \bar{K}^0\mu^+\nu$ &  $16516\pm 130$ & $(8.72\pm 0.07\pm 0.18)\%$ & \raisebox{1.5ex}[0pt]{$R^+_K=1.014\pm 0.011\pm 0.013$} \\
      \hline
      $D^+\to \pi^0  e^+\nu$     &  $ 3402\pm  70$ & $(0.363\pm 0.008\pm 0.005)\%$ &  \\
      $D^+\to \pi^0\mu^+\nu$     &  $ 1335\pm  42$ & $(0.350\pm 0.011\pm 0.010)\%$ & \raisebox{1.5ex}[0pt]{$R^+_{\pi}=0.964\pm 0.037\pm 0.026$} \\
    \hline\hline
    \end{tabular}
\end{table}

Although BESIII also measured many other semi-leptonic decays of
$D^0$, $D^+$, and $D_s^+$ into final states with a light scalar
hadron~\cite{Ablikim:2015mjo,Ablikim:2018ffp,Ablikim:2018lmn,Ablikim:2018qzz,Ablikim:2018upe},
a light vector
meson~\cite{Ablikim:2015mjo,Ablikim:2015gyp,Ablikim:2017omq,Ablikim:2018lmn,Ablikim:2018qzz,Ablikim:2018upe},
and other light pseudoscalar
particles~\cite{Ablikim:2015qgt,Ablikim:2016rqq,Ablikim:2017omq},
many of these were first measurements that lack sufficient
precision to extract meaningful information on the CKM matrix
elements or form factors. For many of these modes, there are still
no LQCD form-factor calculations for comparison.

\subsection{\boldmath Strong phases in hadronic $D$-meson decays}

Violations of $CP$ invariance in the SM are  characterized by the
$\alpha$, $\beta$, and $\gamma$ internal angles of the CKM
unitarity triangle (see below) that are determined from various
$B$ decay processes~\cite{Chau:1984fp}. The SM does not specify
the values of these angles, but strictly requires that $\alpha +
\beta + \gamma=180^{\circ}$ in order to satisfy unitarity. The sum
of the current experimental values is $(180.6\pm 6.8)^{\circ}$,
with precision limited by the $\alpha$ and $\gamma$ determinations
that are both known to within $\sim$$\pm
5^{\circ}$~\cite{Amhis:2018udz}. Improved measurements of these
two angles are important.

An improved measurement of the angle $\gamma$ is also needed
because it is the only $CP$-violating angle that can be measured
in processes that are not mediated by quantum loop diagrams and,
thus, not susceptible to the influence of virtual heavy particles
that might be associated with new physics. Thus, $\gamma$ is a
benchmark reference point for the SM $CP$-violation ($CPV$)
mechanism in searches for non-SM sources of $CPV$, and refined
measurements of it are top priorities for the upgraded
LHCb~\cite{Alves:2008zz} and the Belle~II~\cite{Kou:2018nap}
experiments.

Precision determinations of $\gamma$ rely on the measurements of
the interference effects between the $B\to D^{(*)}K^{(*)}$ and
${\bar D}^{(*)}K^{(*)}$ decay amplitudes, and these require as
input a precise knowledge of the ($CP$-conserving)
strong-interaction phase differences between the Cabibbo-favoured
(transitions between quarks in the same generation which are
dominant) and doubly Cabibbo-suppressed (transitions between
quarks in different generations which are strongly suppressed)
amplitudes for the decays of quantum-correlated $D^0$-$\bar{D}^0$
meson systems. BESIII measurements with $D^0\bar{D}^0$ meson pairs
produced at the $\psi(3770)$ resonance peak are uniquely well
suited for determining these strong-interaction phase
differences~\cite{Asner:2008nq}.

Recent determinations of $\gamma$ from measurements of $B\to
D^{(*)}K^{(*)}$ with $D\to K^0_S\pi^+\pi^-$, which is the most
promising channel for future high-precision results, use measured
values of the strong-interaction phase differences determined by
the CLEO-c experiment~\cite{Libby:2010nu}. The contribution to the
final uncertainty from uncertainties of the strong phases'
measured values is $\approx \pm 2^{\circ}$~\cite{Aaij:2018uns}.
The current BESIII $\psi(3770)$ data set is three and a half times
larger than that of CLEO-c and, when fully analyzed, is expected
to contribute an uncertainty on $\gamma$ that is of order
$1^\circ$~\cite{Malde:2223391}. This precision should be adequate
for the LHCb Run-2 measurement that has an expected sensitivity of
$4^\circ$ and the Belle~II measurement that will ultimately have
an expected precision of $1.5^\circ$~\cite{Kou:2018nap}.

\subsection{\boldmath Measurement of absolute $\Lambda^+_c$ decay branching fractions}

Although the $\Lambda^+_c$ baryon was discovered forty years
ago~\cite{Abrams:1979iu}, its decay properties are still poorly
understood. The first model-independent measurement of a
$\Lambda^+_c$ decay branching fraction was a 2014 Belle
result\cite{Zupanc:2013iki} $\BR(\Lambda^+_c\to
pK^-\pi^+)=(6.84\pm 0.36)$\%. In 2014, BEPCII operated at
$\ECM=4.60$~GeV, which is 27~MeV above threshold for $\EE\to
\Lambda^+_c\bar\Lambda^-_c$ production. At this energy, when a
$\Lambda^{\mp}_c$ is reconstructed, the accompanying
$\Lambda^{\pm}_c$ is tagged. With a sample of about fourteen
thousand tagged $\Lambda^+_c\bar\Lambda^-_c$ events, BESIII
reported the first model-independent branching fraction
measurements for the semi-leptonic modes $\Lambda^+_c\to\Lambda
e^+\nu$~\cite{Ablikim:2015prg} and
$\Lambda\mu^+\nu$~\cite{Ablikim:2016vqd}, and for twelve different
hadronic $\Lambda_c$ decay modes, including $\BR(\Lambda^+_c\to
pK^-\pi^+)=(5.84\pm 0.35)$\%~\cite{Ablikim:2015flg}, which is two
standard deviations below the Belle value.

The BESIII $\Lambda_c$ measurements were carried out at
$\ECM=4.60$~GeV, the maximum BEPCII energy, where $\sigma(\EE\to
\Lambda^+_c\bar\Lambda^-_c)= 0.24\pm
0.02$~nb~\cite{Ablikim:2017lct}. The Belle group reported that
this cross-section has a $0.44\pm 0.18$~nb peak at $\ECM=4.64\sim
4.66$~GeV~\cite{Pakhlova:2008vn}, which is likely a
$\Lambda^+_c\bar\Lambda^-_c$ decay mode of the $Y(4660)$, a
charmonium-like $\pp\psip$ resonance found by
Belle~\cite{Wang:2007ea} with mass $4643\pm
9$~MeV/$c^2$~\cite{Tanabashi:2018oca}. A BEPCII upgrade that is
currently underway will increase its maximum energy to 4.9~GeV
that will easily cover the $Y(4660)$ peak region, and future
BESIII $\Lambda_c$ measurements will be done at an energy
corresponding to the peak cross-section value with twice the event
rate.

\section{Discovery and study of charmonium-like states}

All studies of $\xyz$ mesons at $\EE$ $B$-factories suffer from
low event rates and limited statistical precision. In contrast,
BESIII can tune the $\EE$ CM energy to match the peaks of the
vector charmonium-like resonances (the $Y$ mesons), where event
rates are high enough to facilitate precise, detailed measurements
of their resonance parameters and also search for new states among
their decay products.

\subsection{\boldmath Discovery of $\zc$ and $\zcp$}

\begin{figure*}[htbp]
\centering
 \includegraphics[height=5cm]{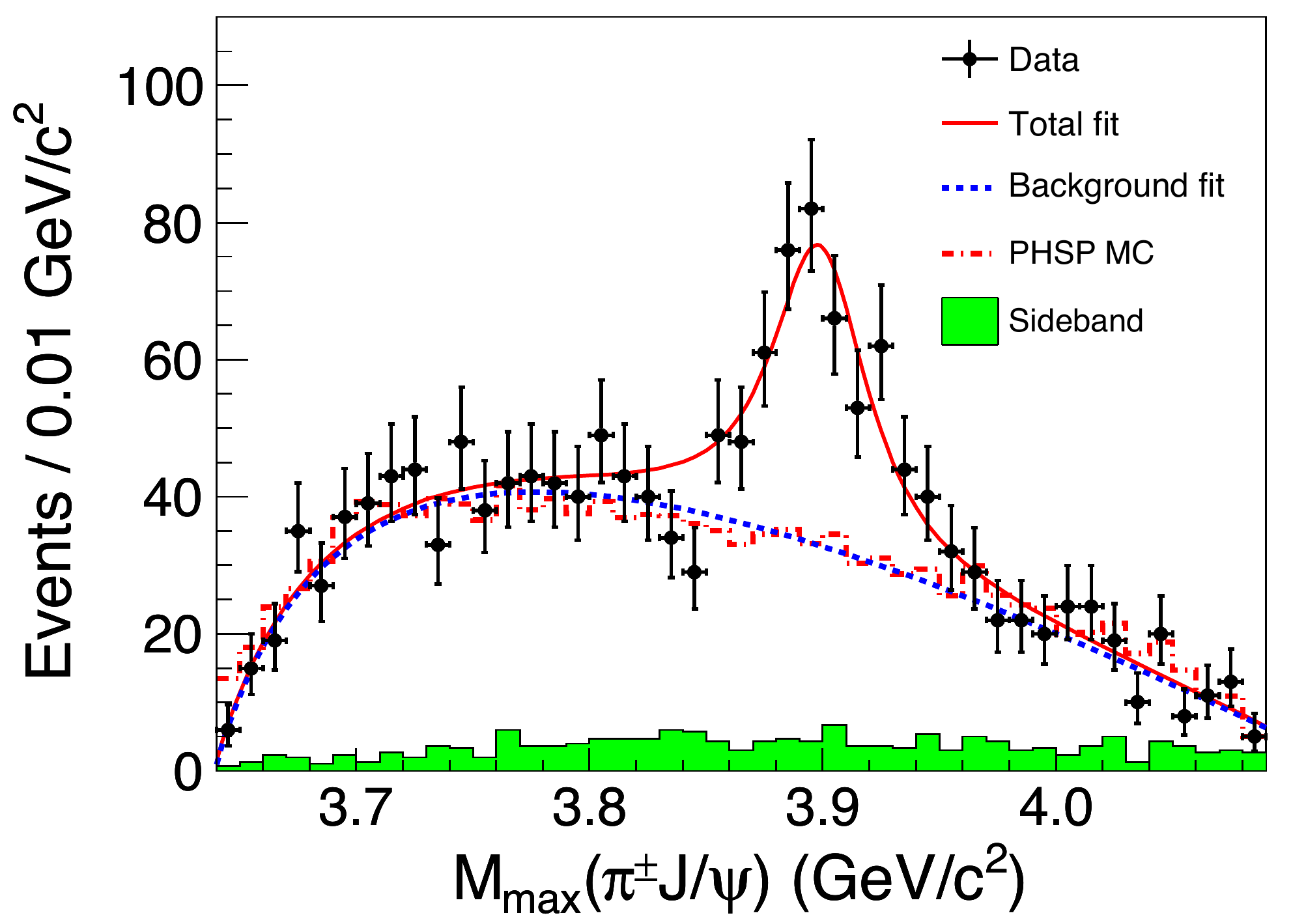}
 \includegraphics[height=5cm]{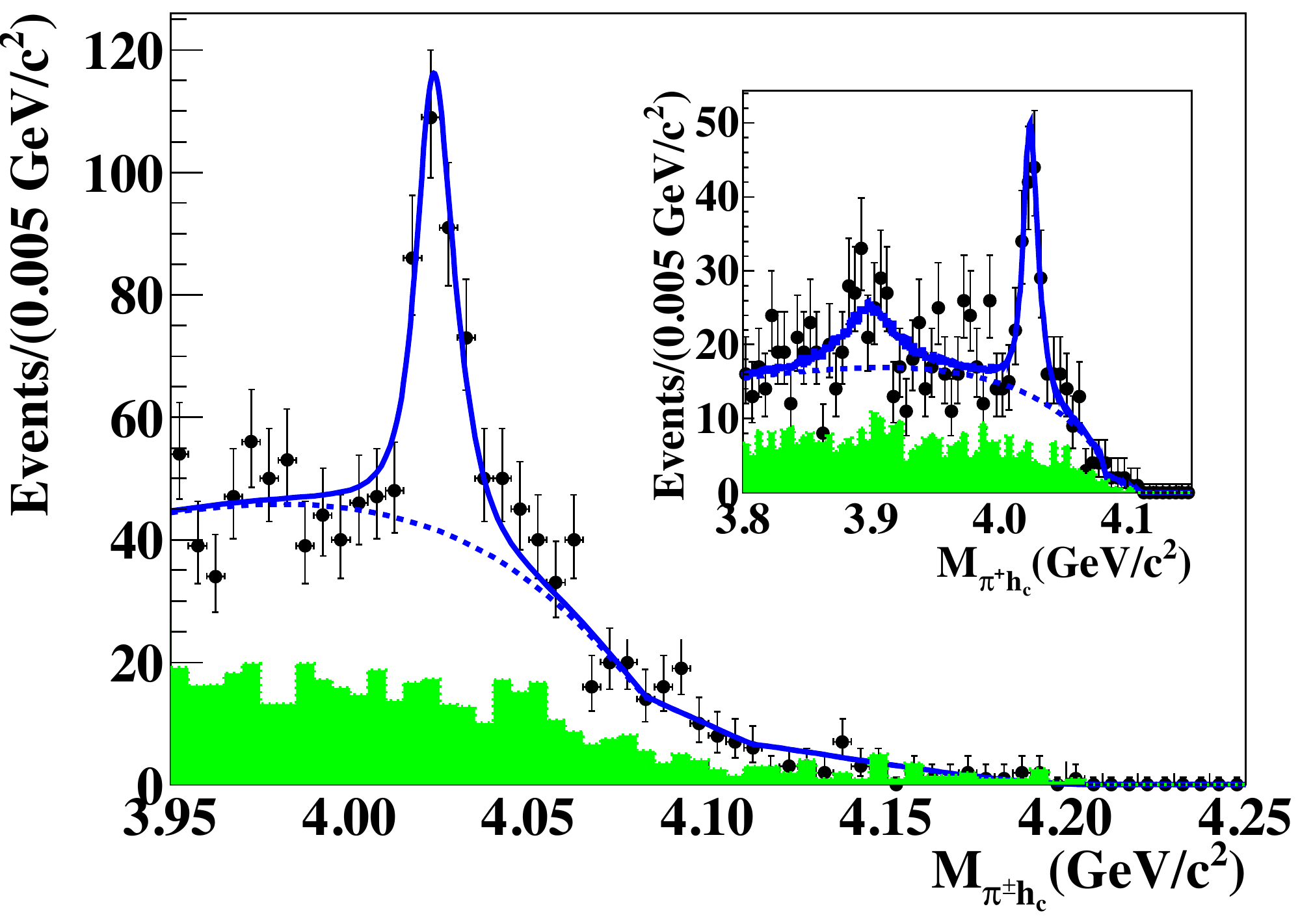}\\
 \includegraphics[height=5cm]{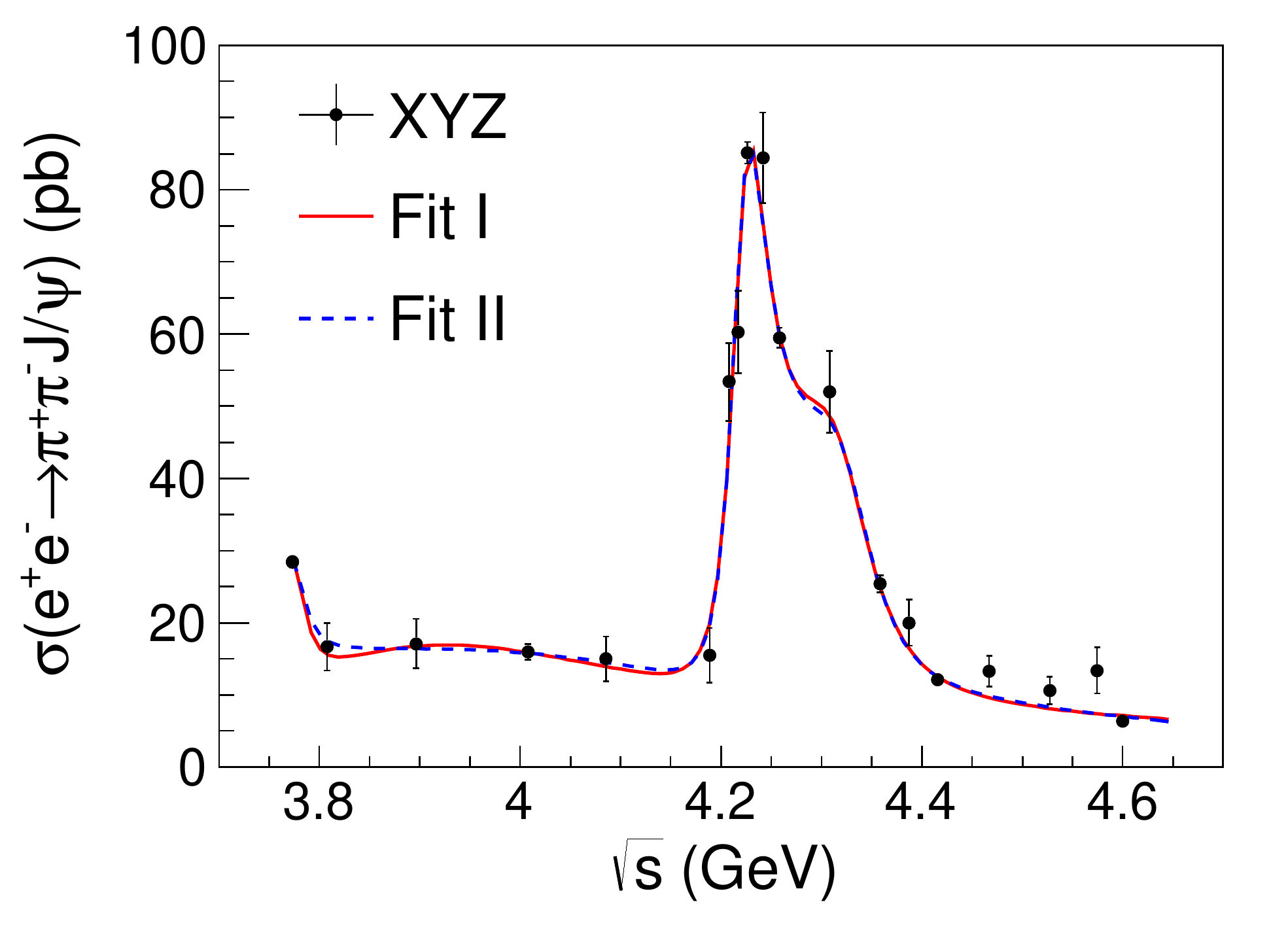}
 \includegraphics[height=5cm]{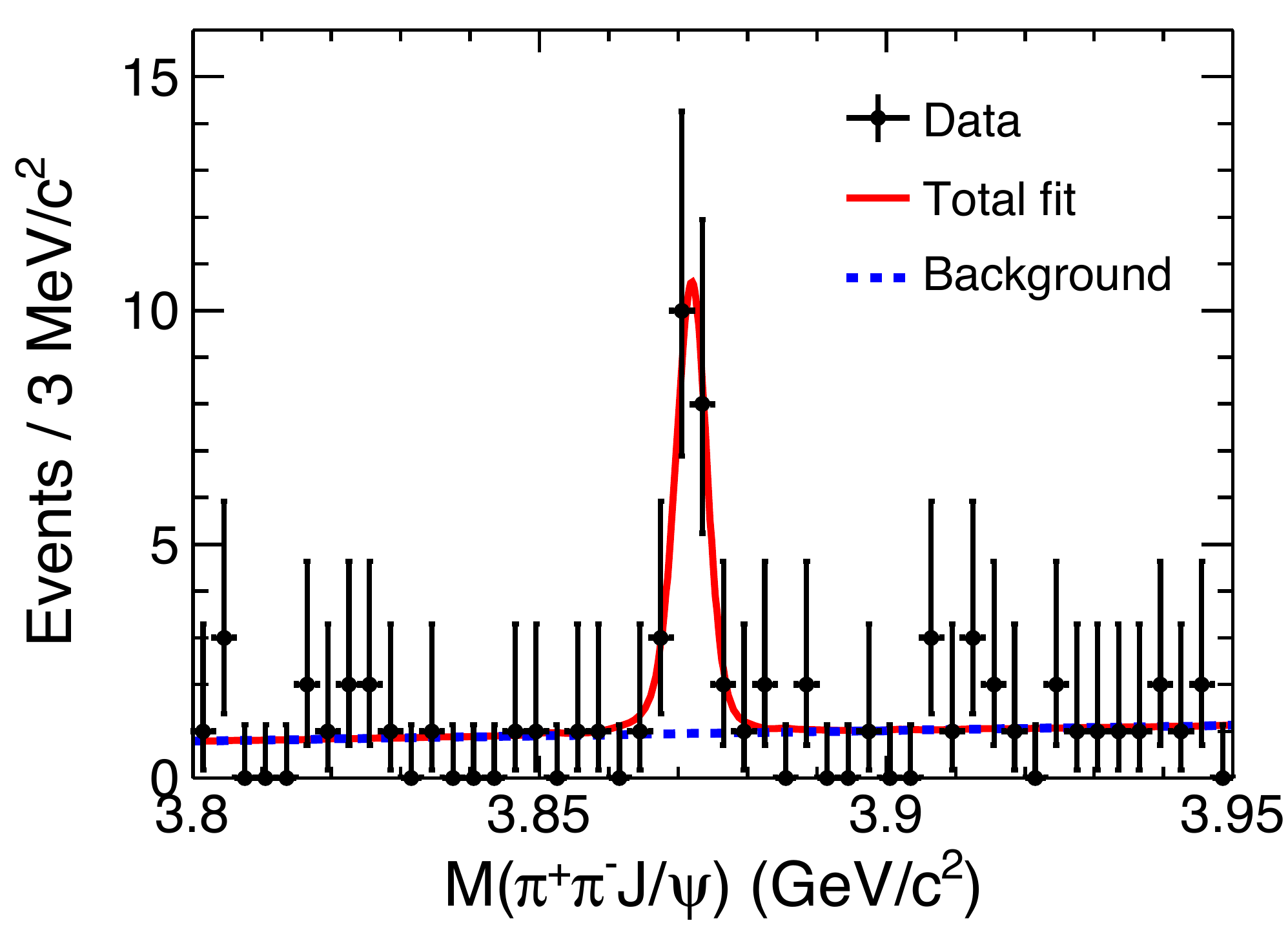}
 \caption{The $\zc$ signal observed in $\EE\to \pp\jpsi$~\cite{Ablikim:2013mio} (top left),
the $\zcp$ signal observed in $\EE\to \pphc$~\cite{Ablikim:2013wzq} (top right), the $\y$ signal
and the fine structure in $\EE\to \ppjpsi$ line shape~\cite{Ablikim:2016qzw} (bottom left), and the
$\x$ signal observed in $\EE\to \gamma \ppjpsi$~\cite{Ablikim:2013dyn} (bottom right). The inset in
the top right panel shows the sum of the data at 4.23 and 4.26~GeV with a hint of
the $\zc$ signal. }\label{plot_xyz}
\end{figure*}

The $\zc$ hadron was first seen by BESIII as a prominent peak (see
Fig.~\ref{plot_xyz}) in the $\pi^\pm\jpsi$ invariant mass spectrum
in a sample of 1.5K $\EE\to \ppjpsi$ events collected at
$\ECM=4.26$~GeV, which is near the peak of the $\y$
resonance~\cite{Ablikim:2013mio}. A fit to this signal with a
Breit-Wigner function found its mass and width to be $M=(3899.0\pm
6.1)~{\rm MeV}/c^2$ and $\Gamma=(46 \pm 22)$~MeV, with a
statistical significance greater than 8 standard deviations. The
state was named $\zc$ following the tradition that uses $Z$ to
designate a charged quarkonium-like state and a subscript $c$ to
indicate it contains charm quarks. The $\zc$ was the first charged
charmonium-like state to be confirmed by other
experiments~\cite{Liu:2013dau,Xiao:2013iha} and is a strong
candidate for a four-quark meson ($Z^+_c =c\bar{c}u\bar{d};\ Z^-_c
=c\bar{c}d\bar{u})$, where the $\ccb$ pair is needed to account
for its decay into a $\jpsi$ charmonium state and the $u$ and $d$
quarks are needed to account for its non-zero electrical charge.

Subsequently, with $5964\pm 83$ $\ppjpsi$ events collected at
$\ECM=4.26$ and $4.23$~GeV (where the $\EE\to \ppjpsi$
cross-section was found to be larger), a full amplitude analysis
of the $\ppjpsi$ system was performed and the spin-parity of the
$\zc$ was determined to be $J^P=1^+$, and its pole mass and width
to be $(3881 \pm 53)$~MeV/$c^2$ and $(52 \pm 36)$~MeV,
respectively~\cite{Collaboration:2017njt}. The shift in mass and
increase in the uncertainties reflect the effects of interference
between the $\zc$ and other amplitudes that were not considered in
the initial analysis. The $\zc$ mass is 5~MeV/$c^2$ above the
energy threshold for the production of $D\bar{D}^*$.

With $\EE\to \pi^\pm(D\bar{D}^*)^{\mp}$ events in the same data
set, BESIII observed a near-threshold $D\bar{D}^*$ mass peak with
$J^P=1^+$ that is consistent with $\zc^\pm\to (D\bar{D}^*)^\pm$
decay~\cite{Ablikim:2013xfr}. The pole mass and width determined
from a fit to the peak are $(3882.2\pm 1.9)$~MeV/$c^2$ and
$(26.5\pm 2.7)$~MeV, respectively, and in agreement with the
values from the $\pi\jpsi$
mode~\cite{Ablikim:2013xfr,Ablikim:2015swa}. The measured
branching fraction for the $\zc\rt D\bar{D}^*$ decay mode is
larger than that for the $\pi\jpsi$ mode by a factor of $6.2\pm
2.9$. In addition to the well established $\pi\jpsi$ and
$D\bar{D}^*$ modes, BESIII has reported evidence for
$\zc\to\rho\eta_c$~\cite{Yuan:2018inv,Ablikim:2019ipd} and
$\pi\hc$~\cite{Ablikim:2013wzq}, and an upper limit on $\zc\to
\pi\psip$~\cite{Ablikim:2017oaf}.

\begin{figure*}[htbp]
\centering
 \includegraphics[width=15cm]{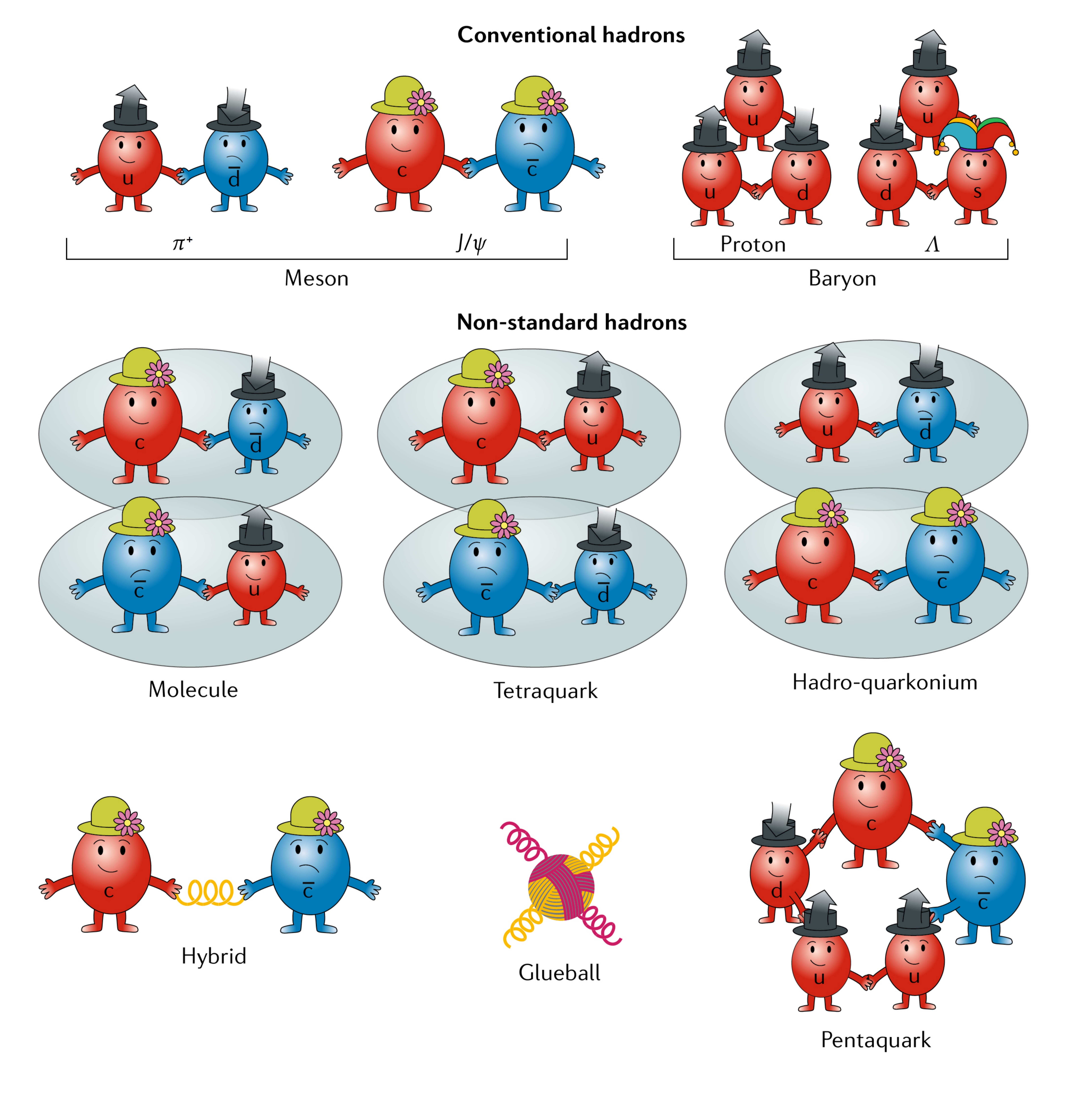}
\caption{Conventional and nonstandard hadrons. Conventional mesons
are composed of one quark (red characters) and one antiquark (blue
characters), conventional baryons are composed of three quarks.
The quarks can have different flavours such as up (u), down (d),
charm (c) or strange (s). Hadrons with other configurations are
referred to as nonstandard. Nonstandard hadrons include
hadron-hadron molecules, diquark-diantiquark tetraquark
mesons,hadro-quarkonia, quark-antiquark-gluon hybrids, multi-gluon
glueballs and pentaquark baryons. } \label{hadrons}
\end{figure*}

Since the $\zc$ mass is near that of the $\x$, a neutral
$\pp\jpsi$ resonance first observed by Belle in
2003~\cite{Choi:2003ue}, the two states have been interpreted as
isovector and isoscalar $D\bar{D}^*$ molecules loosely bound by
Yukawa-like nuclear forces~\cite{Tornqvist:2003na,Wang:2013cya}.
Another possibility is that they are QCD tetraquark states
comprising coloured diquarks and diantiquarks tightly bound by the
exchange of coloured gluons~\cite{Faccini:2013lda,Lebed:2016yvr}
(The quark and gluon configurations of different kind of hadrons
are depicted in Fig.~\ref{hadrons}).

Many of the models that were proposed for the $\zc$ predict the
existence of a similar state near the $D^*\bar{D}^*$ threshold.
Although there are no strong indications of a state near the
$D^*\bar{D}^*$ mass threshold in BESIII's $\EE\to\pp\jpsi$ or $\pi
D\bar{D}^*$ data samples, distinct signals for the predicted
state, the $\zcp$, were discovered in the $\EE\to \pp\hc$ and $\pi
D^*\bar{D}^*$ channels.

An analysis of $\EE\to \pp\hc$ events collected with $\ECM$ at and
near the $\y$ resonance peak found distinct evidence for a
nonstandard charmonium-like structure in the $\pi^\pm\hc$
invariant mass distribution as shown in
Fig.~\ref{plot_xyz}~\cite{Ablikim:2013wzq}. The mass and width of
this structure were determined to be $(4022.9 \pm 2.8)~{\rm
MeV}/c^2$ and $(7.9\pm 3.7)$~MeV, respectively, with a statistical
significance greater than $8.9$ standard deviations. This
discovery was only possible because of the very clean experimental
environment of $\EE$ collisions in the $\tau$-charm threshold
energy region uniquely facilitates the isolation of distinct $\hc$
meson signals. Neither the BaBar and Belle $B$-factory nor the
LHCb experiment has ever seen an $\hc$ signal.

BESIII also observed a strong $\zcp^\pm\to(D^{*}
\bar{D}^{*})^{\pm}$ signal in $\EE \to (D^{*} \bar{D}^{*})^{\pm}
\pi^\mp$ events at $\ECM=4.26$~GeV. The measured mass and width in
this channel are $(4026.3 \pm 4.5)$~MeV/$c^2$ and $(24.8\pm
9.5)$~MeV, respectively, and the statistical significance is $13$
standard deviations~\cite{Ablikim:2013emm}.

The $J^{P}=1^{+}$ $Z_c$ states found by BESIII bear an uncanny
resemblance to analogous $Z_b$ charged bottomonium-like $1^{+}$
states discovered by Belle~\cite{Belle:2011aa} in 2011. The $\zc$
and $\zcp$ are $\sim$5~MeV/$c^2$ above the $D\bar{D}^*$ and
$D^*\bar{D}^*$ thresholds, respectively, whereas the $Z_b(10610)$
and $Z_b(10650)$ are $\sim$3~MeV/$c^2$ above the respective
$B\bar{B}^*$ and $B^*\bar{B}^*$ thresholds (where $B$ denotes the
$B$ meson), and they all have similar widths.  These features are
similar to those expected for kinematic effects that can produce
peaks just above thresholds that look like resonances, but have
nothing to do with bona fide
mesons~\cite{Bugg:2011jr,Chen:2013coa,Swanson:2015bsa}. This
possibility was studied for the specific case of the $\zc$ in
Ref.~\cite{Guo:2014iya}, which concluded that the characteristics
of the narrow $\zc\rt D\bar{D}^*$ signal~\cite{Ablikim:2013xfr}
establishes the presence of a genuine meson-like pole in the
$D\bar{D}^*$ $S$-matrix. A more general discussion of this issue
is provided in Ref.~\cite{Szczepaniak:2015hya}.

In the QCD tetraquark and molecular pictures, the $\zc^{\pm}$ and
$\zcp^{\pm}$ states are the $I_3=\pm 1$ members of isospin
triplets. BESIII confirmed this by observing their neutral,
isospin $I_3 = 0$ partners: the $\zc^0$, in both the
$\piz\jpsi$~\cite{Ablikim:2015tbp} and
$(D\bar{D}^*)^0$~\cite{Ablikim:2015gda} decay modes; and the
$\zcp^0$, in the $\piz\hc$~\cite{Ablikim:2014dxl} and
$(D^*\bar{D}^*)^0$~\cite{Ablikim:2015vvn} decay modes. These
observations establish the $\zc$ and $\zcp$ as isovector states
with even $G$-parity.

\subsection{\boldmath Improved understanding of the $\y$}

The $Y$ charmonium-like states are vector mesons with spin-parity
quantum numbers $J^{PC} = 1^{--}$, the same quantum numbers as the
photon. As a result, they can be produced directly in $\EE$
annihiations via a single virtual photon, that is $\EE\to Y$.
These states, which include the $\y$~\cite{Aubert:2005rm}, the
$Y(4360)$~\cite{Aubert:2007zz,Wang:2007ea}, and the
$Y(4660)$~\cite{Wang:2007ea}, have strong couplings to $\ccb$
charmonium final states, in contrast to conventional vector $\ccb$
charmonium states in the same energy region (such as $\psi(4040)$,
$\psi(4160)$, and $\psi(4415)$) that dominantly couple to pairs of
$D$ mesons~\cite{Tanabashi:2018oca}. These $Y$ states are
candidates for a variety of types of nonstandard hadrons
including:
molecules~\cite{Qiao:2005av,Ding:2008gr,Dai:2012pb,Wang:2013cya};
diquark-diantiquarks~\cite{Maiani:2005pe,Ali:2017wsf};
QCD-hybrids~\cite{Zhu:2005hp,Kou:2005gt}; and
hadrocharmonia~\cite{Dubynskiy:2008mq} (See Fig.~\ref{hadrons}).

A series of high-luminosity data sets taken with $\ECM$ spanning
the $\y$ mass region provided more precise measurements of the
$\ECM$-dependence of the cross-section $\sigma(\EE\to\y\to
\ppjpsi)$~\cite{Ablikim:2016qzw} than any of the seven previous
measurements that were based on samples of ISR events in higher
energy $\EE$ collision data~\cite{Aubert:2005rm,Yuan:2007sj}. With
an order-of-magnitude better statistical precision, the BESIII
measurements revealed a previously unnoticed composite structure
in the $\y$ resonance line-shape, evident in Fig.~\ref{plot_xyz},
that can be described by the overlap of a narrow peak with mass
$(4222.0\pm 3.4)$~MeV/$c^2$ and width $(44.1\pm 4.7)$~MeV, and a
wider one with mass $(4320\pm 13)$~MeV/$c^2$ and width $(101\pm
27)$~MeV. Compared to the previous 2016 world average values for
the $\y$~\cite{Patrignani:2016xqp}, the mass of 4222~MeV$/c^2$
resonance is about 30~MeV/$c^2$ lower and its width is nearly a
factor of 3 narrower.

BESIII also observed additional decay modes of the lower-mass $\y$
peak, including: $\y\to \pp\hc$~\cite{BESIII:2016adj};
$\omega\chi_{c0}$~\cite{Ablikim:2014qwy};
$\pp\psip$~\cite{Ablikim:2017oaf}; and the $\pi
D\bar{D}^*$~\cite{Ablikim:2018vxx} open-charm-meson mode.
Curiously, the $4320$~MeV/$c^2$ peak is not seen in any of these
additional channels. The resonant parameters of the $Y(4260)$
measured in different modes are listed in Table~\ref{tab:Y4260}.

\begin{table}[htbp]
\caption{Resonant parameters of the $Y(4260)$ from different modes
measured at BESIII. The cross sections measured at CM energy
4.226~GeV are also listed.}
    \label{tab:Y4260}
    \centering
    \begin{tabular}{cccc}
    \hline\hline
           Mode              &  Mass (GeV/$c^2$)              & Width (MeV)    & $\sigma$ at $\sqrt{s}=4.226$~GeV (pb) \\\hline
    $\EE\to \pp\jpsi$        & $4222.0\pm 3.1\pm 1.4$         & $44.1\pm 4.3\pm 2.0$         & $85.1\pm 1.5\pm 4.9$  \\
    $\EE\to \pp\hc$          & $4218.4^{+5.5}_{-4.5}\pm 0.9$  & $66.0^{+12.3}_{-8.3}\pm 0.4$ & $55.2\pm 2.6\pm 8.9$  \\
    $\EE\to \omega\chi_{c0}$ & $4218.5\pm 1.6\pm 4.0$         & $28.2\pm 3.9\pm 1.6$         & $55.4\pm 6.0\pm 5.9$  \\
    $\EE\to \pp\psp$         & $4209.5\pm 7.4\pm 1.4$         & $80.1\pm 24.6\pm 2.9$        & $21.3\pm 1.1\pm 1.6$  \\
    $\EE\to \pi^+ D^0 D^{*-}+c.c.$  & $4228.6\pm 4.1\pm 6.3$  & $77.0\pm 6.8\pm 6.3$         & $252\pm 5\pm 15$  \\
    \hline\hline
    \end{tabular}
\end{table}

The $\y$ has attracted a considerable amount of attention ever
since its discovery in 2005. The presence of nearby thresholds for
$D_s^{*+}D_s^{*-}$, $D\bar{D}_1(2420)$, and $\omega\chi_{cJ}$
production, and its mass overlap with the $\psi(4160)$ and
$\psi(4415)$ conventional vector $\ccb$ charmonium states
complicate its interpretation. Whereas BESIII will continue to
supply more results on the $\y$ properties, sophisticated
theoretical treatments are likely needed to understand the nature
of this state.

\subsection{\boldmath Commonality between the $\x$, $\y$, and $\zc$}

With data taken with $\ECM$ at and near the $\y$ resonance peak,
BESIII discovered a clear signal for $\x$ production in
association with a $\gamma$-ray~\cite{Ablikim:2013dyn}, as shown
in Fig.~\ref{plot_xyz}. The $\ECM$-dependence of the $\EE\to
\gamma\x$ cross-section is suggestive of a $\y\to \gamma\x$ decay
process, which indicates that there might be some common features
to the  internal structures of the $\zc$, $\y$, and $\x$.

All the final state particles in $\EE\to \gamma\x$ followed by
$\x$ decays are recorded in the detector and there is no other
particles in an event. This makes it well suited for studies of
$\x$ decays to final states that include $\gamma$ rays and $\piz$
mesons. BESIII exploited this to make the first observation of the
$X(3872)\to\piz\chi_{c1}$ decay mode~\cite{Ablikim:2019soz}, a
process that would be strongly suppressed for a $\ccb$ state, but
allowed for a four-quark state~\cite{Dubynskiy:2007tj}. This
process was also used to determine the branching fraction for
$X(3872)\to \omega \jpsi$ with a two-fold improvement in precision
over previous measurements~\cite{Ablikim:2019zio}.

\section{\boldmath Study of light hadrons in $\jpsi$ decays}

\subsection{Scalar and tensor glueball searches with partial wave analyses}

An intriguing QCD prediction that is yet to be experimentally
confirmed is the existence of mesons comprised only of gluons,
with no quark content, and commonly referred to as
glueballs~\cite{Jaffe:1975fd}. Glueballs are electrically neutral,
zero isospin SU(3) singlets~\cite{Toki:1996si}. Two gluons form
scalar ($0^{++})$, pseudoscalar ($0^{-+}$), tensor ($2^{++}$)
glueballs and so on. Radiative $\jpsi$ decays proceed via the
$\jpsi\rt\gamma gg$ process and are expected to be prolific
sources of glueballs with mass below 3~GeV/$c^2$. BESIII has
performed systematic studies of radiative $\jpsi$ decays using
partial wave analysis (PWA) techniques to search for and
characterize glueball candidates.

The lowest mass scalar glueball is expected to have a mass between
1~and~2~GeV/$c^2$ and decay into meson-antimeson
pairs~\cite{Chen:2005mg}. BESIII found three scalar mesons in this
mass range in the radiative decay processes
$\jpsi\to\gamma\piz\piz$~\cite{Ablikim:2015umt},
$\gamma\ks\ks$~\cite{Ablikim:2018izx}, and $\gamma \eta\eta$,
where $\eta$ stands for the eta meson~\cite{Ablikim:2013hq}: the
$f_0(1370)$, $f_0(1500)$, and $f_0(1710)$. Two conventional
$q\bar{q}$ scalar mesons are needed to account for the $SU(3)$
octet and singlet members of the $1^3P_0$ meson nonet, leaving one
of them as a candidate for a scalar glueball. One interpretation
is that these three mesons mix and, thus, are the three orthogonal
mixtures of $n\bar{n}=(u\bar{u}+d\bar{d})/\sqrt{2}$, $s\bar{s}$,
and ${\mathcal G}$, that are the two $SU(3)$ $q\bar{q}$ states and
a glueball. Of the three candidates, the $f_0(1710)$ has the
highest production rate; the branching fraction for $\jpsi\to
\gamma f_0(1710)$ is nearly ten times higher than those for
$f_0(1370)$ and $f_0(1500)$, and compatible with an LQCD
calculated value for a glueball~\cite{Gui:2012gx}. This suggests
that the $f_0(1710)$ has the largest gluon component. Better
measurements of the couplings of these states to $\pp$,
$K\bar{K}$, $\eta\eta$, and $\eta\etap$ will supply additional
insight into the relative $n\bar{n}$, $s\bar{s}$, and ${\mathcal
G}$ content of these states.

The lowest lying tensor glueball is expected to have a mass above
2~GeV/$c^2$~\cite{Chen:2005mg}, and PWA of $\jpsi\to
\gamma\ks\ks$~\cite{Ablikim:2018izx},
$\gamma\eta\eta$~\cite{Ablikim:2013hq}, and
$\gamma\phi\phi$~\cite{Ablikim:2016hlu} revealed a tensor meson,
the $f_2(2340)$, that decays to each one of these channels.
However, its coupling to the $\pi\pi$ mode has not been
established~\cite{Ablikim:2015umt} and its total production rate
appears to be substantially lower than the LQCD calculated
value~\cite{Yang:2013xba}, which is of order 1\%. This may be
because there are a number of $f_2(2340)$ decay modes that have
not yet been identified, which is reasonable for a state with such
high mass. Thus, much more effort is needed to establish and
characterize the lowest tensor glueball.

A pseudoscalar ($0^{-+}$) glueball could also be produced in
radiative $\jpsi$ decays, but for these the dominant decay modes
would be to three pseudoscalars such as $\eta^{(\prime)}\pi\pi$,
$\eta^{(\prime)}K\bar{K}$, and $K\bar{K}\pi$. As noted below, the
identification of a pseudoscalar glueball is not easy.

\subsection{\boldmath Study of states close to the $\ppb$ mass threshold}

\begin{figure*}[htbp]
  \centering
  \includegraphics[height=8cm]{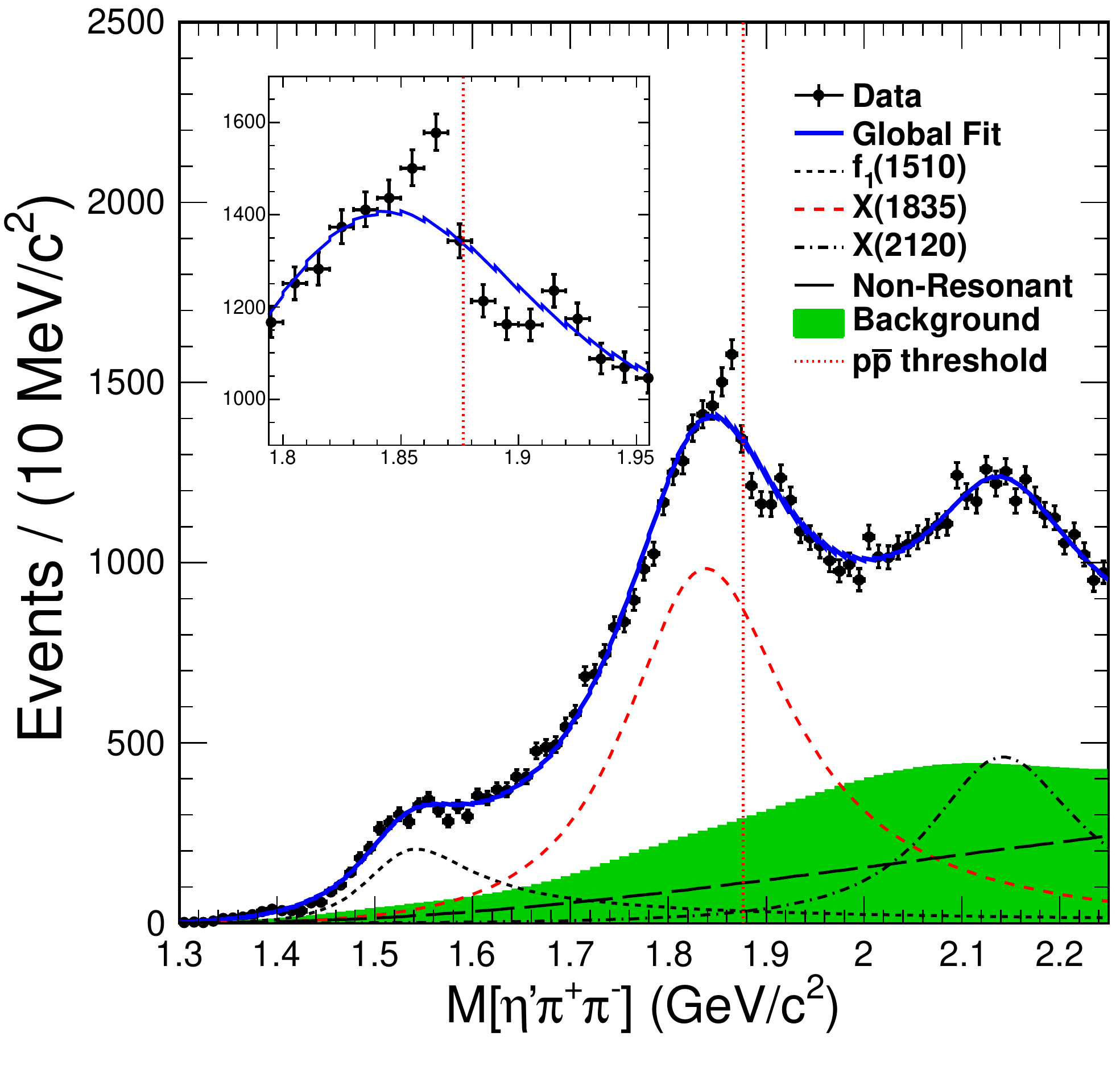}
\caption{The $\etap\pip\pim$ invariant mass spectra in $\jpsi\to
\gamma\etap\pp$ events~\cite{Ablikim:2016itz}. The dotted vertical line shows the
position of $\ppb$ mass threshold; the blue curve shows the
results for a fit that ignores the influence of the $\ppb$
threshold. The inset shows the detail around the $\ppb$ threshold.}
  \label{ppbar}
\end{figure*}

A distinct, narrow peak just at the $\ppb$ threshold was observed
in $\jpsi\to \gamma \ppb$ by BESII, the second phase of the BES
project, in a data set containing 58 million $\jpsi$
decays~\cite{Bai:2003sw}. This peak was subsequently confirmed and
its spin-parity was measured to be $0^{-+}$ by BESIII with 225
million $\jpsi$ events~\cite{BESIII:2011aa}.  BESII also observed
an $\etap\pp$ invariant mass peak in $\jpsi\to \gamma\etap\pp$
decays, the $X(1835)$~\cite{Ablikim:2005um}, that was suggested as
being due to an $\etap\pp$ decay mode of the same state as that
seen in $\ppb$~\cite{Ding:2005ew}. Subsequent BESIII studies of
the $X(1835)\to \etap\pp$ line shape with a 1.3 billion $\jpsi$
event sample revealed an anomalous structure in its line shape,
that is centered at the $\ppb$ threshold  (see Fig.~\ref{ppbar}),
which could be equally well described as the interference with a
new narrow resonance that has a mass nearly equal to $2m_p$ or a
wide resonance with an anomalously strong coupling to the $\ppb$
final state~\cite{Ablikim:2016itz}.

States in this mass region were also reported in
$\jpsi\to\omega\eta\pp$~\cite{Ablikim:2011pu} and $\gamma
3(\pp)$~\cite{Ablikim:2013spp}, and these also have line shape
distortions near the $\ppb$ mass threshold. However, no
near-threshold $\ppb$ structures are observed in $\jpsi$ decays to
$\omega\ppb$~\cite{Ablikim:2013cif} or
$\phi\ppb$~\cite{Ablikim:2015pkc}.

The difficulty in drawing solid conclusions about the states in
this mass region is that the extracted resonance parameters are
strongly model-dependent. They depend on the parameterization of
the resonant amplitudes, the way different amplitudes interfere
with each other, and how they are affected by the opening of
nucleon-antinucleon decay channels at the $2m_N$ thresholds, where
$m_N$ is the nucleon mass. As a result, the inferred masses and
widths of these states can vary over a wide range of values, with
correspondingly large uncertainties. In addition, ambiguities
inherent to the effects of interference preclude precise
production rate measurements.

Very elaborate, coupled-channel PWA of the different processes may
be the key for extracting reliable information from the data and
for understanding the nature of meson states in this mass
region~\cite{Battaglieri:2014gca}. This is an important future
direction for BESIII studies of light meson spectroscopy.

\subsection{\boldmath First observation of $a^0_0(980)\lrarrow f_0(980)$ mixing}

Five decades have passed since the
discovery~\cite{Armenteros:1965zz,Protopopescu:1973sh} of the
$a^{0,\pm}_0(980)$ and $f_0(980)$ scalar mesons (with
$J^{PC}=0^{++}$), but their nature remains controversial. These
two nearly-equal mass states, with different isospin and decay
modes, have defied attempts to classify them as conventional
$q\bar{q}$ mesons~\cite{Achasov:2017ozk}. They have long been
considered candidates for $K\bar{K}$
molecules~\cite{Weinstein:1990gu}, QCD
tetraquarks~\cite{Jaffe:1976ig,Black:1998wt,Oller:1998zr,Maiani:2004uc,Hooft:2008we},
or QCD hybrids~\cite{Ishida:1995km}. In 1979, measurements of
$a_0$-$f_0$ mixing were proposed as sensitive probes into the
nature of these states~\cite{Achasov:1979xc}.

Forty years later, BESIII made the first experimental observations
of this process~\cite{Ablikim:2018pik}. Using 1.3 billion $\jpsi$
and 0.45 billion $\psi(3686)$ events, BESIII detected distinct
signals for $\jpsi\to\phi f_0(980)$ events in which the $f_0$
mixes into an $a^0_0(980)$ that decays into $\eta\piz$, and
$\chi_{c1}\to \piz a^0_0(980)$ events where the $a^0_0$ mixes into
an $f_0(980)$ that decays to $\pp$. The statistical significances
are $7.2$ and $5.5$ standard deviations, respectively. The
extracted mixing intensities favour the QCD tetraquark scenario.

\section{\boldmath Baryon polarization in $\EE$ annihilation and $\jpsi$ decays}

Baryons produced directly via $\EE$ annihilation into
baryon-antibaryon ($B\bar{B}$) pairs or $\jpsi\to B\bar{B}$ decays
can be polarized transversely due to a non-zero relative phase
$\Delta\Phi$ between the two complex amplitudes that govern this
process~\cite{Dubnickova:1992ii,Faldt:2017kgy}. This polarization
was generally expected to be small and, thus, it was big surprise
when BESIII discovered that, in fact, the polarization of
$\Lambda$ (and $\bar{\Lambda}$) hyperons produced in
$\jpsi\to\LLb$ decays is substantial~\cite{Ablikim:2018zay}. A
non-zero polarization $\vec{P}_{\Lambda}$ enables separate
measurements of $\alpha_{-}$ and $\alpha_{+}$, the
parity-violating parameters that characterize the final state
$\pi^-$ and $\pi^+$ angular distributions for polarized
$\Lambda\to p\pi^-$ and $\bar{\Lambda}\to\bar{p}\pi^+$ decays:
$dn_{\pm}/d\cos\theta_{\pi^{\pm}}\propto
1-\alpha_{\pm}|\vec{P}_{\Lambda}|\cos\theta_{\pi^\pm}$, where
$\theta_{\pi^{\pm}}$ is the $\pi^{\pm}$ direction relative to
$\vec{P}_{\Lambda}$.
 Prior to the BESIII discovery, it was thought that testing
$CP$ with $\jpsi\to\LLb$ pairs produced via $\EE$ collisions would
require a polarized $e^-$ beam.

\begin{figure*}[htbp]
\centering
 \includegraphics[height=5cm]{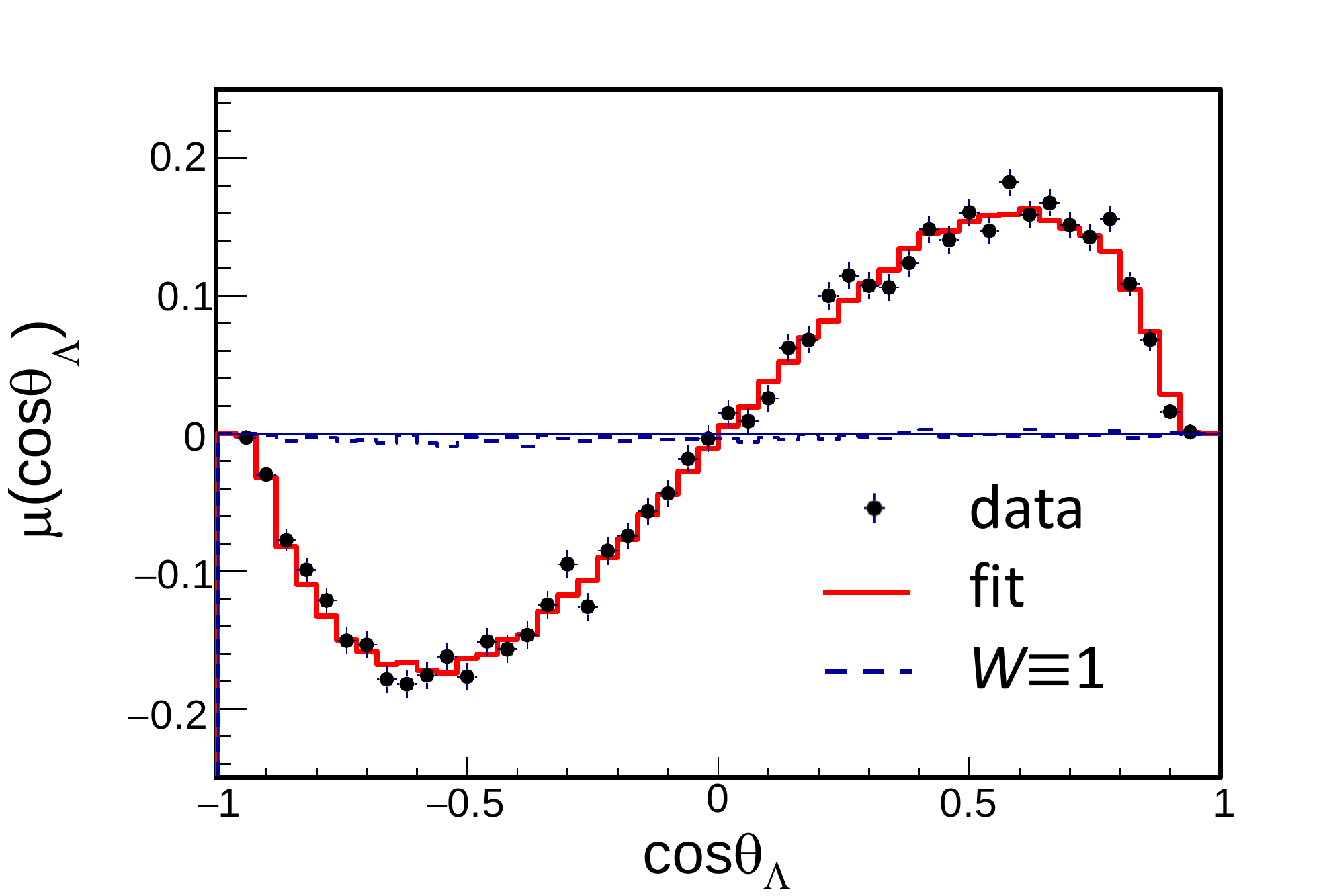}
 \includegraphics[height=5cm]{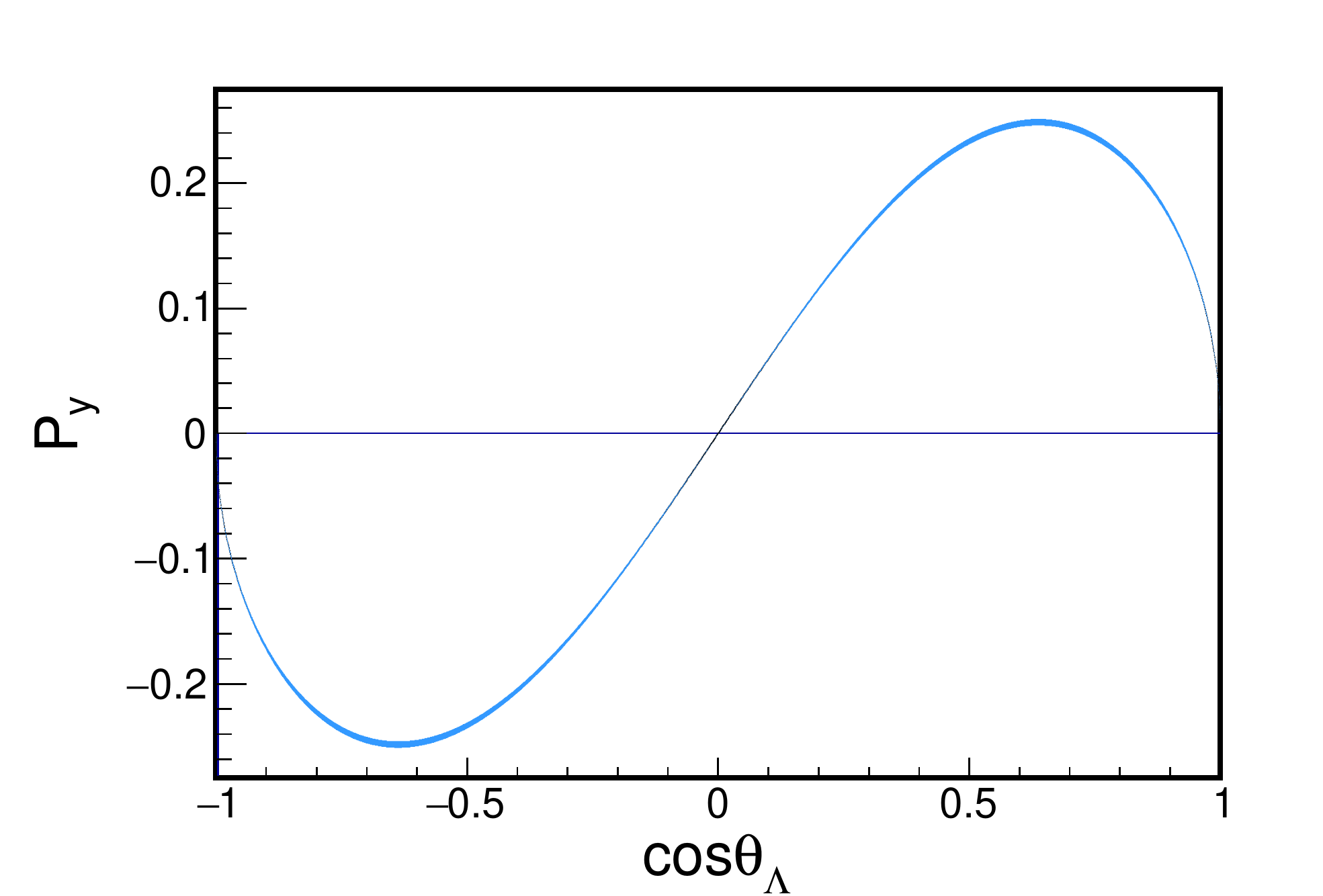} \\
 \includegraphics[height=5cm]{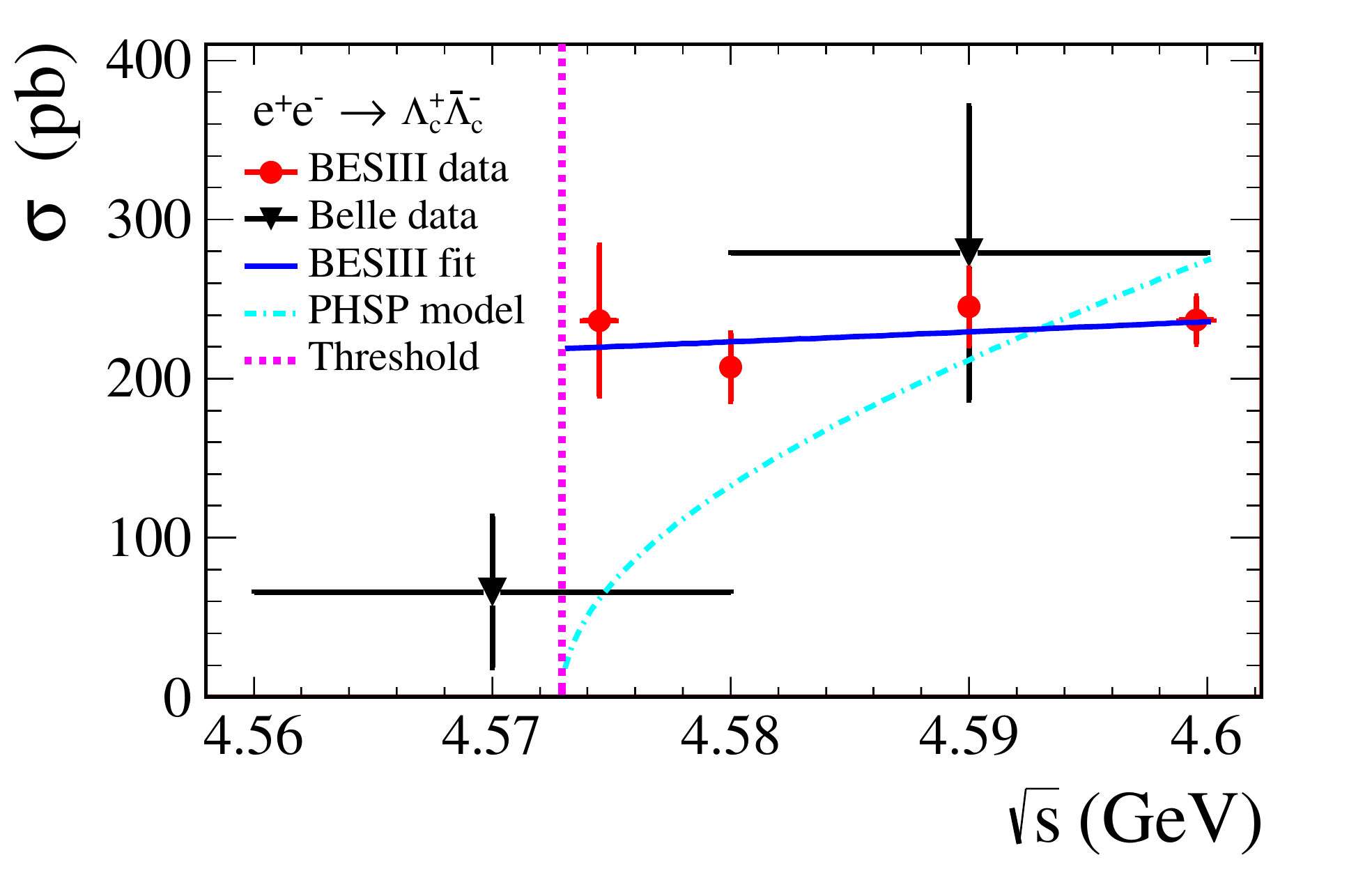}
 \includegraphics[height=5cm]{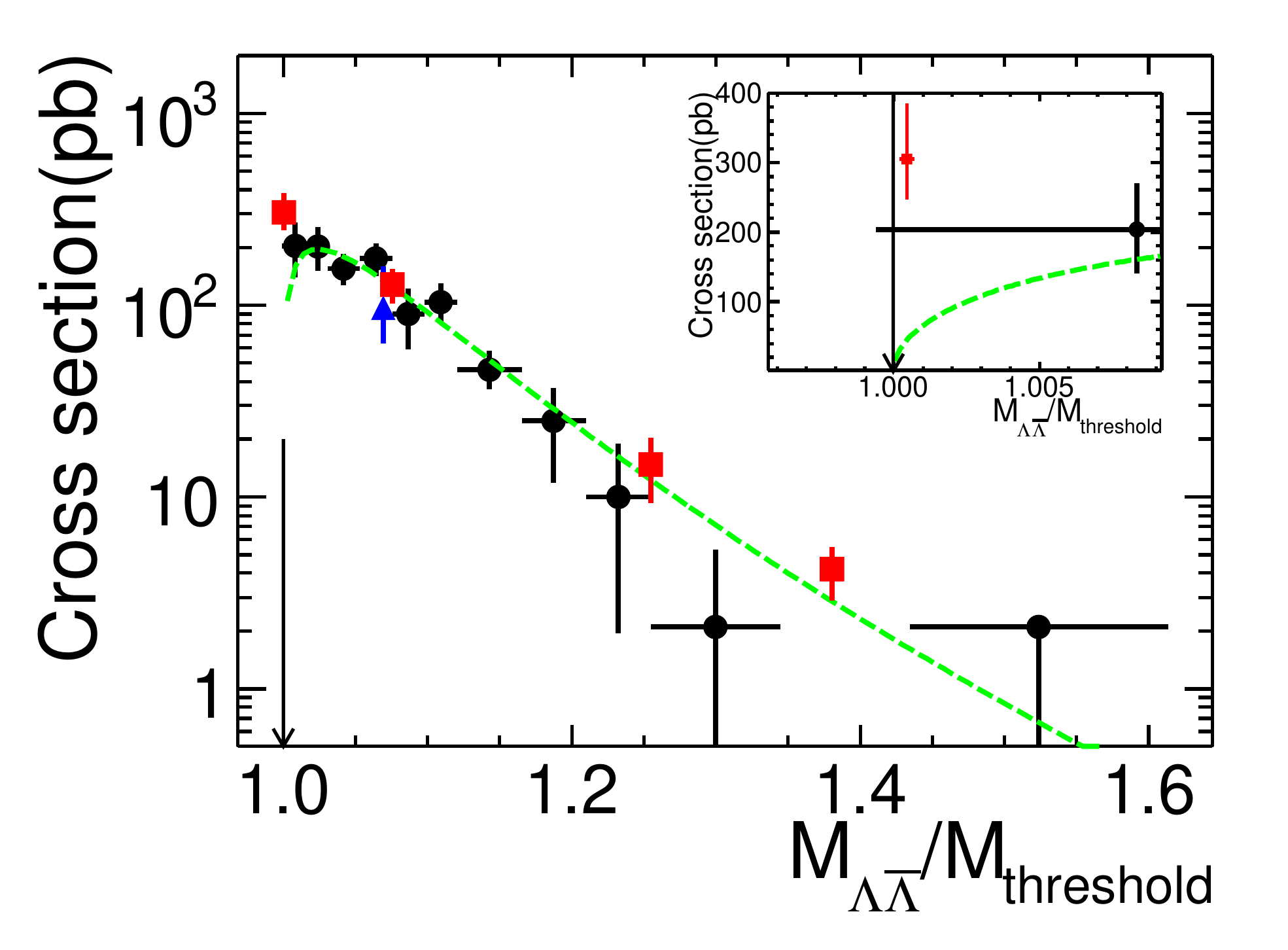}
\caption{BESIII measurements on baryon properties. (Top left) The fitted
results and (top right) the transverse polarization $P_y$ as a function of
$\cos\theta_\Lambda$ for $\jpsi\to \Lambda\bar{\Lambda}\to
p\pi^-\bar{p}\pi^+$ events~\cite{Ablikim:2018zay}. The dashed histogram in 
the top left panel shows the
no polarization scenario. (Bottom left) The near-threshold $\EE\to
\Lambda^+_c\bar\Lambda^-_c$ cross sections measured by BESIII (red
points) and Belle (black points) experiments and comparison with
phase space model (dash-dot cyan curve)~\cite{Ablikim:2017lct}. The vertical dashed line
indicates the threshold. (Bottom right) The near-threshold $\EE\to\LLb$ cross
section measured by BESIII (red squares), BaBar (black dots), and
DM2 (blue triangles)~\cite{Ablikim:2017pyl}. The inset is an expanded view on a linear
scale near the threshold. The green dashed lines are a
phenomenological fit result based on a perturbative quantum
chromodynamics prediction~\cite{Pacetti:2015iqa}, the black arrows
indicate the threshold. }
\label{hyperon}
\end{figure*}

With 1.3 billion $\jpsi$ events collected in 2009 and 2012, a
total of 420593 fully reconstructed $\jpsi\to \LLb$ events with
$\Lambda\to p\pi^-$ and $\bar{\Lambda}\to \bar{p}\pi^+$ were
isolated with a nearly negligible, 399-event background. The data
is described well with a large relative phase $\Delta\Phi=(42.4\pm
0.8)^\circ$ as shown in Fig.~\ref{hyperon}, whereas
$\Delta\Phi=0$ is clearly excluded by the data. The transverse
polarization of the $\Lambda$ ($\bar{\Lambda}$), which depends on
$\cos\theta_{\Lambda}$, the $\Lambda$ direction relative to the
$\EE$ beam axis, is shown in Fig.~\ref{hyperon}, the values
range between $-25\%$ and $+25\%$.

For $\Lambda\to p\pi^-$, BESIII measured $\alpha_-=0.750\pm
0.010$, which is more than five standard deviations higher than
the previous world average value of $\alpha_-=0.642\pm 0.013$ that
was based entirely on pre-1974
measurements~\cite{Cronin:1963zb,Overseth:1967zz,Dauber:1969hg,Cleland:1972fa,Astbury:1975hn}.
The measured $\bar{\Lambda}\to \bar{p}\pi^+$  asymmetry parameter,
$\alpha_+=-0.758\pm 0.012$, is also high, and consistent, within
uncertainties, with the $\Lambda$ result. The $CP$-violating
asymmetry, $\ACP\equiv {(\alpha_-+\alpha_+)}/{(\alpha_--\alpha_+)
} =-0.006\pm 0.014$, is compatible with zero and a factor of two
more sensitive than the best previous $\Lambda$-based measurement,
$\ACP=+0.013 \pm 0.022$~\cite{Barnes:1996si}. This is still well
above the SM expectation of $\ACP^{\rm SM}\sim 10^{-4}$ based on
the CKM mechanism~\cite{Donoghue:1986hh}.

BESIII used 47009 $\jpsi\to \Lambda\ (p\pi^-)\ \bar{\Lambda}\
(\bar{n}\piz)$ signal events with a negligible 66-event background
to measure $\bar{\alpha}_0$, the parity violating parameter for
$\bar{\Lambda}\to \bar{n}\piz$.  The result,
$\bar{\alpha}_0=-0.692\pm 0.017$, is a threefold improvement on
the only previous measurement~\cite{Olsen:1970vb}; the
$3$~standard deviations difference from $\alpha_+$ could reflect
the presence of an isospin=3/2 component of the $\pi \bar{N}$
final state.

The $\ACP$ measurement is based on a ratio of asymmetries in which
many detector-related systematic effects cancel, and the current
quoted precision is limited by statistical uncertainties. BESIII's
accumulated $\jpsi$ data set has recently grown to 10 billion
events, and this should provide a threefold reduction in the
$\ACP$ uncertainty. In addition, BESIII is extending similar
analyses to  $\jpsi\to \Xi\bar{\Xi}$ and $\psip\to
\Omega^-\bar\Omega^+$ hyperon pairs. For these, the decay final
states include a $\Lambda$ ($\bar{\Lambda}$) whose polarization
can be determined from its decay asymmetry. This feature provides
the capability of additional $CP$ tests~\cite{Gonzalez:1994zc}
that are potentially more sensitive to new physics than $\ACP$.

The CKM mechanism for $CPV$ in the SM fails to explain the
matter-antimatter asymmetry of the Universe by more than 10
orders-of-magnitude~\cite{Morrissey:2012db}. This suggests that
additional, heretofore undiscovered, $CP$ violating processes
occur, which has motivated intensive searches for new sources of
$CPV$ in bottom-quark
decays~\cite{Aushev:2010bq,Bediaga:2012uyd,Kou:2018nap,Bediaga:2018lhg}
and neutrino oscillations~\cite{Acciarri:2016crz,Abe:2018uyc}.
BESIII's capability to test $CP$ symmetry in the decays of
polarized, quantum-entangled hyperon pairs produced via $\jpsi\to
\BBb$ adds an exciting new dimension to the study of $CP$
violations.

Searches for new sources of $CPV$ have been elevated to a new
level of interest by the recent LHCb discovery of $CPV$ in $D^0\to
\kk$ and $D^0\to \pp$ decays. They measure the time-integrated
$CPV$ asymmetry
\begin{eqnarray}
  \Delta A_{\rm CP}&=&\frac{\Gamma(D\to K^+K^-)-\Gamma(\bar{D}\to K^+K^-)}
                     {\Gamma(D\to K^+K^-)+\Gamma(\bar{D}\to K^+K^-)}
               -\frac{\Gamma(D\to \pp)-\Gamma(\bar{D}\to \pp)}
                     {\Gamma(D\to \pp)+\Gamma(\bar{D}\to \pp)}\\\nonumber
             &=&(-0.154\pm 0.029)\%,
\end{eqnarray}
where $D$ ($\bar{D}$) is a $D^0$ ($\bar{D}^0$) at time
$t=0$~\cite{Aaij:2019kcg}. The significance of the $\Delta a_{\rm
CP}$ deviation from zero is $5.3$ standard deviations, making this
the first observation of $CPV$ in the charm-quark sector. The
measured $\Delta A_{\rm CP}$ value is at the high end of
theoretical estimates for its SM value, which range from
$10^{-4}$~\cite{Khodjamirian:2017zdu} to
$10^{-3}$~\cite{Golden:1989qx,Buccella:1994nf,Bianco:2003vb,Grossman:2006jg}.
The LHCb result is intriguing, because it may be a sign of the
long-sought-for non-SM mechanism for $CPV$. However, uncertainties
in the calculations of the SM expectation for $\Delta a_{\rm CP}$
make it impossible to either establish or rule out this
possibility. However, BESIII's current $\sim 10^{-2}$ level of
sensitivity on $\ACP$ is still two orders-of-magnitude above the
highest conceivable SM effects~\cite{Donoghue:1986hh}. Any
non-zero measurement of $\ACP$ in the intervening range would be
an unambiguous signature for new physics.

\section{Measurements of time-like baryon electromagnetic form factors}

The electric charge and magnetization distributions inside baryons
are described by electric and magnetic form factors (FFs),
$G_E(q^2)$ and $G_M(q^2)$, where $q^2$ is the square of the
four-momentum transferred in the process. Historically, huge
efforts have been made to determine space-like ($q^2<0$) nucleon
FFs by electron-nucleon elastic scattering experiments. In
contrast, time-like FFs ($q^2>4m^2_B$), which are measured in
$\EE\to \BBb$ or $\ppb\to \EE$ annihilation reactions, with
$q^2=\ECMsqr$, are less well studied~\cite{Pacetti:2015iqa}.

The behaviour of the time-like FFs near the $\BBb$ mass threshold
is particularly interesting. In $\EE\to \BBb$, the baryon pair is
produced in a short-distance interaction mediated by a hard,
$q^2\ge 4m^2_B$, virtual photon. However, the final state $B$ and
$\bar{B}$ are nearly at rest relative to each other in a
non-relativistic, long-distance system. In the case where $B$ and
$\bar{B}$ are electrically charged, their final state interactions
are modified by their mutual Coulomb force. For a point-like
charged particle, this Coulomb interaction produces an abrupt jump
in the production cross-section right at threshold of
$\Delta\sigma=\pi^2\alpha^3/2m^2_B$; for $m_B=m_p$, $\Delta\sigma
= 0.85$~nb~\cite{Sakharov:1948yq,Arbuzov:2011ff}. Amazingly,
BaBar~\cite{Lees:2013ebn} and CMD-3~\cite{CMD-3:2018kql}
measurements of $\EE\rt\ppb$ have seen an abrupt jump in the
$\EE\to\ppb$ cross-section at the $\ECM=2m_p$ threshold that is
consistent with the 0.85~nb expectation for a point-like charged
particle. BESIII measurements of
$\sigma(\EE\to\Lambda^+_c\bar\Lambda^-_c)$, shown in
Fig.~\ref{hyperon}, have a $\Delta\sigma = 0.236 \pm 0.047$~nb
threshold jump~\cite{Ablikim:2017lct} that is larger, although
marginally consistent with the $\Delta\sigma=0.14$~nb expectation
for an $m_B=m_{\Lambda^+_c}$ point-like charged particle.

In the case of neutral particles, there is no Coulomb interaction
and no mechanism for an abrupt jump at the threshold; the
cross-section is expected to grow  as the available phase space,
which is proportional to $\beta_f$, the relative velocity of the
two final-state particles. However, in another big surprise,
$\EE\to n\bar{n}$ measurements did not confirm a
$\sigma\propto\beta_n$ behaviour, instead they also showed an
abrupt threshold jump~\cite{Antonelli:1998fv,Achasov:2014ncd} that
is consistent (within large uncertainties) with the 0.85~nb jump
seen for $\EE\to\ppb$.

Early measurements of $\EE\to\LLb$ showed some evidence for a
non-zero, near-threshold cross-section, but these had large
uncertainties and were confined to $\ECM$ values corresponding to
$\beta_\Lambda\ge 0.3$~\cite{Bisello:1990rf,Aubert:2007uf}. In
2017, BESIII reported measurements of $\EE\to\LLb$ for energies
ranging from $\ECM=2.2324$~GeV (1~MeV above the $2m_{\Lambda}$
threshold with $\beta_\Lambda\approx 0.03$) to $\ECM=
3.080$~GeV~\cite{Ablikim:2017pyl}. The measured cross-section is
maximum right at the $\LLb$ threshold and falls off at higher
energies contrary to theoretical expectations (see
Fig.~\ref{hyperon}). The cross-section value at threshold is
$0.305\pm 0.058$~nb, about one half the size of the 0.60~nb jump
calculated for a (hypothetical) charged point-like baryon with
mass $m_B=m_{\Lambda}$. One theoretical analysis concluded that
such a threshold jump in the $\EE\to\LLb$ cross-section is a
strong indication of the presence of a very narrow, $^3S_1$ $\LLb$
resonance with a mass close to
$2m_{\Lambda}$~\cite{Haidenbauer:2016won}.

Measurements of near-threshold $\EE\to \BBb$ pair production
cross-sections uncovered some intriguing discrepancies with
theoretical expectations. To help clarify the underlying physics
scenario, BESIII will perform precise cross-section measurements
at all of the stable $\BBb$ thresholds that are accessible in the
BEPCII CM energy range, including the three $\Sigma\bar{\Sigma}$
and two $\Xi\bar{\Xi}$ thresholds, as well as those for
$\Omega^-\bar\Omega^+$ and $\Lambda^+_c\bar\Lambda^-_c$.

\section{Outlook}

Prior to the start of BESIII operation, the BES scientific
community, including theorists and experimentalists, prepared an
800-page report that mapped out a diverse ten-year-long physics
research program for the project~\cite{Asner:2008nq}. During its
first decade of operation, BESIII has, in accordance with this
plan: improved on its measurement of $m_{\tau}$ and $\sigma(\EE\to
{\rm hadrons})$; produced world's best measurements of charmed
particle decay constants, form-factors and the $|V_{cs}|$ and
$|V_{cd}|$ CKM matrix elements; reported definitive measurements
of the $\eta_c$~\cite{BESIII:2011ab} and
$h_c$~\cite{Ablikim:2012ur} masses and widths plus many other
precision charmonium results; and made a number contributions to
the understanding of light hadron physics, such as confirming that
the $X(1835)$ structure has a strong coupling to the
nucleon-antinucleon final state and identifying strong candidates
for the scalar and tensor glueballs.

In addition, during this period, BESIII produced a number of
results that were not anticipated in the 2008 plan. These include:
discoveries of the $Z_c(3900)$ and $Z_c(4020)$ charged
charmonium-like states; a complex structure of the $\y$ line
shape; $X(3872)$ production via radiative $\y$ transitions;
substantial signals for $a^0_0(980)\lrarrow f_0(980)$ mixing;
large transverse $\Lambda$ polarization in $\jpsi\to\LLb$ decays
and world's best limits on $CPV$ asymmetries in $\Lambda$ hyperon
decays; an anomalous threshold behaviour for $\sigma(\EE\to\LLb)$
that strongly contradicts theoretical expectations; and
measurements of the strong phase in quantum-correlated $D^0\to
K^0\pp$ decays.

Each of the items listed in the previous paragraph beg for
additional, high-statistics investigations that were not
considered in the original BESIII plan. The puzzles associated
with the $\xyz$ mesons call for high-statistics data runs that
span the 4.0~to~4.7~GeV energy region in small increments to
support advanced, $\ECM$-dependent coupled-channel PWA studies in
addition to a unique set of $X(3872)$-related measurements that
could be done with a large data set accumulated at the peak energy
of the $\y$ resonance.  The $\Lambda$ transverse polarization in
$\jpsi\to\LLb$ decays provides singular, and previously
unexpected, opportunities to search for new sources of $CPV$ in
hyperon decays that could use many times the currently available
$\jpsi$ data. Measurements of strong phases in quantum-correlated
$D^0$ decays with the precision required for the interpretation of
future, high profile LHCb and Belle~II measurements of the
$\gamma$ $CPV$ angle are only possible with BESIII, and will
require two or three years of dedicated data-taking at the peak of
the $\psi(3770)$. The anomalous $\EE\to\LLb$ threshold behaviour
motivates the accumulation of high-statistics data sets in the
vicinities of each of the hyperon-antihyperon thresholds.  This
menu of measurements will, at minimum, require operating the
program for another decade~\cite{Ablikim:2019hff}.

In support of this program, the luminosity performance of BEPCII
is being upgraded and the maximum CM energy is being increased. In
the current mode of operation, the beams are accumulated in the
storage rings and then made to collide until the luminosity
decreases to about two-thirds of its initial value, at which time
collisions are stopped and the beams restored to their maximum
currents. A system that is currently being implemented  will
continuously maintain the beam currents near their peak values,
thereby increasing the time-integrated luminosity by about 30\%. A
two-year-long program that involves upgrading the ring magnet
power supplies and replacing some critical magnets will extend the
maximum $\ECM$ from 4.6 to 4.9~GeV. This will provide full
coverage of the  $Y(4660)$, a $1^{--}$ charmonium-like resonance
that peaks near $\ECM=4660$~MeV, improve the production rate of
$\Lambda_c$ baryons, and provide access to $\Sigma_c$ and excited
$\Lambda_c$ baryon states.

In the longer-term future, two upgraded versions of the
BESIII/BEPCII facility, so-called super tau-charm factories, have
been proposed: the STCF in China~\cite{Luo:2018njj} and the SCTF
in Russia~\cite{Levichev:2018}. Both machines would run at CM
energies that reach 6~GeV or higher, with peak luminosities of
$10^{35}$~cm$^{-2}$s$^{-1}$, corresponding to a factor of 100
improvement over BEPCII. These improvements would enable
systematic studies of charm particle decays, $\XYZ$ mesons,
searches for new physics sources of $CPV$ in hyperon decays, and
the elucidation of many issues in light-hadron physics with
unprecedented precision.

%%\section*{Proposed display items}

\appendix

\section{The BESIII experiment at BEPCII}
\label{appA}

The construction of the Beijing Electron-Positron Collider (BEPC)
started in 1984 and the first operation for high-energy physics
and synchrotron radiation applications occurred in 1989. In 2008,
BEPC was upgraded to BEPCII, a double-ring collider with a peak
luminosity of $1.0\times 10^{33}$~cm$^{-2}$s$^{-1}$, about two
orders of magnitude higher than the maximum achieved with BEPC.
For about six months of each year, BEPCII operates at
center-of-mass energies that range between 2.0 and 4.6~GeV for
particle physics experiments with the Beijing Spectrometer III
(BESIII). For three months, it operates in a single beam mode
with an energy of 2.5~GeV to provide high fluxes of synchrotron
X-rays to 14 separate beam lines for material and biological
science applications.

\begin{figure*}[htbp]
\centering
 \includegraphics[width=15cm]{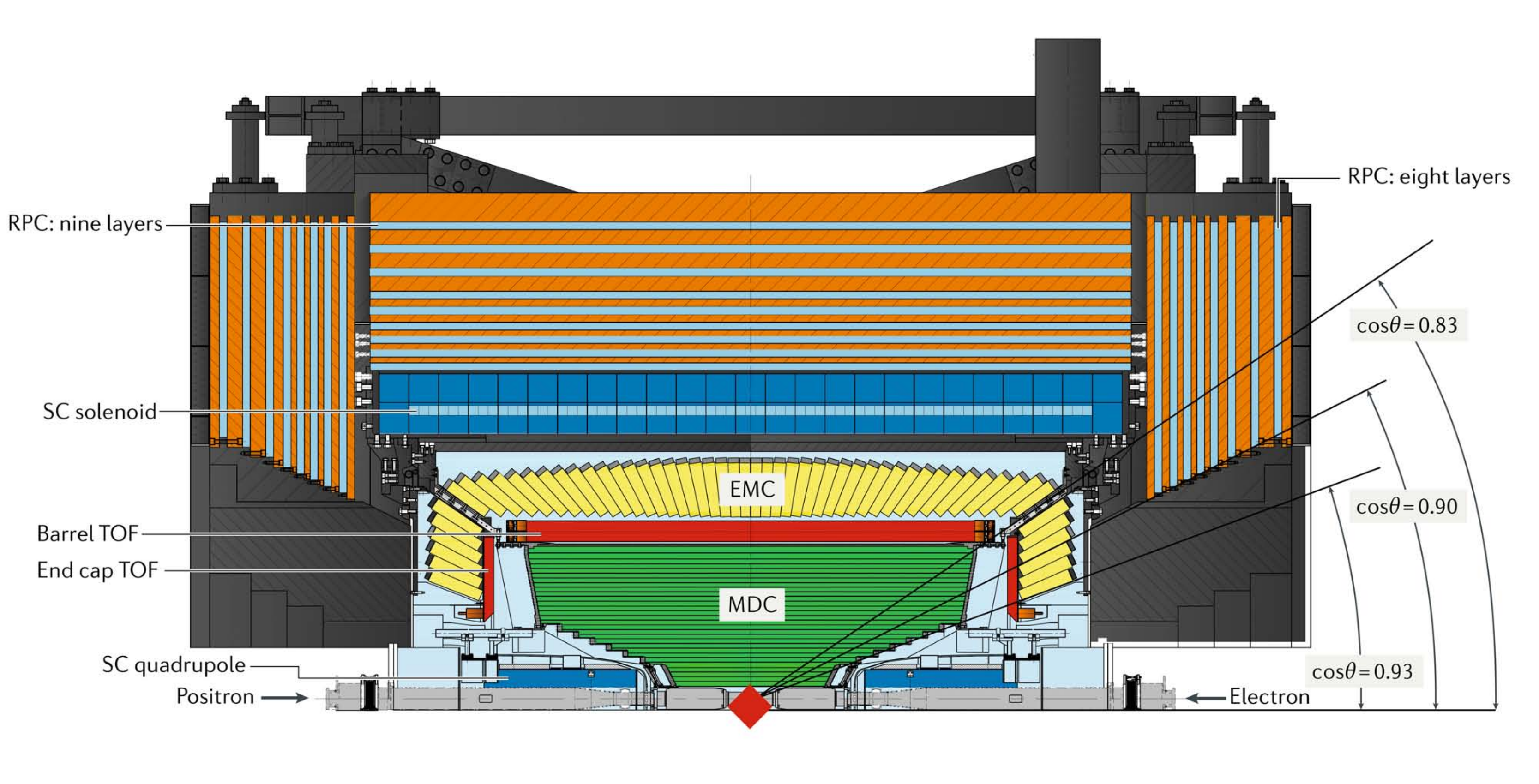}
 \caption{Cutaway view of the top half of the BESIII detector.
 (Produced by Dr. Xiao-Yan Ma.)}
\label{bes3}
\end{figure*}

The BESIII experiment~\cite{Ablikim:2009aa} recorded its first
collisions in July 2008. The detector, shown in Fig.~\ref{bes3}, is a
general-purpose spectrometer based on a 1~Tesla superconducting
(SC) solenoid magnet with a geometrical acceptance that covers
93\% of $4\pi$ steradians. It consists of a layered arrangement of
nested instruments, including: a small cell helium-based
multi-layered drift chamber (MDC) that provides momentum
measurements of charged particles; a cylinder of plastic
scintillators that are used to identify charged particles based on
their time-of-flight (ToF); an electromagnetic calorimeter that is
a mosaic array of CsI(Tl) crystals that are used to detect and
measure the energies of $\gamma$-rays and provide trigger signals;
a surrounding array of Resistive Plate Chambers (RPC) to detect
and identify muons. The MDC provides $0.5$\% momentum resolution
for 1~GeV/$c$ charged particle tracks and energy-loss ($dE/dx$)
measurements with a resolution of 6\%. The electromagnetic
calorimeter energy resolution for 1~GeV $\gamma$ rays is $2.5$\%
in the polar angle region $|\cos\theta|< 0.83$ (barrel) and $5$\%
for $0.83\le |\cos\theta|\le 0.93$ (endcaps). The TOF system time
resolution is $68$~ps in the barrel region and $110$~ps in the
endcaps. In 2015, the endcap TOF scintillators were replaced with
a system of multi-layer RPCs that improved the time resolution to
60~ps~\cite{Wang:2016bzv}.

%%%%%%%%%%%%%%%%%%%%%%%%%%%%%%%%%%%%%
%%%% Acknowledge           %%%%%%%%%%
%%%%%%%%%%%%%%%%%%%%%%%%%%%%%%%%%%%%%
\acknowledgments This work is supported in part by National
Natural Science Foundation of China (NSFC) under contract Nos.
11835012, 11521505, and 11475187; the Ministry of Science and
Technology of China under Contract No. 2015CB856701; Key Research
Program of Frontier Sciences, CAS, Grant No. QYZDJ-SSW-SLH011; the
CAS Center for Excellence in Particle Physics (CCEPP); and the CAS
President International Fellowship Initiative (PIFI) program.

%%%%%%%%%%%%%%%%%%%%%%%%%%%%%%%%%%%%%
%%\bibliographystyle{elsarticle-num-names}%
%%\bibliographystyle{plain}%
\bibliography{nature_besiii}

\begin{thebibliography}{204}
\expandafter\ifx\csname natexlab\endcsname\relax\def\natexlab#1{#1}\fi
\expandafter\ifx\csname bibnamefont\endcsname\relax
  \def\bibnamefont#1{#1}\fi
\expandafter\ifx\csname bibfnamefont\endcsname\relax
  \def\bibfnamefont#1{#1}\fi
\expandafter\ifx\csname citenamefont\endcsname\relax
  \def\citenamefont#1{#1}\fi
\expandafter\ifx\csname url\endcsname\relax
  \def\url#1{\texttt{#1}}\fi
\expandafter\ifx\csname urlprefix\endcsname\relax\def\urlprefix{URL }\fi
\providecommand{\bibinfo}[2]{#2}
\providecommand{\eprint}[2][]{\url{#2}}

\bibitem[{\citenamefont{Weinberg}(2018)}]{Weinberg:2018apv}
\bibinfo{author}{\bibfnamefont{S.}~\bibnamefont{Weinberg}},
  \bibinfo{journal}{Phys. Rev. Lett.} \textbf{\bibinfo{volume}{121}},
  \bibinfo{pages}{220001} (\bibinfo{year}{2018}).

\bibitem[{\citenamefont{Aubert et~al.}(2002)}]{Aubert:2001tu}
\bibinfo{author}{\bibfnamefont{B.}~\bibnamefont{Aubert}} \bibnamefont{et~al.}
  (\bibinfo{collaboration}{BaBar}), \bibinfo{journal}{Nucl. Instrum. Meth.}
  \textbf{\bibinfo{volume}{A479}}, \bibinfo{pages}{1} (\bibinfo{year}{2002}),
  \eprint{hep-ex/0105044}.

\bibitem[{\citenamefont{Abashian et~al.}(2002)}]{Abashian:2000cg}
\bibinfo{author}{\bibfnamefont{A.}~\bibnamefont{Abashian}} \bibnamefont{et~al.}
  (\bibinfo{collaboration}{Belle}), \bibinfo{journal}{Nucl. Instrum. Meth.}
  \textbf{\bibinfo{volume}{A479}}, \bibinfo{pages}{117} (\bibinfo{year}{2002}).

\bibitem[{\citenamefont{Ablikim et~al.}(2010)}]{Ablikim:2009aa}
\bibinfo{author}{\bibfnamefont{M.}~\bibnamefont{Ablikim}} \bibnamefont{et~al.}
  (\bibinfo{collaboration}{BESIII}), \bibinfo{journal}{Nucl. Instrum. Meth.}
  \textbf{\bibinfo{volume}{A614}}, \bibinfo{pages}{345} (\bibinfo{year}{2010}),
  \eprint{0911.4960}.

\bibitem[{\citenamefont{Alves et~al.}(2008)}]{Alves:2008zz}
\bibinfo{author}{\bibfnamefont{A.~A.} \bibnamefont{Alves}, \bibfnamefont{Jr.}}
  \bibnamefont{et~al.} (\bibinfo{collaboration}{LHCb}),
  \bibinfo{journal}{JINST} \textbf{\bibinfo{volume}{3}},
  \bibinfo{pages}{S08005} (\bibinfo{year}{2008}).

\bibitem[{\citenamefont{Altmannshofer et~al.}(2018)}]{Kou:2018nap}
\bibinfo{author}{\bibfnamefont{W.}~\bibnamefont{Altmannshofer}}
  \bibnamefont{et~al.} (\bibinfo{collaboration}{Belle II})
  (\bibinfo{year}{2018}), \eprint{1808.10567}.

\bibitem[{\citenamefont{Luo and Xu}(2018)}]{Luo:2018njj}
\bibinfo{author}{\bibfnamefont{Q.}~\bibnamefont{Luo}} \bibnamefont{and}
  \bibinfo{author}{\bibfnamefont{D.}~\bibnamefont{Xu}}, in
  \emph{\bibinfo{booktitle}{{Proceedings, 9th International Particle
  Accelerator Conference (IPAC 2018): Vancouver, BC Canada}}}
  (\bibinfo{year}{2018}), p. \bibinfo{pages}{MOPML013}.

\bibitem[{\citenamefont{Levichev et~al.}(2018)\citenamefont{Levichev,
  Skrinskii, Tumaikin, and Shatunov}}]{Levichev:2018}
\bibinfo{author}{\bibfnamefont{E.~B.} \bibnamefont{Levichev}},
  \bibinfo{author}{\bibfnamefont{A.~N.} \bibnamefont{Skrinskii}},
  \bibinfo{author}{\bibfnamefont{G.~M.} \bibnamefont{Tumaikin}},
  \bibnamefont{and} \bibinfo{author}{\bibfnamefont{Y.~M.}
  \bibnamefont{Shatunov}}, \bibinfo{journal}{Phys. Usp.}
  \textbf{\bibinfo{volume}{61}}, \bibinfo{pages}{405} (\bibinfo{year}{2018}),
  \urlprefix\url{https://ufn.ru/en/articles/2018/5/b/}.

\bibitem[{\citenamefont{Bevan et~al.}(2014)}]{Bevan:2014iga}
\bibinfo{author}{\bibfnamefont{A.~J.} \bibnamefont{Bevan}} \bibnamefont{et~al.}
  (\bibinfo{collaboration}{BaBar, Belle}), \bibinfo{journal}{Eur. Phys. J.}
  \textbf{\bibinfo{volume}{C74}}, \bibinfo{pages}{3026} (\bibinfo{year}{2014}),
  \eprint{1406.6311}.

\bibitem[{\citenamefont{Kobayashi and Maskawa}(1973)}]{Kobayashi:1973fv}
\bibinfo{author}{\bibfnamefont{M.}~\bibnamefont{Kobayashi}} \bibnamefont{and}
  \bibinfo{author}{\bibfnamefont{T.}~\bibnamefont{Maskawa}},
  \bibinfo{journal}{Prog. Theor. Phys.} \textbf{\bibinfo{volume}{49}},
  \bibinfo{pages}{652} (\bibinfo{year}{1973}).

\bibitem[{\citenamefont{Brambilla et~al.}(2011)}]{Brambilla:2010cs}
\bibinfo{author}{\bibfnamefont{N.}~\bibnamefont{Brambilla}}
  \bibnamefont{et~al.}, \bibinfo{journal}{Eur. Phys. J.}
  \textbf{\bibinfo{volume}{C71}}, \bibinfo{pages}{1534} (\bibinfo{year}{2011}),
  \eprint{1010.5827}.

\bibitem[{\citenamefont{Chen et~al.}(2016)\citenamefont{Chen, Chen, Liu, and
  Zhu}}]{Chen:2016qju}
\bibinfo{author}{\bibfnamefont{H.-X.} \bibnamefont{Chen}},
  \bibinfo{author}{\bibfnamefont{W.}~\bibnamefont{Chen}},
  \bibinfo{author}{\bibfnamefont{X.}~\bibnamefont{Liu}}, \bibnamefont{and}
  \bibinfo{author}{\bibfnamefont{S.-L.} \bibnamefont{Zhu}},
  \bibinfo{journal}{Phys. Rept.} \textbf{\bibinfo{volume}{639}},
  \bibinfo{pages}{1} (\bibinfo{year}{2016}), \eprint{1601.02092}.

\bibitem[{\citenamefont{Olsen et~al.}(2018)\citenamefont{Olsen, Skwarnicki, and
  Zieminska}}]{Olsen:2017bmm}
\bibinfo{author}{\bibfnamefont{S.~L.} \bibnamefont{Olsen}},
  \bibinfo{author}{\bibfnamefont{T.}~\bibnamefont{Skwarnicki}},
  \bibnamefont{and}
  \bibinfo{author}{\bibfnamefont{D.}~\bibnamefont{Zieminska}},
  \bibinfo{journal}{Rev. Mod. Phys.} \textbf{\bibinfo{volume}{90}},
  \bibinfo{pages}{015003} (\bibinfo{year}{2018}), \eprint{1708.04012}.

\bibitem[{\citenamefont{Guo et~al.}(2018)\citenamefont{Guo, Hanhart,
  Mei{\ss}ner, Wang, Zhao, and Zou}}]{Guo:2017jvc}
\bibinfo{author}{\bibfnamefont{F.-K.} \bibnamefont{Guo}},
  \bibinfo{author}{\bibfnamefont{C.}~\bibnamefont{Hanhart}},
  \bibinfo{author}{\bibfnamefont{U.-G.} \bibnamefont{Mei{\ss}ner}},
  \bibinfo{author}{\bibfnamefont{Q.}~\bibnamefont{Wang}},
  \bibinfo{author}{\bibfnamefont{Q.}~\bibnamefont{Zhao}}, \bibnamefont{and}
  \bibinfo{author}{\bibfnamefont{B.-S.} \bibnamefont{Zou}},
  \bibinfo{journal}{Rev. Mod. Phys.} \textbf{\bibinfo{volume}{90}},
  \bibinfo{pages}{015004} (\bibinfo{year}{2018}), \eprint{1705.00141}.

\bibitem[{\citenamefont{Brambilla et~al.}(2019)\citenamefont{Brambilla,
  Eidelman, Hanhart, Nefediev, Shen, Thomas, Vairo, and
  Yuan}}]{Brambilla:2019esw}
\bibinfo{author}{\bibfnamefont{N.}~\bibnamefont{Brambilla}},
  \bibinfo{author}{\bibfnamefont{S.}~\bibnamefont{Eidelman}},
  \bibinfo{author}{\bibfnamefont{C.}~\bibnamefont{Hanhart}},
  \bibinfo{author}{\bibfnamefont{A.}~\bibnamefont{Nefediev}},
  \bibinfo{author}{\bibfnamefont{C.-P.} \bibnamefont{Shen}},
  \bibinfo{author}{\bibfnamefont{C.~E.} \bibnamefont{Thomas}},
  \bibinfo{author}{\bibfnamefont{A.}~\bibnamefont{Vairo}}, \bibnamefont{and}
  \bibinfo{author}{\bibfnamefont{C.-Z.} \bibnamefont{Yuan}}
  (\bibinfo{year}{2019}), \eprint{1907.07583}.

\bibitem[{\citenamefont{Gell-Mann}(1964)}]{GellMann:1964nj}
\bibinfo{author}{\bibfnamefont{M.}~\bibnamefont{Gell-Mann}},
  \bibinfo{journal}{Phys. Lett.} \textbf{\bibinfo{volume}{8}},
  \bibinfo{pages}{214} (\bibinfo{year}{1964}).

\bibitem[{\citenamefont{Zweig}(1964)}]{Zweig:1981pd}
\bibinfo{author}{\bibfnamefont{G.}~\bibnamefont{Zweig}} (\bibinfo{year}{1964}),
  \bibinfo{note}{{An SU(3) model for strong interaction symmetry and its
  breaking. CERN-TH-401}}.

\bibitem[{\citenamefont{Bai et~al.}(1994)}]{Bai:1994zm}
\bibinfo{author}{\bibfnamefont{J.~Z.} \bibnamefont{Bai}} \bibnamefont{et~al.}
  (\bibinfo{collaboration}{BES}), \bibinfo{journal}{Nucl. Instrum. Meth.}
  \textbf{\bibinfo{volume}{A344}}, \bibinfo{pages}{319} (\bibinfo{year}{1994}).

\bibitem[{\citenamefont{Bai et~al.}(2001)}]{Bai:2001dw}
\bibinfo{author}{\bibfnamefont{J.~Z.} \bibnamefont{Bai}} \bibnamefont{et~al.}
  (\bibinfo{collaboration}{BES}), \bibinfo{journal}{Nucl. Instrum. Meth.}
  \textbf{\bibinfo{volume}{A458}}, \bibinfo{pages}{627} (\bibinfo{year}{2001}).

\bibitem[{\citenamefont{Qin et~al.}(2010)\citenamefont{Qin, Ma, Wang, and
  Zhang}}]{Qin:2010zzd}
\bibinfo{author}{\bibfnamefont{Q.}~\bibnamefont{Qin}},
  \bibinfo{author}{\bibfnamefont{L.}~\bibnamefont{Ma}},
  \bibinfo{author}{\bibfnamefont{J.}~\bibnamefont{Wang}}, \bibnamefont{and}
  \bibinfo{author}{\bibfnamefont{C.}~\bibnamefont{Zhang}},
  \bibinfo{journal}{Conf. Proc.} \textbf{\bibinfo{volume}{C100523}},
  \bibinfo{pages}{WEXMH01} (\bibinfo{year}{2010}).

\bibitem[{\citenamefont{Bai et~al.}(1992)}]{Bai:1992bu}
\bibinfo{author}{\bibfnamefont{J.~Z.} \bibnamefont{Bai}} \bibnamefont{et~al.}
  (\bibinfo{collaboration}{BES}), \bibinfo{journal}{Phys. Rev. Lett.}
  \textbf{\bibinfo{volume}{69}}, \bibinfo{pages}{3021} (\bibinfo{year}{1992}).

\bibitem[{\citenamefont{Bai et~al.}(1995)}]{Bai:1994qz}
\bibinfo{author}{\bibfnamefont{J.~Z.} \bibnamefont{Bai}} \bibnamefont{et~al.}
  (\bibinfo{collaboration}{BES}), \bibinfo{journal}{Phys. Rev. Lett.}
  \textbf{\bibinfo{volume}{74}}, \bibinfo{pages}{4599} (\bibinfo{year}{1995}).

\bibitem[{\citenamefont{Bai et~al.}(2002)}]{Bai:2001ct}
\bibinfo{author}{\bibfnamefont{J.~Z.} \bibnamefont{Bai}} \bibnamefont{et~al.}
  (\bibinfo{collaboration}{BES}), \bibinfo{journal}{Phys. Rev. Lett.}
  \textbf{\bibinfo{volume}{88}}, \bibinfo{pages}{101802}
  (\bibinfo{year}{2002}), \eprint{hep-ex/0102003}.

\bibitem[{\citenamefont{Ablikim et~al.}(2004)}]{Ablikim:2004qna}
\bibinfo{author}{\bibfnamefont{M.}~\bibnamefont{Ablikim}} \bibnamefont{et~al.}
  (\bibinfo{collaboration}{BES}), \bibinfo{journal}{Phys. Lett.}
  \textbf{\bibinfo{volume}{B598}}, \bibinfo{pages}{149} (\bibinfo{year}{2004}),
  \eprint{hep-ex/0406038}.

\bibitem[{\citenamefont{Ablikim et~al.}(2006)}]{Ablikim:2005ni}
\bibinfo{author}{\bibfnamefont{M.}~\bibnamefont{Ablikim}} \bibnamefont{et~al.}
  (\bibinfo{collaboration}{BES}), \bibinfo{journal}{Phys. Lett.}
  \textbf{\bibinfo{volume}{B633}}, \bibinfo{pages}{681} (\bibinfo{year}{2006}),
  \eprint{hep-ex/0506055}.

\bibitem[{\citenamefont{Mo et~al.}(2007)\citenamefont{Mo, Yuan, and
  Wang}}]{Mo:2006cy}
\bibinfo{author}{\bibfnamefont{X.-H.} \bibnamefont{Mo}},
  \bibinfo{author}{\bibfnamefont{C.-Z.} \bibnamefont{Yuan}}, \bibnamefont{and}
  \bibinfo{author}{\bibfnamefont{P.}~\bibnamefont{Wang}},
  \bibinfo{journal}{High Energy Phys. \& Nucl. Phys.}
  \textbf{\bibinfo{volume}{31}}, \bibinfo{pages}{686} (\bibinfo{year}{2007}),
  \eprint{hep-ph/0611214}.

\bibitem[{\citenamefont{Jaffe}(2005)}]{Jaffe:2004ph}
\bibinfo{author}{\bibfnamefont{R.~L.} \bibnamefont{Jaffe}},
  \bibinfo{journal}{Phys. Rept.} \textbf{\bibinfo{volume}{409}},
  \bibinfo{pages}{1} (\bibinfo{year}{2005}), \eprint{hep-ph/0409065}.

\bibitem[{\citenamefont{Deng et~al.}(2017)\citenamefont{Deng, Liu, Gui, and
  Zhong}}]{Deng:2016stx}
\bibinfo{author}{\bibfnamefont{W.-J.} \bibnamefont{Deng}},
  \bibinfo{author}{\bibfnamefont{H.}~\bibnamefont{Liu}},
  \bibinfo{author}{\bibfnamefont{L.-C.} \bibnamefont{Gui}}, \bibnamefont{and}
  \bibinfo{author}{\bibfnamefont{X.-H.} \bibnamefont{Zhong}},
  \bibinfo{journal}{Phys. Rev.} \textbf{\bibinfo{volume}{D95}},
  \bibinfo{pages}{034026} (\bibinfo{year}{2017}), \eprint{1608.00287}.

\bibitem[{\citenamefont{Brodsky et~al.}(2014)\citenamefont{Brodsky, Hwang, and
  Lebed}}]{Brodsky:2014xia}
\bibinfo{author}{\bibfnamefont{S.~J.} \bibnamefont{Brodsky}},
  \bibinfo{author}{\bibfnamefont{D.~S.} \bibnamefont{Hwang}}, \bibnamefont{and}
  \bibinfo{author}{\bibfnamefont{R.~F.} \bibnamefont{Lebed}},
  \bibinfo{journal}{Phys. Rev. Lett.} \textbf{\bibinfo{volume}{113}},
  \bibinfo{pages}{112001} (\bibinfo{year}{2014}), \eprint{1406.7281}.

\bibitem[{\citenamefont{Bondar et~al.}(2012)}]{Belle:2011aa}
\bibinfo{author}{\bibfnamefont{A.}~\bibnamefont{Bondar}} \bibnamefont{et~al.}
  (\bibinfo{collaboration}{Belle}), \bibinfo{journal}{Phys. Rev. Lett.}
  \textbf{\bibinfo{volume}{108}}, \bibinfo{pages}{122001}
  (\bibinfo{year}{2012}), \eprint{1110.2251}.

\bibitem[{\citenamefont{Aaij et~al.}(2015)}]{Aaij:2015tga}
\bibinfo{author}{\bibfnamefont{R.}~\bibnamefont{Aaij}} \bibnamefont{et~al.}
  (\bibinfo{collaboration}{LHCb}), \bibinfo{journal}{Phys. Rev. Lett.}
  \textbf{\bibinfo{volume}{115}}, \bibinfo{pages}{072001}
  (\bibinfo{year}{2015}), \eprint{1507.03414}.

\bibitem[{\citenamefont{Ablikim et~al.}(2014{\natexlab{a}})}]{Ablikim:2014uzh}
\bibinfo{author}{\bibfnamefont{M.}~\bibnamefont{Ablikim}} \bibnamefont{et~al.}
  (\bibinfo{collaboration}{BESIII}), \bibinfo{journal}{Phys. Rev.}
  \textbf{\bibinfo{volume}{D90}}, \bibinfo{pages}{012001}
  (\bibinfo{year}{2014}{\natexlab{a}}), \eprint{1405.1076}.

\bibitem[{\citenamefont{Abakumova et~al.}(2011)}]{Abakumova:2011rp}
\bibinfo{author}{\bibfnamefont{E.~V.} \bibnamefont{Abakumova}}
  \bibnamefont{et~al.}, \bibinfo{journal}{Nucl. Instrum. Meth.}
  \textbf{\bibinfo{volume}{A659}}, \bibinfo{pages}{21} (\bibinfo{year}{2011}),
  \eprint{1109.5771}.

\bibitem[{\citenamefont{Zhang}(2018)}]{Zhang:2018gol}
\bibinfo{author}{\bibfnamefont{J.~Y.} \bibnamefont{Zhang}}
  (\bibinfo{collaboration}{BESIII}) (\bibinfo{year}{2018}),
  \eprint{1812.10056}.

\bibitem[{\citenamefont{Bennett et~al.}(2006)}]{Bennett:2006fi}
\bibinfo{author}{\bibfnamefont{G.~W.} \bibnamefont{Bennett}}
  \bibnamefont{et~al.} (\bibinfo{collaboration}{Muon g-2}),
  \bibinfo{journal}{Phys. Rev.} \textbf{\bibinfo{volume}{D73}},
  \bibinfo{pages}{072003} (\bibinfo{year}{2006}), \eprint{hep-ex/0602035}.

\bibitem[{\citenamefont{Davier et~al.}(2017)\citenamefont{Davier, Hoecker,
  Malaescu, and Zhang}}]{Davier:2017zfy}
\bibinfo{author}{\bibfnamefont{M.}~\bibnamefont{Davier}},
  \bibinfo{author}{\bibfnamefont{A.}~\bibnamefont{Hoecker}},
  \bibinfo{author}{\bibfnamefont{B.}~\bibnamefont{Malaescu}}, \bibnamefont{and}
  \bibinfo{author}{\bibfnamefont{Z.}~\bibnamefont{Zhang}},
  \bibinfo{journal}{Eur. Phys. J.} \textbf{\bibinfo{volume}{C77}},
  \bibinfo{pages}{827} (\bibinfo{year}{2017}), \eprint{1706.09436}.

\bibitem[{\citenamefont{Keshavarzi et~al.}(2018)\citenamefont{Keshavarzi,
  Nomura, and Teubner}}]{Keshavarzi:2018mgv}
\bibinfo{author}{\bibfnamefont{A.}~\bibnamefont{Keshavarzi}},
  \bibinfo{author}{\bibfnamefont{D.}~\bibnamefont{Nomura}}, \bibnamefont{and}
  \bibinfo{author}{\bibfnamefont{T.}~\bibnamefont{Teubner}},
  \bibinfo{journal}{Phys. Rev.} \textbf{\bibinfo{volume}{D97}},
  \bibinfo{pages}{114025} (\bibinfo{year}{2018}), \eprint{1802.02995}.

\bibitem[{\citenamefont{Grange et~al.}(2015)}]{Grange:2015fou}
\bibinfo{author}{\bibfnamefont{J.}~\bibnamefont{Grange}} \bibnamefont{et~al.}
  (\bibinfo{collaboration}{Muon g-2}) (\bibinfo{year}{2015}),
  \eprint{1501.06858}.

\bibitem[{\citenamefont{Otani}(2015)}]{Otani:2015jra}
\bibinfo{author}{\bibfnamefont{M.}~\bibnamefont{Otani}}
  (\bibinfo{collaboration}{E34}), \bibinfo{journal}{JPS Conf. Proc.}
  \textbf{\bibinfo{volume}{8}}, \bibinfo{pages}{025008} (\bibinfo{year}{2015}).

\bibitem[{\citenamefont{Miura}(2019)}]{Miura:2019xtd}
\bibinfo{author}{\bibfnamefont{K.}~\bibnamefont{Miura}}, in
  \emph{\bibinfo{booktitle}{{36th International Symposium on Lattice Field
  Theory (Lattice 2018) East Lansing, MI, United States, July 22-28, 2018}}}
  (\bibinfo{year}{2019}), \eprint{1901.09052}.

\bibitem[{\citenamefont{Davies et~al.}(2019)}]{Davies:2019efs}
\bibinfo{author}{\bibfnamefont{C.~T.~H.} \bibnamefont{Davies}}
  \bibnamefont{et~al.} (\bibinfo{collaboration}{Fermilab Lattice,
  LATTICE-HPQCD, MILC}) (\bibinfo{year}{2019}), \eprint{1902.04223}.

\bibitem[{\citenamefont{Achasov et~al.}(2006)}]{Achasov:2006vp}
\bibinfo{author}{\bibfnamefont{M.~N.} \bibnamefont{Achasov}}
  \bibnamefont{et~al.}, \bibinfo{journal}{J. Exp. Theor. Phys.}
  \textbf{\bibinfo{volume}{103}}, \bibinfo{pages}{380} (\bibinfo{year}{2006}),
  \bibinfo{note}{[Zh. Eksp. Teor. Fiz.130,437(2006)]}, \eprint{hep-ex/0605013}.

\bibitem[{\citenamefont{Lees et~al.}(2012)}]{Lees:2012cj}
\bibinfo{author}{\bibfnamefont{J.~P.} \bibnamefont{Lees}} \bibnamefont{et~al.}
  (\bibinfo{collaboration}{BaBar}), \bibinfo{journal}{Phys. Rev.}
  \textbf{\bibinfo{volume}{D86}}, \bibinfo{pages}{032013}
  (\bibinfo{year}{2012}), \eprint{1205.2228}.

\bibitem[{\citenamefont{Ablikim et~al.}(2016{\natexlab{a}})}]{Ablikim:2015orh}
\bibinfo{author}{\bibfnamefont{M.}~\bibnamefont{Ablikim}} \bibnamefont{et~al.}
  (\bibinfo{collaboration}{BESIII}), \bibinfo{journal}{Phys. Lett.}
  \textbf{\bibinfo{volume}{B753}}, \bibinfo{pages}{629}
  (\bibinfo{year}{2016}{\natexlab{a}}), \eprint{1507.08188}.

\bibitem[{\citenamefont{Akhmetshin et~al.}(2004)}]{Akhmetshin:2003zn}
\bibinfo{author}{\bibfnamefont{R.~R.} \bibnamefont{Akhmetshin}}
  \bibnamefont{et~al.} (\bibinfo{collaboration}{CMD-2}),
  \bibinfo{journal}{Phys. Lett.} \textbf{\bibinfo{volume}{B578}},
  \bibinfo{pages}{285} (\bibinfo{year}{2004}), \eprint{hep-ex/0308008}.

\bibitem[{\citenamefont{Akhmetshin et~al.}(2007)}]{Akhmetshin:2006bx}
\bibinfo{author}{\bibfnamefont{R.~R.} \bibnamefont{Akhmetshin}}
  \bibnamefont{et~al.} (\bibinfo{collaboration}{CMD-2}),
  \bibinfo{journal}{Phys. Lett.} \textbf{\bibinfo{volume}{B648}},
  \bibinfo{pages}{28} (\bibinfo{year}{2007}), \eprint{hep-ex/0610021}.

\bibitem[{\citenamefont{Anastasi et~al.}(2018)}]{Anastasi:2017eio}
\bibinfo{author}{\bibfnamefont{A.}~\bibnamefont{Anastasi}} \bibnamefont{et~al.}
  (\bibinfo{collaboration}{KLOE-2}), \bibinfo{journal}{JHEP}
  \textbf{\bibinfo{volume}{03}}, \bibinfo{pages}{173} (\bibinfo{year}{2018}),
  \eprint{1711.03085}.

\bibitem[{\citenamefont{Akhmetshin et~al.}(2017)}]{Akhmetshin:2016dtr}
\bibinfo{author}{\bibfnamefont{R.~R.} \bibnamefont{Akhmetshin}}
  \bibnamefont{et~al.}, \bibinfo{journal}{Phys. Lett.}
  \textbf{\bibinfo{volume}{B768}}, \bibinfo{pages}{345} (\bibinfo{year}{2017}),
  \eprint{1612.04483}.

\bibitem[{\citenamefont{Ablikim et~al.}(2019{\natexlab{a}})}]{Ablikim:2019sjw}
\bibinfo{author}{\bibfnamefont{M.}~\bibnamefont{Ablikim}} \bibnamefont{et~al.}
  (\bibinfo{collaboration}{BESIII}) (\bibinfo{year}{2019}{\natexlab{a}}),
  \eprint{1912.11208}.

\bibitem[{\citenamefont{Ripka}(2018)}]{Ripka:2018jsj}
\bibinfo{author}{\bibfnamefont{M.}~\bibnamefont{Ripka}}
  (\bibinfo{collaboration}{BESIII}), \bibinfo{journal}{Nucl. Part. Phys. Proc.}
  \textbf{\bibinfo{volume}{294-296}}, \bibinfo{pages}{158}
  (\bibinfo{year}{2018}).

\bibitem[{\citenamefont{Ablikim et~al.}(2017{\natexlab{a}})}]{Ablikim:2017wlt}
\bibinfo{author}{\bibfnamefont{M.}~\bibnamefont{Ablikim}} \bibnamefont{et~al.}
  (\bibinfo{collaboration}{BESIII}), \bibinfo{journal}{Chin. Phys.}
  \textbf{\bibinfo{volume}{C41}}, \bibinfo{pages}{063001}
  (\bibinfo{year}{2017}{\natexlab{a}}), \eprint{1702.04977}.

\bibitem[{\citenamefont{Bazavov et~al.}(2018)}]{Bazavov:2017lyh}
\bibinfo{author}{\bibfnamefont{A.}~\bibnamefont{Bazavov}} \bibnamefont{et~al.},
  \bibinfo{journal}{Phys. Rev.} \textbf{\bibinfo{volume}{D98}},
  \bibinfo{pages}{074512} (\bibinfo{year}{2018}), \eprint{1712.09262}.

\bibitem[{\citenamefont{Ablikim et~al.}(2014{\natexlab{b}})}]{Ablikim:2013uvu}
\bibinfo{author}{\bibfnamefont{M.}~\bibnamefont{Ablikim}} \bibnamefont{et~al.}
  (\bibinfo{collaboration}{BESIII}), \bibinfo{journal}{Phys. Rev.}
  \textbf{\bibinfo{volume}{D89}}, \bibinfo{pages}{051104}
  (\bibinfo{year}{2014}{\natexlab{b}}), \eprint{1312.0374}.

\bibitem[{\citenamefont{Tanabashi et~al.}(2018)}]{Tanabashi:2018oca}
\bibinfo{author}{\bibfnamefont{M.}~\bibnamefont{Tanabashi}}
  \bibnamefont{et~al.} (\bibinfo{collaboration}{Particle Data Group}),
  \bibinfo{journal}{Phys. Rev.} \textbf{\bibinfo{volume}{D98}},
  \bibinfo{pages}{030001} (\bibinfo{year}{2018}).

\bibitem[{\citenamefont{Ablikim et~al.}(2019{\natexlab{b}})}]{Ablikim:2018jun}
\bibinfo{author}{\bibfnamefont{M.}~\bibnamefont{Ablikim}} \bibnamefont{et~al.}
  (\bibinfo{collaboration}{BESIII}), \bibinfo{journal}{Phys. Rev. Lett.}
  \textbf{\bibinfo{volume}{122}}, \bibinfo{pages}{071802}
  (\bibinfo{year}{2019}{\natexlab{b}}), \eprint{1811.10890}.

\bibitem[{\citenamefont{Carrasco et~al.}(2015)}]{Carrasco:2014poa}
\bibinfo{author}{\bibfnamefont{N.}~\bibnamefont{Carrasco}}
  \bibnamefont{et~al.}, \bibinfo{journal}{Phys. Rev.}
  \textbf{\bibinfo{volume}{D91}}, \bibinfo{pages}{054507}
  (\bibinfo{year}{2015}), \eprint{1411.7908}.

\bibitem[{\citenamefont{Besson et~al.}(2009)}]{Besson:2009uv}
\bibinfo{author}{\bibfnamefont{D.}~\bibnamefont{Besson}} \bibnamefont{et~al.}
  (\bibinfo{collaboration}{CLEO}), \bibinfo{journal}{Phys. Rev.}
  \textbf{\bibinfo{volume}{D80}}, \bibinfo{pages}{032005}
  (\bibinfo{year}{2009}), \eprint{0906.2983}.

\bibitem[{\citenamefont{Ablikim et~al.}(2015{\natexlab{a}})}]{Ablikim:2015ixa}
\bibinfo{author}{\bibfnamefont{M.}~\bibnamefont{Ablikim}} \bibnamefont{et~al.}
  (\bibinfo{collaboration}{BESIII}), \bibinfo{journal}{Phys. Rev.}
  \textbf{\bibinfo{volume}{D92}}, \bibinfo{pages}{072012}
  (\bibinfo{year}{2015}{\natexlab{a}}), \eprint{1508.07560}.

\bibitem[{\citenamefont{Wang et~al.}(2003)\citenamefont{Wang, Wu, and
  Zhong}}]{Wang:2002zba}
\bibinfo{author}{\bibfnamefont{W.~Y.} \bibnamefont{Wang}},
  \bibinfo{author}{\bibfnamefont{Y.~L.} \bibnamefont{Wu}}, \bibnamefont{and}
  \bibinfo{author}{\bibfnamefont{M.}~\bibnamefont{Zhong}},
  \bibinfo{journal}{Phys. Rev.} \textbf{\bibinfo{volume}{D67}},
  \bibinfo{pages}{014024} (\bibinfo{year}{2003}), \eprint{hep-ph/0205157}.

\bibitem[{\citenamefont{Ablikim et~al.}(2019{\natexlab{c}})}]{Ablikim:2018evp}
\bibinfo{author}{\bibfnamefont{M.}~\bibnamefont{Ablikim}} \bibnamefont{et~al.}
  (\bibinfo{collaboration}{BESIII}), \bibinfo{journal}{Phys. Rev. Lett.}
  \textbf{\bibinfo{volume}{122}}, \bibinfo{pages}{011804}
  (\bibinfo{year}{2019}{\natexlab{c}}), \eprint{1810.03127}.

\bibitem[{\citenamefont{Ablikim et~al.}(2018{\natexlab{a}})}]{Ablikim:2018frk}
\bibinfo{author}{\bibfnamefont{M.}~\bibnamefont{Ablikim}} \bibnamefont{et~al.}
  (\bibinfo{collaboration}{BESIII}), \bibinfo{journal}{Phys. Rev. Lett.}
  \textbf{\bibinfo{volume}{121}}, \bibinfo{pages}{171803}
  (\bibinfo{year}{2018}{\natexlab{a}}), \eprint{1802.05492}.

\bibitem[{\citenamefont{Ablikim et~al.}(2017{\natexlab{b}})}]{Ablikim:2017lks}
\bibinfo{author}{\bibfnamefont{M.}~\bibnamefont{Ablikim}} \bibnamefont{et~al.}
  (\bibinfo{collaboration}{BESIII}), \bibinfo{journal}{Phys. Rev.}
  \textbf{\bibinfo{volume}{D96}}, \bibinfo{pages}{012002}
  (\bibinfo{year}{2017}{\natexlab{b}}), \eprint{1703.09084}.

\bibitem[{\citenamefont{Ablikim et~al.}(2016{\natexlab{b}})}]{Ablikim:2016sqt}
\bibinfo{author}{\bibfnamefont{M.}~\bibnamefont{Ablikim}} \bibnamefont{et~al.}
  (\bibinfo{collaboration}{BESIII}), \bibinfo{journal}{Eur. Phys. J.}
  \textbf{\bibinfo{volume}{C76}}, \bibinfo{pages}{369}
  (\bibinfo{year}{2016}{\natexlab{b}}), \eprint{1605.00068}.

\bibitem[{\citenamefont{Bifani et~al.}(2019)\citenamefont{Bifani,
  Descotes-Genon, Romero~Vidal, and Schune}}]{Bifani:2018zmi}
\bibinfo{author}{\bibfnamefont{S.}~\bibnamefont{Bifani}},
  \bibinfo{author}{\bibfnamefont{S.}~\bibnamefont{Descotes-Genon}},
  \bibinfo{author}{\bibfnamefont{A.}~\bibnamefont{Romero~Vidal}},
  \bibnamefont{and} \bibinfo{author}{\bibfnamefont{M.-H.}
  \bibnamefont{Schune}}, \bibinfo{journal}{J. Phys.}
  \textbf{\bibinfo{volume}{G46}}, \bibinfo{pages}{023001}
  (\bibinfo{year}{2019}), \eprint{1809.06229}.

\bibitem[{\citenamefont{Riggio et~al.}(2018)\citenamefont{Riggio, Salerno, and
  Simula}}]{Riggio:2017zwh}
\bibinfo{author}{\bibfnamefont{L.}~\bibnamefont{Riggio}},
  \bibinfo{author}{\bibfnamefont{G.}~\bibnamefont{Salerno}}, \bibnamefont{and}
  \bibinfo{author}{\bibfnamefont{S.}~\bibnamefont{Simula}},
  \bibinfo{journal}{Eur. Phys. J.} \textbf{\bibinfo{volume}{C78}},
  \bibinfo{pages}{501} (\bibinfo{year}{2018}), \eprint{1706.03657}.

\bibitem[{\citenamefont{Soni and Pandya}(2017)}]{Soni:2017eug}
\bibinfo{author}{\bibfnamefont{N.~R.} \bibnamefont{Soni}} \bibnamefont{and}
  \bibinfo{author}{\bibfnamefont{J.~N.} \bibnamefont{Pandya}},
  \bibinfo{journal}{Phys. Rev.} \textbf{\bibinfo{volume}{D96}},
  \bibinfo{pages}{016017} (\bibinfo{year}{2017}), \eprint{1706.01190}.

\bibitem[{\citenamefont{Ablikim et~al.}(2016{\natexlab{c}})}]{Ablikim:2015mjo}
\bibinfo{author}{\bibfnamefont{M.}~\bibnamefont{Ablikim}} \bibnamefont{et~al.}
  (\bibinfo{collaboration}{BESIII}), \bibinfo{journal}{Phys. Rev.}
  \textbf{\bibinfo{volume}{D94}}, \bibinfo{pages}{032001}
  (\bibinfo{year}{2016}{\natexlab{c}}), \eprint{1512.08627}.

\bibitem[{\citenamefont{Ablikim et~al.}(2018{\natexlab{b}})}]{Ablikim:2018ffp}
\bibinfo{author}{\bibfnamefont{M.}~\bibnamefont{Ablikim}} \bibnamefont{et~al.}
  (\bibinfo{collaboration}{BESIII}), \bibinfo{journal}{Phys. Rev. Lett.}
  \textbf{\bibinfo{volume}{121}}, \bibinfo{pages}{081802}
  (\bibinfo{year}{2018}{\natexlab{b}}), \eprint{1803.02166}.

\bibitem[{\citenamefont{Ablikim et~al.}(2019{\natexlab{d}})}]{Ablikim:2018lmn}
\bibinfo{author}{\bibfnamefont{M.}~\bibnamefont{Ablikim}} \bibnamefont{et~al.}
  (\bibinfo{collaboration}{BESIII}), \bibinfo{journal}{Phys. Rev.}
  \textbf{\bibinfo{volume}{D99}}, \bibinfo{pages}{011103}
  (\bibinfo{year}{2019}{\natexlab{d}}), \eprint{1811.11349}.

\bibitem[{\citenamefont{Ablikim et~al.}(2019{\natexlab{e}})}]{Ablikim:2018qzz}
\bibinfo{author}{\bibfnamefont{M.}~\bibnamefont{Ablikim}} \bibnamefont{et~al.}
  (\bibinfo{collaboration}{BESIII}), \bibinfo{journal}{Phys. Rev. Lett.}
  \textbf{\bibinfo{volume}{122}}, \bibinfo{pages}{062001}
  (\bibinfo{year}{2019}{\natexlab{e}}), \eprint{1809.06496}.

\bibitem[{\citenamefont{Ablikim et~al.}(2019{\natexlab{f}})}]{Ablikim:2018upe}
\bibinfo{author}{\bibfnamefont{M.}~\bibnamefont{Ablikim}} \bibnamefont{et~al.}
  (\bibinfo{collaboration}{BESIII}), \bibinfo{journal}{Phys. Rev. Lett.}
  \textbf{\bibinfo{volume}{122}}, \bibinfo{pages}{061801}
  (\bibinfo{year}{2019}{\natexlab{f}}), \eprint{1811.02911}.

\bibitem[{\citenamefont{Ablikim et~al.}(2015{\natexlab{b}})}]{Ablikim:2015gyp}
\bibinfo{author}{\bibfnamefont{M.}~\bibnamefont{Ablikim}} \bibnamefont{et~al.}
  (\bibinfo{collaboration}{BESIII}), \bibinfo{journal}{Phys. Rev.}
  \textbf{\bibinfo{volume}{D92}}, \bibinfo{pages}{071101}
  (\bibinfo{year}{2015}{\natexlab{b}}), \eprint{1508.00151}.

\bibitem[{\citenamefont{Ablikim et~al.}(2018{\natexlab{c}})}]{Ablikim:2017omq}
\bibinfo{author}{\bibfnamefont{M.}~\bibnamefont{Ablikim}} \bibnamefont{et~al.}
  (\bibinfo{collaboration}{BESIII}), \bibinfo{journal}{Phys. Rev.}
  \textbf{\bibinfo{volume}{D97}}, \bibinfo{pages}{012006}
  (\bibinfo{year}{2018}{\natexlab{c}}), \eprint{1709.03680}.

\bibitem[{\citenamefont{Ablikim et~al.}(2015{\natexlab{c}})}]{Ablikim:2015qgt}
\bibinfo{author}{\bibfnamefont{M.}~\bibnamefont{Ablikim}} \bibnamefont{et~al.}
  (\bibinfo{collaboration}{BESIII}), \bibinfo{journal}{Phys. Rev.}
  \textbf{\bibinfo{volume}{D92}}, \bibinfo{pages}{112008}
  (\bibinfo{year}{2015}{\natexlab{c}}), \eprint{1510.00308}.

\bibitem[{\citenamefont{Ablikim et~al.}(2016{\natexlab{d}})}]{Ablikim:2016rqq}
\bibinfo{author}{\bibfnamefont{M.}~\bibnamefont{Ablikim}} \bibnamefont{et~al.}
  (\bibinfo{collaboration}{BESIII}), \bibinfo{journal}{Phys. Rev.}
  \textbf{\bibinfo{volume}{D94}}, \bibinfo{pages}{112003}
  (\bibinfo{year}{2016}{\natexlab{d}}), \eprint{1608.06484}.

\bibitem[{\citenamefont{Chau and Keung}(1984)}]{Chau:1984fp}
\bibinfo{author}{\bibfnamefont{L.-L.} \bibnamefont{Chau}} \bibnamefont{and}
  \bibinfo{author}{\bibfnamefont{W.-Y.} \bibnamefont{Keung}},
  \bibinfo{journal}{Phys. Rev. Lett.} \textbf{\bibinfo{volume}{53}},
  \bibinfo{pages}{1802} (\bibinfo{year}{1984}).

\bibitem[{\citenamefont{Amhis et~al.}(2018)}]{Amhis:2018udz}
\bibinfo{author}{\bibfnamefont{Y.}~\bibnamefont{Amhis}} \bibnamefont{et~al.}
  (\bibinfo{collaboration}{HFLAV}) (\bibinfo{year}{2018}), \eprint{1812.07461}.

\bibitem[{\citenamefont{Asner et~al.}(2009)}]{Asner:2008nq}
\bibinfo{author}{\bibfnamefont{D.~M.} \bibnamefont{Asner}}
  \bibnamefont{et~al.}, \bibinfo{journal}{Int. J. Mod. Phys.}
  \textbf{\bibinfo{volume}{A24}}, \bibinfo{pages}{S1} (\bibinfo{year}{2009}),
  \eprint{0809.1869}.

\bibitem[{\citenamefont{Libby et~al.}(2010)}]{Libby:2010nu}
\bibinfo{author}{\bibfnamefont{J.}~\bibnamefont{Libby}} \bibnamefont{et~al.}
  (\bibinfo{collaboration}{CLEO}), \bibinfo{journal}{Phys. Rev.}
  \textbf{\bibinfo{volume}{D82}}, \bibinfo{pages}{112006}
  (\bibinfo{year}{2010}), \eprint{1010.2817}.

\bibitem[{\citenamefont{Aaij et~al.}(2018{\natexlab{a}})}]{Aaij:2018uns}
\bibinfo{author}{\bibfnamefont{R.}~\bibnamefont{Aaij}} \bibnamefont{et~al.}
  (\bibinfo{collaboration}{LHCb}), \bibinfo{journal}{JHEP}
  \textbf{\bibinfo{volume}{08}}, \bibinfo{pages}{176}
  (\bibinfo{year}{2018}{\natexlab{a}}), \bibinfo{note}{[JHEP18,176(2020)]},
  \eprint{1806.01202}.

\bibitem[{\citenamefont{Malde}(2016)}]{Malde:2223391}
\bibinfo{author}{\bibfnamefont{S.~S.} \bibnamefont{Malde}},
  \bibinfo{type}{Tech. Rep.} \bibinfo{number}{LHCb-PUB-2016-025.
  CERN-LHCb-PUB-2016-025}, \bibinfo{institution}{CERN},
  \bibinfo{address}{Geneva} (\bibinfo{year}{2016}),
  \urlprefix\url{http://cds.cern.ch/record/2223391}.

\bibitem[{\citenamefont{Abrams et~al.}(1980)}]{Abrams:1979iu}
\bibinfo{author}{\bibfnamefont{G.~S.} \bibnamefont{Abrams}}
  \bibnamefont{et~al.}, \bibinfo{journal}{Phys. Rev. Lett.}
  \textbf{\bibinfo{volume}{44}}, \bibinfo{pages}{10} (\bibinfo{year}{1980}).

\bibitem[{\citenamefont{Zupanc et~al.}(2014)}]{Zupanc:2013iki}
\bibinfo{author}{\bibfnamefont{A.}~\bibnamefont{Zupanc}} \bibnamefont{et~al.}
  (\bibinfo{collaboration}{Belle}), \bibinfo{journal}{Phys. Rev. Lett.}
  \textbf{\bibinfo{volume}{113}}, \bibinfo{pages}{042002}
  (\bibinfo{year}{2014}), \eprint{1312.7826}.

\bibitem[{\citenamefont{Ablikim et~al.}(2015{\natexlab{d}})}]{Ablikim:2015prg}
\bibinfo{author}{\bibfnamefont{M.}~\bibnamefont{Ablikim}} \bibnamefont{et~al.}
  (\bibinfo{collaboration}{BESIII}), \bibinfo{journal}{Phys. Rev. Lett.}
  \textbf{\bibinfo{volume}{115}}, \bibinfo{pages}{221805}
  (\bibinfo{year}{2015}{\natexlab{d}}), \eprint{1510.02610}.

\bibitem[{\citenamefont{Ablikim et~al.}(2017{\natexlab{c}})}]{Ablikim:2016vqd}
\bibinfo{author}{\bibfnamefont{M.}~\bibnamefont{Ablikim}} \bibnamefont{et~al.}
  (\bibinfo{collaboration}{BESIII}), \bibinfo{journal}{Phys. Lett.}
  \textbf{\bibinfo{volume}{B767}}, \bibinfo{pages}{42}
  (\bibinfo{year}{2017}{\natexlab{c}}), \eprint{1611.04382}.

\bibitem[{\citenamefont{Ablikim et~al.}(2016{\natexlab{e}})}]{Ablikim:2015flg}
\bibinfo{author}{\bibfnamefont{M.}~\bibnamefont{Ablikim}} \bibnamefont{et~al.}
  (\bibinfo{collaboration}{BESIII}), \bibinfo{journal}{Phys. Rev. Lett.}
  \textbf{\bibinfo{volume}{116}}, \bibinfo{pages}{052001}
  (\bibinfo{year}{2016}{\natexlab{e}}), \eprint{1511.08380}.

\bibitem[{\citenamefont{Ablikim et~al.}(2018{\natexlab{d}})}]{Ablikim:2017lct}
\bibinfo{author}{\bibfnamefont{M.}~\bibnamefont{Ablikim}} \bibnamefont{et~al.}
  (\bibinfo{collaboration}{BESIII}), \bibinfo{journal}{Phys. Rev. Lett.}
  \textbf{\bibinfo{volume}{120}}, \bibinfo{pages}{132001}
  (\bibinfo{year}{2018}{\natexlab{d}}), \eprint{1710.00150}.

\bibitem[{\citenamefont{Pakhlova et~al.}(2008)}]{Pakhlova:2008vn}
\bibinfo{author}{\bibfnamefont{G.}~\bibnamefont{Pakhlova}} \bibnamefont{et~al.}
  (\bibinfo{collaboration}{Belle}), \bibinfo{journal}{Phys. Rev. Lett.}
  \textbf{\bibinfo{volume}{101}}, \bibinfo{pages}{172001}
  (\bibinfo{year}{2008}), \eprint{0807.4458}.

\bibitem[{\citenamefont{Wang et~al.}(2007)}]{Wang:2007ea}
\bibinfo{author}{\bibfnamefont{X.~L.} \bibnamefont{Wang}} \bibnamefont{et~al.}
  (\bibinfo{collaboration}{Belle}), \bibinfo{journal}{Phys. Rev. Lett.}
  \textbf{\bibinfo{volume}{99}}, \bibinfo{pages}{142002}
  (\bibinfo{year}{2007}), \eprint{0707.3699}.

\bibitem[{\citenamefont{Ablikim et~al.}(2013{\natexlab{a}})}]{Ablikim:2013mio}
\bibinfo{author}{\bibfnamefont{M.}~\bibnamefont{Ablikim}} \bibnamefont{et~al.}
  (\bibinfo{collaboration}{BESIII}), \bibinfo{journal}{Phys. Rev. Lett.}
  \textbf{\bibinfo{volume}{110}}, \bibinfo{pages}{252001}
  (\bibinfo{year}{2013}{\natexlab{a}}), \eprint{1303.5949}.

\bibitem[{\citenamefont{Liu et~al.}(2013)}]{Liu:2013dau}
\bibinfo{author}{\bibfnamefont{Z.~Q.} \bibnamefont{Liu}} \bibnamefont{et~al.}
  (\bibinfo{collaboration}{Belle}), \bibinfo{journal}{Phys. Rev. Lett.}
  \textbf{\bibinfo{volume}{110}}, \bibinfo{pages}{252002}
  (\bibinfo{year}{2013}), \eprint{1304.0121}.

\bibitem[{\citenamefont{Xiao et~al.}(2013)\citenamefont{Xiao, Dobbs, Tomaradze,
  and Seth}}]{Xiao:2013iha}
\bibinfo{author}{\bibfnamefont{T.}~\bibnamefont{Xiao}},
  \bibinfo{author}{\bibfnamefont{S.}~\bibnamefont{Dobbs}},
  \bibinfo{author}{\bibfnamefont{A.}~\bibnamefont{Tomaradze}},
  \bibnamefont{and} \bibinfo{author}{\bibfnamefont{K.~K.} \bibnamefont{Seth}},
  \bibinfo{journal}{Phys. Lett.} \textbf{\bibinfo{volume}{B727}},
  \bibinfo{pages}{366} (\bibinfo{year}{2013}), \eprint{1304.3036}.

\bibitem[{\citenamefont{Ablikim et~al.}(2013{\natexlab{b}})}]{Ablikim:2013wzq}
\bibinfo{author}{\bibfnamefont{M.}~\bibnamefont{Ablikim}} \bibnamefont{et~al.}
  (\bibinfo{collaboration}{BESIII}), \bibinfo{journal}{Phys. Rev. Lett.}
  \textbf{\bibinfo{volume}{111}}, \bibinfo{pages}{242001}
  (\bibinfo{year}{2013}{\natexlab{b}}), \eprint{1309.1896}.

\bibitem[{\citenamefont{Ablikim et~al.}(2017{\natexlab{d}})}]{Ablikim:2016qzw}
\bibinfo{author}{\bibfnamefont{M.}~\bibnamefont{Ablikim}} \bibnamefont{et~al.}
  (\bibinfo{collaboration}{BESIII}), \bibinfo{journal}{Phys. Rev. Lett.}
  \textbf{\bibinfo{volume}{118}}, \bibinfo{pages}{092001}
  (\bibinfo{year}{2017}{\natexlab{d}}), \eprint{1611.01317}.

\bibitem[{\citenamefont{Ablikim et~al.}(2014{\natexlab{c}})}]{Ablikim:2013dyn}
\bibinfo{author}{\bibfnamefont{M.}~\bibnamefont{Ablikim}} \bibnamefont{et~al.}
  (\bibinfo{collaboration}{BESIII}), \bibinfo{journal}{Phys. Rev. Lett.}
  \textbf{\bibinfo{volume}{112}}, \bibinfo{pages}{092001}
  (\bibinfo{year}{2014}{\natexlab{c}}), \eprint{1310.4101}.

\bibitem[{\citenamefont{Ablikim
  et~al.}(2017{\natexlab{e}})}]{Collaboration:2017njt}
\bibinfo{author}{\bibfnamefont{M.}~\bibnamefont{Ablikim}} \bibnamefont{et~al.}
  (\bibinfo{collaboration}{BESIII}), \bibinfo{journal}{Phys. Rev. Lett.}
  \textbf{\bibinfo{volume}{119}}, \bibinfo{pages}{072001}
  (\bibinfo{year}{2017}{\natexlab{e}}), \eprint{1706.04100}.

\bibitem[{\citenamefont{Ablikim et~al.}(2014{\natexlab{d}})}]{Ablikim:2013xfr}
\bibinfo{author}{\bibfnamefont{M.}~\bibnamefont{Ablikim}} \bibnamefont{et~al.}
  (\bibinfo{collaboration}{BESIII}), \bibinfo{journal}{Phys. Rev. Lett.}
  \textbf{\bibinfo{volume}{112}}, \bibinfo{pages}{022001}
  (\bibinfo{year}{2014}{\natexlab{d}}), \eprint{1310.1163}.

\bibitem[{\citenamefont{Ablikim et~al.}(2015{\natexlab{e}})}]{Ablikim:2015swa}
\bibinfo{author}{\bibfnamefont{M.}~\bibnamefont{Ablikim}} \bibnamefont{et~al.}
  (\bibinfo{collaboration}{BESIII}), \bibinfo{journal}{Phys. Rev.}
  \textbf{\bibinfo{volume}{D92}}, \bibinfo{pages}{092006}
  (\bibinfo{year}{2015}{\natexlab{e}}), \eprint{1509.01398}.

\bibitem[{\citenamefont{Yuan}(2018)}]{Yuan:2018inv}
\bibinfo{author}{\bibfnamefont{C.-Z.} \bibnamefont{Yuan}},
  \bibinfo{journal}{Int. J. Mod. Phys.} \textbf{\bibinfo{volume}{A33}},
  \bibinfo{pages}{1830018} (\bibinfo{year}{2018}), \eprint{1808.01570}.

\bibitem[{\citenamefont{Ablikim et~al.}(2019{\natexlab{g}})}]{Ablikim:2019ipd}
\bibinfo{author}{\bibfnamefont{M.}~\bibnamefont{Ablikim}} \bibnamefont{et~al.}
  (\bibinfo{collaboration}{BESIII}), \bibinfo{journal}{Phys. Rev.}
  \textbf{\bibinfo{volume}{D100}}, \bibinfo{pages}{111102}
  (\bibinfo{year}{2019}{\natexlab{g}}), \eprint{1906.00831}.

\bibitem[{\citenamefont{Ablikim et~al.}(2017{\natexlab{f}})}]{Ablikim:2017oaf}
\bibinfo{author}{\bibfnamefont{M.}~\bibnamefont{Ablikim}} \bibnamefont{et~al.}
  (\bibinfo{collaboration}{BESIII}), \bibinfo{journal}{Phys. Rev.}
  \textbf{\bibinfo{volume}{D96}}, \bibinfo{pages}{032004}
  (\bibinfo{year}{2017}{\natexlab{f}}), \eprint{1703.08787}.

\bibitem[{\citenamefont{Choi et~al.}(2003)}]{Choi:2003ue}
\bibinfo{author}{\bibfnamefont{S.~K.} \bibnamefont{Choi}} \bibnamefont{et~al.}
  (\bibinfo{collaboration}{Belle}), \bibinfo{journal}{Phys. Rev. Lett.}
  \textbf{\bibinfo{volume}{91}}, \bibinfo{pages}{262001}
  (\bibinfo{year}{2003}), \eprint{hep-ex/0309032}.

\bibitem[{\citenamefont{Tornqvist}(2003)}]{Tornqvist:2003na}
\bibinfo{author}{\bibfnamefont{N.~A.} \bibnamefont{Tornqvist}}
  (\bibinfo{year}{2003}), \eprint{hep-ph/0308277}.

\bibitem[{\citenamefont{Wang et~al.}(2013)\citenamefont{Wang, Hanhart, and
  Zhao}}]{Wang:2013cya}
\bibinfo{author}{\bibfnamefont{Q.}~\bibnamefont{Wang}},
  \bibinfo{author}{\bibfnamefont{C.}~\bibnamefont{Hanhart}}, \bibnamefont{and}
  \bibinfo{author}{\bibfnamefont{Q.}~\bibnamefont{Zhao}},
  \bibinfo{journal}{Phys. Rev. Lett.} \textbf{\bibinfo{volume}{111}},
  \bibinfo{pages}{132003} (\bibinfo{year}{2013}), \eprint{1303.6355}.

\bibitem[{\citenamefont{Maiani et~al.}(2013)\citenamefont{Maiani, Riquer,
  Faccini, Piccinini, Pilloni, and Polosa}}]{Faccini:2013lda}
\bibinfo{author}{\bibfnamefont{L.}~\bibnamefont{Maiani}},
  \bibinfo{author}{\bibfnamefont{V.}~\bibnamefont{Riquer}},
  \bibinfo{author}{\bibfnamefont{R.}~\bibnamefont{Faccini}},
  \bibinfo{author}{\bibfnamefont{F.}~\bibnamefont{Piccinini}},
  \bibinfo{author}{\bibfnamefont{A.}~\bibnamefont{Pilloni}}, \bibnamefont{and}
  \bibinfo{author}{\bibfnamefont{A.~D.} \bibnamefont{Polosa}},
  \bibinfo{journal}{Phys. Rev.} \textbf{\bibinfo{volume}{D87}},
  \bibinfo{pages}{111102} (\bibinfo{year}{2013}), \eprint{1303.6857}.

\bibitem[{\citenamefont{Lebed and Polosa}(2016)}]{Lebed:2016yvr}
\bibinfo{author}{\bibfnamefont{R.~F.} \bibnamefont{Lebed}} \bibnamefont{and}
  \bibinfo{author}{\bibfnamefont{A.~D.} \bibnamefont{Polosa}},
  \bibinfo{journal}{Phys. Rev.} \textbf{\bibinfo{volume}{D93}},
  \bibinfo{pages}{094024} (\bibinfo{year}{2016}), \eprint{1602.08421}.

\bibitem[{\citenamefont{Ablikim et~al.}(2014{\natexlab{e}})}]{Ablikim:2013emm}
\bibinfo{author}{\bibfnamefont{M.}~\bibnamefont{Ablikim}} \bibnamefont{et~al.}
  (\bibinfo{collaboration}{BESIII}), \bibinfo{journal}{Phys. Rev. Lett.}
  \textbf{\bibinfo{volume}{112}}, \bibinfo{pages}{132001}
  (\bibinfo{year}{2014}{\natexlab{e}}), \eprint{1308.2760}.

\bibitem[{\citenamefont{Bugg}(2011)}]{Bugg:2011jr}
\bibinfo{author}{\bibfnamefont{D.~V.} \bibnamefont{Bugg}},
  \bibinfo{journal}{EPL} \textbf{\bibinfo{volume}{96}}, \bibinfo{pages}{11002}
  (\bibinfo{year}{2011}), \eprint{1105.5492}.

\bibitem[{\citenamefont{Chen et~al.}(2013)\citenamefont{Chen, Liu, and
  Matsuki}}]{Chen:2013coa}
\bibinfo{author}{\bibfnamefont{D.-Y.} \bibnamefont{Chen}},
  \bibinfo{author}{\bibfnamefont{X.}~\bibnamefont{Liu}}, \bibnamefont{and}
  \bibinfo{author}{\bibfnamefont{T.}~\bibnamefont{Matsuki}},
  \bibinfo{journal}{Phys. Rev.} \textbf{\bibinfo{volume}{D88}},
  \bibinfo{pages}{036008} (\bibinfo{year}{2013}), \eprint{1304.5845}.

\bibitem[{\citenamefont{Swanson}(2016)}]{Swanson:2015bsa}
\bibinfo{author}{\bibfnamefont{E.~S.} \bibnamefont{Swanson}},
  \bibinfo{journal}{Int. J. Mod. Phys.} \textbf{\bibinfo{volume}{E25}},
  \bibinfo{pages}{1642010} (\bibinfo{year}{2016}), \eprint{1504.07952}.

\bibitem[{\citenamefont{Guo et~al.}(2015)\citenamefont{Guo, Hanhart, Wang, and
  Zhao}}]{Guo:2014iya}
\bibinfo{author}{\bibfnamefont{F.-K.} \bibnamefont{Guo}},
  \bibinfo{author}{\bibfnamefont{C.}~\bibnamefont{Hanhart}},
  \bibinfo{author}{\bibfnamefont{Q.}~\bibnamefont{Wang}}, \bibnamefont{and}
  \bibinfo{author}{\bibfnamefont{Q.}~\bibnamefont{Zhao}},
  \bibinfo{journal}{Phys. Rev.} \textbf{\bibinfo{volume}{D91}},
  \bibinfo{pages}{051504} (\bibinfo{year}{2015}), \eprint{1411.5584}.

\bibitem[{\citenamefont{Szczepaniak}(2016)}]{Szczepaniak:2015hya}
\bibinfo{author}{\bibfnamefont{A.~P.} \bibnamefont{Szczepaniak}},
  \bibinfo{journal}{Phys. Lett.} \textbf{\bibinfo{volume}{B757}},
  \bibinfo{pages}{61} (\bibinfo{year}{2016}), \eprint{1510.01789}.

\bibitem[{\citenamefont{Ablikim et~al.}(2015{\natexlab{f}})}]{Ablikim:2015tbp}
\bibinfo{author}{\bibfnamefont{M.}~\bibnamefont{Ablikim}} \bibnamefont{et~al.}
  (\bibinfo{collaboration}{BESIII}), \bibinfo{journal}{Phys. Rev. Lett.}
  \textbf{\bibinfo{volume}{115}}, \bibinfo{pages}{112003}
  (\bibinfo{year}{2015}{\natexlab{f}}), \eprint{1506.06018}.

\bibitem[{\citenamefont{Ablikim et~al.}(2015{\natexlab{g}})}]{Ablikim:2015gda}
\bibinfo{author}{\bibfnamefont{M.}~\bibnamefont{Ablikim}} \bibnamefont{et~al.}
  (\bibinfo{collaboration}{BESIII}), \bibinfo{journal}{Phys. Rev. Lett.}
  \textbf{\bibinfo{volume}{115}}, \bibinfo{pages}{222002}
  (\bibinfo{year}{2015}{\natexlab{g}}), \eprint{1509.05620}.

\bibitem[{\citenamefont{Ablikim et~al.}(2014{\natexlab{f}})}]{Ablikim:2014dxl}
\bibinfo{author}{\bibfnamefont{M.}~\bibnamefont{Ablikim}} \bibnamefont{et~al.}
  (\bibinfo{collaboration}{BESIII}), \bibinfo{journal}{Phys. Rev. Lett.}
  \textbf{\bibinfo{volume}{113}}, \bibinfo{pages}{212002}
  (\bibinfo{year}{2014}{\natexlab{f}}), \eprint{1409.6577}.

\bibitem[{\citenamefont{Ablikim et~al.}(2015{\natexlab{h}})}]{Ablikim:2015vvn}
\bibinfo{author}{\bibfnamefont{M.}~\bibnamefont{Ablikim}} \bibnamefont{et~al.}
  (\bibinfo{collaboration}{BESIII}), \bibinfo{journal}{Phys. Rev. Lett.}
  \textbf{\bibinfo{volume}{115}}, \bibinfo{pages}{182002}
  (\bibinfo{year}{2015}{\natexlab{h}}), \eprint{1507.02404}.

\bibitem[{\citenamefont{Aubert et~al.}(2005)}]{Aubert:2005rm}
\bibinfo{author}{\bibfnamefont{B.}~\bibnamefont{Aubert}} \bibnamefont{et~al.}
  (\bibinfo{collaboration}{BaBar}), \bibinfo{journal}{Phys. Rev. Lett.}
  \textbf{\bibinfo{volume}{95}}, \bibinfo{pages}{142001}
  (\bibinfo{year}{2005}), \eprint{hep-ex/0506081}.

\bibitem[{\citenamefont{Aubert et~al.}(2007{\natexlab{a}})}]{Aubert:2007zz}
\bibinfo{author}{\bibfnamefont{B.}~\bibnamefont{Aubert}} \bibnamefont{et~al.}
  (\bibinfo{collaboration}{BaBar}), \bibinfo{journal}{Phys. Rev. Lett.}
  \textbf{\bibinfo{volume}{98}}, \bibinfo{pages}{212001}
  (\bibinfo{year}{2007}{\natexlab{a}}), \eprint{hep-ex/0610057}.

\bibitem[{\citenamefont{Qiao}(2006)}]{Qiao:2005av}
\bibinfo{author}{\bibfnamefont{C.-F.} \bibnamefont{Qiao}},
  \bibinfo{journal}{Phys. Lett.} \textbf{\bibinfo{volume}{B639}},
  \bibinfo{pages}{263} (\bibinfo{year}{2006}), \eprint{hep-ph/0510228}.

\bibitem[{\citenamefont{Ding}(2009)}]{Ding:2008gr}
\bibinfo{author}{\bibfnamefont{G.-J.} \bibnamefont{Ding}},
  \bibinfo{journal}{Phys. Rev.} \textbf{\bibinfo{volume}{D79}},
  \bibinfo{pages}{014001} (\bibinfo{year}{2009}), \eprint{0809.4818}.

\bibitem[{\citenamefont{Dai et~al.}(2015)\citenamefont{Dai, Shi, Tang, and
  Zheng}}]{Dai:2012pb}
\bibinfo{author}{\bibfnamefont{L.~Y.} \bibnamefont{Dai}},
  \bibinfo{author}{\bibfnamefont{M.}~\bibnamefont{Shi}},
  \bibinfo{author}{\bibfnamefont{G.-Y.} \bibnamefont{Tang}}, \bibnamefont{and}
  \bibinfo{author}{\bibfnamefont{H.~Q.} \bibnamefont{Zheng}},
  \bibinfo{journal}{Phys. Rev.} \textbf{\bibinfo{volume}{D92}},
  \bibinfo{pages}{014020} (\bibinfo{year}{2015}), \eprint{1206.6911}.

\bibitem[{\citenamefont{Maiani et~al.}(2005)\citenamefont{Maiani, Riquer,
  Piccinini, and Polosa}}]{Maiani:2005pe}
\bibinfo{author}{\bibfnamefont{L.}~\bibnamefont{Maiani}},
  \bibinfo{author}{\bibfnamefont{V.}~\bibnamefont{Riquer}},
  \bibinfo{author}{\bibfnamefont{F.}~\bibnamefont{Piccinini}},
  \bibnamefont{and} \bibinfo{author}{\bibfnamefont{A.~D.}
  \bibnamefont{Polosa}}, \bibinfo{journal}{Phys. Rev.}
  \textbf{\bibinfo{volume}{D72}}, \bibinfo{pages}{031502}
  (\bibinfo{year}{2005}), \eprint{hep-ph/0507062}.

\bibitem[{\citenamefont{Ali et~al.}(2018)\citenamefont{Ali, Maiani, Borisov,
  Ahmed, Jamil~Aslam, Parkhomenko, Polosa, and Rehman}}]{Ali:2017wsf}
\bibinfo{author}{\bibfnamefont{A.}~\bibnamefont{Ali}},
  \bibinfo{author}{\bibfnamefont{L.}~\bibnamefont{Maiani}},
  \bibinfo{author}{\bibfnamefont{A.~V.} \bibnamefont{Borisov}},
  \bibinfo{author}{\bibfnamefont{I.}~\bibnamefont{Ahmed}},
  \bibinfo{author}{\bibfnamefont{M.}~\bibnamefont{Jamil~Aslam}},
  \bibinfo{author}{\bibfnamefont{A.~{\relax Ya}.} \bibnamefont{Parkhomenko}},
  \bibinfo{author}{\bibfnamefont{A.~D.} \bibnamefont{Polosa}},
  \bibnamefont{and} \bibinfo{author}{\bibfnamefont{A.}~\bibnamefont{Rehman}},
  \bibinfo{journal}{Eur. Phys. J.} \textbf{\bibinfo{volume}{C78}},
  \bibinfo{pages}{29} (\bibinfo{year}{2018}), \eprint{1708.04650}.

\bibitem[{\citenamefont{Zhu}(2005)}]{Zhu:2005hp}
\bibinfo{author}{\bibfnamefont{S.-L.} \bibnamefont{Zhu}},
  \bibinfo{journal}{Phys. Lett.} \textbf{\bibinfo{volume}{B625}},
  \bibinfo{pages}{212} (\bibinfo{year}{2005}), \eprint{hep-ph/0507025}.

\bibitem[{\citenamefont{Kou and Pene}(2005)}]{Kou:2005gt}
\bibinfo{author}{\bibfnamefont{E.}~\bibnamefont{Kou}} \bibnamefont{and}
  \bibinfo{author}{\bibfnamefont{O.}~\bibnamefont{Pene}},
  \bibinfo{journal}{Phys. Lett.} \textbf{\bibinfo{volume}{B631}},
  \bibinfo{pages}{164} (\bibinfo{year}{2005}), \eprint{hep-ph/0507119}.

\bibitem[{\citenamefont{Dubynskiy and
  Voloshin}(2008{\natexlab{a}})}]{Dubynskiy:2008mq}
\bibinfo{author}{\bibfnamefont{S.}~\bibnamefont{Dubynskiy}} \bibnamefont{and}
  \bibinfo{author}{\bibfnamefont{M.~B.} \bibnamefont{Voloshin}},
  \bibinfo{journal}{Phys. Lett.} \textbf{\bibinfo{volume}{B666}},
  \bibinfo{pages}{344} (\bibinfo{year}{2008}{\natexlab{a}}),
  \eprint{0803.2224}.

\bibitem[{\citenamefont{Yuan et~al.}(2007)}]{Yuan:2007sj}
\bibinfo{author}{\bibfnamefont{C.~Z.} \bibnamefont{Yuan}} \bibnamefont{et~al.}
  (\bibinfo{collaboration}{Belle}), \bibinfo{journal}{Phys. Rev. Lett.}
  \textbf{\bibinfo{volume}{99}}, \bibinfo{pages}{182004}
  (\bibinfo{year}{2007}), \eprint{0707.2541}.

\bibitem[{\citenamefont{Patrignani et~al.}(2016)}]{Patrignani:2016xqp}
\bibinfo{author}{\bibfnamefont{C.}~\bibnamefont{Patrignani}}
  \bibnamefont{et~al.} (\bibinfo{collaboration}{Particle Data Group}),
  \bibinfo{journal}{Chin. Phys.} \textbf{\bibinfo{volume}{C40}},
  \bibinfo{pages}{100001} (\bibinfo{year}{2016}).

\bibitem[{\citenamefont{Ablikim et~al.}(2017{\natexlab{g}})}]{BESIII:2016adj}
\bibinfo{author}{\bibfnamefont{M.}~\bibnamefont{Ablikim}} \bibnamefont{et~al.}
  (\bibinfo{collaboration}{BESIII}), \bibinfo{journal}{Phys. Rev. Lett.}
  \textbf{\bibinfo{volume}{118}}, \bibinfo{pages}{092002}
  (\bibinfo{year}{2017}{\natexlab{g}}), \eprint{1610.07044}.

\bibitem[{\citenamefont{Ablikim et~al.}(2015{\natexlab{i}})}]{Ablikim:2014qwy}
\bibinfo{author}{\bibfnamefont{M.}~\bibnamefont{Ablikim}} \bibnamefont{et~al.}
  (\bibinfo{collaboration}{BESIII}), \bibinfo{journal}{Phys. Rev. Lett.}
  \textbf{\bibinfo{volume}{114}}, \bibinfo{pages}{092003}
  (\bibinfo{year}{2015}{\natexlab{i}}), \eprint{1410.6538}.

\bibitem[{\citenamefont{Ablikim et~al.}(2019{\natexlab{h}})}]{Ablikim:2018vxx}
\bibinfo{author}{\bibfnamefont{M.}~\bibnamefont{Ablikim}} \bibnamefont{et~al.}
  (\bibinfo{collaboration}{BESIII}), \bibinfo{journal}{Phys. Rev. Lett.}
  \textbf{\bibinfo{volume}{122}}, \bibinfo{pages}{102002}
  (\bibinfo{year}{2019}{\natexlab{h}}), \eprint{1808.02847}.

\bibitem[{\citenamefont{Ablikim et~al.}(2019{\natexlab{i}})}]{Ablikim:2019soz}
\bibinfo{author}{\bibfnamefont{M.}~\bibnamefont{Ablikim}} \bibnamefont{et~al.}
  (\bibinfo{collaboration}{BESIII}) (\bibinfo{year}{2019}{\natexlab{i}}),
  \eprint{1901.03992}.

\bibitem[{\citenamefont{Dubynskiy and
  Voloshin}(2008{\natexlab{b}})}]{Dubynskiy:2007tj}
\bibinfo{author}{\bibfnamefont{S.}~\bibnamefont{Dubynskiy}} \bibnamefont{and}
  \bibinfo{author}{\bibfnamefont{M.~B.} \bibnamefont{Voloshin}},
  \bibinfo{journal}{Phys. Rev.} \textbf{\bibinfo{volume}{D77}},
  \bibinfo{pages}{014013} (\bibinfo{year}{2008}{\natexlab{b}}),
  \eprint{0709.4474}.

\bibitem[{\citenamefont{Ablikim et~al.}(2019{\natexlab{j}})}]{Ablikim:2019zio}
\bibinfo{author}{\bibfnamefont{M.}~\bibnamefont{Ablikim}} \bibnamefont{et~al.}
  (\bibinfo{collaboration}{BESIII}) (\bibinfo{year}{2019}{\natexlab{j}}),
  \eprint{1903.04695}.

\bibitem[{\citenamefont{Jaffe and Johnson}(1976)}]{Jaffe:1975fd}
\bibinfo{author}{\bibfnamefont{R.~L.} \bibnamefont{Jaffe}} \bibnamefont{and}
  \bibinfo{author}{\bibfnamefont{K.}~\bibnamefont{Johnson}},
  \bibinfo{journal}{Phys. Lett.} \textbf{\bibinfo{volume}{60B}},
  \bibinfo{pages}{201} (\bibinfo{year}{1976}).

\bibitem[{\citenamefont{Toki}(1996)}]{Toki:1996si}
\bibinfo{author}{\bibfnamefont{W.}~\bibnamefont{Toki}}, in
  \emph{\bibinfo{booktitle}{{The Strong interaction, from hadrons to partons:
  Proceedings, 24th SLAC Summer Institute on Particle Physics (SSI 96),
  Stanford, Calif., 19-30 Aug 1996}}} (\bibinfo{year}{1996}), pp.
  \bibinfo{pages}{141--169},
  \urlprefix\url{http://www.slac.stanford.edu/pubs/confproc/ssi96/ssi96-006.ht%
ml}.

\bibitem[{\citenamefont{Chen et~al.}(2006)}]{Chen:2005mg}
\bibinfo{author}{\bibfnamefont{Y.}~\bibnamefont{Chen}} \bibnamefont{et~al.},
  \bibinfo{journal}{Phys. Rev.} \textbf{\bibinfo{volume}{D73}},
  \bibinfo{pages}{014516} (\bibinfo{year}{2006}), \eprint{hep-lat/0510074}.

\bibitem[{\citenamefont{Ablikim et~al.}(2015{\natexlab{j}})}]{Ablikim:2015umt}
\bibinfo{author}{\bibfnamefont{M.}~\bibnamefont{Ablikim}} \bibnamefont{et~al.}
  (\bibinfo{collaboration}{BESIII}), \bibinfo{journal}{Phys. Rev.}
  \textbf{\bibinfo{volume}{D92}}, \bibinfo{pages}{052003}
  (\bibinfo{year}{2015}{\natexlab{j}}), \bibinfo{note}{[Erratum: Phys. Rev.
  D93, no.3, 039906 (2016)]}, \eprint{1506.00546}.

\bibitem[{\citenamefont{Ablikim et~al.}(2018{\natexlab{e}})}]{Ablikim:2018izx}
\bibinfo{author}{\bibfnamefont{M.}~\bibnamefont{Ablikim}} \bibnamefont{et~al.}
  (\bibinfo{collaboration}{BESIII}), \bibinfo{journal}{Phys. Rev.}
  \textbf{\bibinfo{volume}{D98}}, \bibinfo{pages}{072003}
  (\bibinfo{year}{2018}{\natexlab{e}}), \eprint{1808.06946}.

\bibitem[{\citenamefont{Ablikim et~al.}(2013{\natexlab{c}})}]{Ablikim:2013hq}
\bibinfo{author}{\bibfnamefont{M.}~\bibnamefont{Ablikim}} \bibnamefont{et~al.}
  (\bibinfo{collaboration}{BESIII}), \bibinfo{journal}{Phys. Rev.}
  \textbf{\bibinfo{volume}{D87}}, \bibinfo{pages}{092009}
  (\bibinfo{year}{2013}{\natexlab{c}}), \bibinfo{note}{[Erratum: Phys. Rev.
  D87, no.11, 119901 (2013)]}, \eprint{1301.0053}.

\bibitem[{\citenamefont{Gui et~al.}(2013)\citenamefont{Gui, Chen, Li, Liu, Liu,
  Ma, Yang, and Zhang}}]{Gui:2012gx}
\bibinfo{author}{\bibfnamefont{L.-C.} \bibnamefont{Gui}},
  \bibinfo{author}{\bibfnamefont{Y.}~\bibnamefont{Chen}},
  \bibinfo{author}{\bibfnamefont{G.}~\bibnamefont{Li}},
  \bibinfo{author}{\bibfnamefont{C.}~\bibnamefont{Liu}},
  \bibinfo{author}{\bibfnamefont{Y.-B.} \bibnamefont{Liu}},
  \bibinfo{author}{\bibfnamefont{J.-P.} \bibnamefont{Ma}},
  \bibinfo{author}{\bibfnamefont{Y.-B.} \bibnamefont{Yang}}, \bibnamefont{and}
  \bibinfo{author}{\bibfnamefont{J.-B.} \bibnamefont{Zhang}}
  (\bibinfo{collaboration}{CLQCD}), \bibinfo{journal}{Phys. Rev. Lett.}
  \textbf{\bibinfo{volume}{110}}, \bibinfo{pages}{021601}
  (\bibinfo{year}{2013}), \eprint{1206.0125}.

\bibitem[{\citenamefont{Ablikim et~al.}(2016{\natexlab{f}})}]{Ablikim:2016hlu}
\bibinfo{author}{\bibfnamefont{M.}~\bibnamefont{Ablikim}} \bibnamefont{et~al.}
  (\bibinfo{collaboration}{BESIII}), \bibinfo{journal}{Phys. Rev.}
  \textbf{\bibinfo{volume}{D93}}, \bibinfo{pages}{112011}
  (\bibinfo{year}{2016}{\natexlab{f}}), \eprint{1602.01523}.

\bibitem[{\citenamefont{Yang et~al.}(2013)\citenamefont{Yang, Gui, Chen, Liu,
  Liu, Ma, and Zhang}}]{Yang:2013xba}
\bibinfo{author}{\bibfnamefont{Y.-B.} \bibnamefont{Yang}},
  \bibinfo{author}{\bibfnamefont{L.-C.} \bibnamefont{Gui}},
  \bibinfo{author}{\bibfnamefont{Y.}~\bibnamefont{Chen}},
  \bibinfo{author}{\bibfnamefont{C.}~\bibnamefont{Liu}},
  \bibinfo{author}{\bibfnamefont{Y.-B.} \bibnamefont{Liu}},
  \bibinfo{author}{\bibfnamefont{J.-P.} \bibnamefont{Ma}}, \bibnamefont{and}
  \bibinfo{author}{\bibfnamefont{J.-B.} \bibnamefont{Zhang}}
  (\bibinfo{collaboration}{CLQCD}), \bibinfo{journal}{Phys. Rev. Lett.}
  \textbf{\bibinfo{volume}{111}}, \bibinfo{pages}{091601}
  (\bibinfo{year}{2013}), \eprint{1304.3807}.

\bibitem[{\citenamefont{Bai et~al.}(2003)}]{Bai:2003sw}
\bibinfo{author}{\bibfnamefont{J.~Z.} \bibnamefont{Bai}} \bibnamefont{et~al.}
  (\bibinfo{collaboration}{BES}), \bibinfo{journal}{Phys. Rev. Lett.}
  \textbf{\bibinfo{volume}{91}}, \bibinfo{pages}{022001}
  (\bibinfo{year}{2003}), \eprint{hep-ex/0303006}.

\bibitem[{\citenamefont{Ablikim et~al.}(2012{\natexlab{a}})}]{BESIII:2011aa}
\bibinfo{author}{\bibfnamefont{M.}~\bibnamefont{Ablikim}} \bibnamefont{et~al.}
  (\bibinfo{collaboration}{BESIII}), \bibinfo{journal}{Phys. Rev. Lett.}
  \textbf{\bibinfo{volume}{108}}, \bibinfo{pages}{112003}
  (\bibinfo{year}{2012}{\natexlab{a}}), \eprint{1112.0942}.

\bibitem[{\citenamefont{Ablikim et~al.}(2005)}]{Ablikim:2005um}
\bibinfo{author}{\bibfnamefont{M.}~\bibnamefont{Ablikim}} \bibnamefont{et~al.}
  (\bibinfo{collaboration}{BES}), \bibinfo{journal}{Phys. Rev. Lett.}
  \textbf{\bibinfo{volume}{95}}, \bibinfo{pages}{262001}
  (\bibinfo{year}{2005}), \eprint{hep-ex/0508025}.

\bibitem[{\citenamefont{Ding and Yan}(2005)}]{Ding:2005ew}
\bibinfo{author}{\bibfnamefont{G.-J.} \bibnamefont{Ding}} \bibnamefont{and}
  \bibinfo{author}{\bibfnamefont{M.-L.} \bibnamefont{Yan}},
  \bibinfo{journal}{Phys. Rev.} \textbf{\bibinfo{volume}{C72}},
  \bibinfo{pages}{015208} (\bibinfo{year}{2005}), \eprint{hep-ph/0502127}.

\bibitem[{\citenamefont{Ablikim et~al.}(2016{\natexlab{g}})}]{Ablikim:2016itz}
\bibinfo{author}{\bibfnamefont{M.}~\bibnamefont{Ablikim}} \bibnamefont{et~al.}
  (\bibinfo{collaboration}{BESIII}), \bibinfo{journal}{Phys. Rev. Lett.}
  \textbf{\bibinfo{volume}{117}}, \bibinfo{pages}{042002}
  (\bibinfo{year}{2016}{\natexlab{g}}), \eprint{1603.09653}.

\bibitem[{\citenamefont{Ablikim et~al.}(2011)}]{Ablikim:2011pu}
\bibinfo{author}{\bibfnamefont{M.}~\bibnamefont{Ablikim}} \bibnamefont{et~al.}
  (\bibinfo{collaboration}{BESIII}), \bibinfo{journal}{Phys. Rev. Lett.}
  \textbf{\bibinfo{volume}{107}}, \bibinfo{pages}{182001}
  (\bibinfo{year}{2011}), \eprint{1107.1806}.

\bibitem[{\citenamefont{Ablikim et~al.}(2013{\natexlab{d}})}]{Ablikim:2013spp}
\bibinfo{author}{\bibfnamefont{M.}~\bibnamefont{Ablikim}} \bibnamefont{et~al.}
  (\bibinfo{collaboration}{BESIII}), \bibinfo{journal}{Phys. Rev.}
  \textbf{\bibinfo{volume}{D88}}, \bibinfo{pages}{091502}
  (\bibinfo{year}{2013}{\natexlab{d}}), \eprint{1305.5333}.

\bibitem[{\citenamefont{Ablikim et~al.}(2013{\natexlab{e}})}]{Ablikim:2013cif}
\bibinfo{author}{\bibfnamefont{M.}~\bibnamefont{Ablikim}} \bibnamefont{et~al.}
  (\bibinfo{collaboration}{BESIII}), \bibinfo{journal}{Phys. Rev.}
  \textbf{\bibinfo{volume}{D87}}, \bibinfo{pages}{112004}
  (\bibinfo{year}{2013}{\natexlab{e}}), \eprint{1303.3108}.

\bibitem[{\citenamefont{Ablikim et~al.}(2016{\natexlab{h}})}]{Ablikim:2015pkc}
\bibinfo{author}{\bibfnamefont{M.}~\bibnamefont{Ablikim}} \bibnamefont{et~al.}
  (\bibinfo{collaboration}{BESIII}), \bibinfo{journal}{Phys. Rev.}
  \textbf{\bibinfo{volume}{D93}}, \bibinfo{pages}{052010}
  (\bibinfo{year}{2016}{\natexlab{h}}), \eprint{1512.08197}.

\bibitem[{\citenamefont{Battaglieri et~al.}(2015)}]{Battaglieri:2014gca}
\bibinfo{author}{\bibfnamefont{M.}~\bibnamefont{Battaglieri}}
  \bibnamefont{et~al.}, \bibinfo{journal}{Acta Phys. Polon.}
  \textbf{\bibinfo{volume}{B46}}, \bibinfo{pages}{257} (\bibinfo{year}{2015}),
  \eprint{1412.6393}.

\bibitem[{\citenamefont{Armenteros et~al.}(1965)\citenamefont{Armenteros,
  Edwards, and Jacobsen}}]{Armenteros:1965zz}
\bibinfo{author}{\bibfnamefont{R.}~\bibnamefont{Armenteros}},
  \bibinfo{author}{\bibfnamefont{D.~N.} \bibnamefont{Edwards}},
  \bibnamefont{and} \bibinfo{author}{\bibfnamefont{T.}~\bibnamefont{Jacobsen}},
  \bibinfo{journal}{Phys. Lett.} \textbf{\bibinfo{volume}{17}},
  \bibinfo{pages}{344} (\bibinfo{year}{1965}).

\bibitem[{\citenamefont{Protopopescu et~al.}(1973)\citenamefont{Protopopescu,
  Alston-Garnjost, Barbaro-Galtieri, Flatte, Friedman, Lasinski, Lynch, Rabin,
  and Solmitz}}]{Protopopescu:1973sh}
\bibinfo{author}{\bibfnamefont{S.~D.} \bibnamefont{Protopopescu}},
  \bibinfo{author}{\bibfnamefont{M.}~\bibnamefont{Alston-Garnjost}},
  \bibinfo{author}{\bibfnamefont{A.}~\bibnamefont{Barbaro-Galtieri}},
  \bibinfo{author}{\bibfnamefont{S.~M.} \bibnamefont{Flatte}},
  \bibinfo{author}{\bibfnamefont{J.~H.} \bibnamefont{Friedman}},
  \bibinfo{author}{\bibfnamefont{T.~A.} \bibnamefont{Lasinski}},
  \bibinfo{author}{\bibfnamefont{G.~R.} \bibnamefont{Lynch}},
  \bibinfo{author}{\bibfnamefont{M.~S.} \bibnamefont{Rabin}}, \bibnamefont{and}
  \bibinfo{author}{\bibfnamefont{F.~T.} \bibnamefont{Solmitz}},
  \bibinfo{journal}{Phys. Rev.} \textbf{\bibinfo{volume}{D7}},
  \bibinfo{pages}{1279} (\bibinfo{year}{1973}).

\bibitem[{\citenamefont{Achasov}(2017)}]{Achasov:2017ozk}
\bibinfo{author}{\bibfnamefont{N.~N.} \bibnamefont{Achasov}},
  \bibinfo{journal}{Phys. Part. Nucl.} \textbf{\bibinfo{volume}{48}},
  \bibinfo{pages}{681} (\bibinfo{year}{2017}).

\bibitem[{\citenamefont{Weinstein and Isgur}(1990)}]{Weinstein:1990gu}
\bibinfo{author}{\bibfnamefont{J.~D.} \bibnamefont{Weinstein}}
  \bibnamefont{and} \bibinfo{author}{\bibfnamefont{N.}~\bibnamefont{Isgur}},
  \bibinfo{journal}{Phys. Rev.} \textbf{\bibinfo{volume}{D41}},
  \bibinfo{pages}{2236} (\bibinfo{year}{1990}).

\bibitem[{\citenamefont{Jaffe}(1977)}]{Jaffe:1976ig}
\bibinfo{author}{\bibfnamefont{R.~L.} \bibnamefont{Jaffe}},
  \bibinfo{journal}{Phys. Rev.} \textbf{\bibinfo{volume}{D15}},
  \bibinfo{pages}{267} (\bibinfo{year}{1977}).

\bibitem[{\citenamefont{Black et~al.}(1999)\citenamefont{Black, Fariborz,
  Sannino, and Schechter}}]{Black:1998wt}
\bibinfo{author}{\bibfnamefont{D.}~\bibnamefont{Black}},
  \bibinfo{author}{\bibfnamefont{A.~H.} \bibnamefont{Fariborz}},
  \bibinfo{author}{\bibfnamefont{F.}~\bibnamefont{Sannino}}, \bibnamefont{and}
  \bibinfo{author}{\bibfnamefont{J.}~\bibnamefont{Schechter}},
  \bibinfo{journal}{Phys. Rev.} \textbf{\bibinfo{volume}{D59}},
  \bibinfo{pages}{074026} (\bibinfo{year}{1999}), \eprint{hep-ph/9808415}.

\bibitem[{\citenamefont{Oller and Oset}(1999)}]{Oller:1998zr}
\bibinfo{author}{\bibfnamefont{J.~A.} \bibnamefont{Oller}} \bibnamefont{and}
  \bibinfo{author}{\bibfnamefont{E.}~\bibnamefont{Oset}},
  \bibinfo{journal}{Phys. Rev.} \textbf{\bibinfo{volume}{D60}},
  \bibinfo{pages}{074023} (\bibinfo{year}{1999}), \eprint{hep-ph/9809337}.

\bibitem[{\citenamefont{Maiani et~al.}(2004)\citenamefont{Maiani, Piccinini,
  Polosa, and Riquer}}]{Maiani:2004uc}
\bibinfo{author}{\bibfnamefont{L.}~\bibnamefont{Maiani}},
  \bibinfo{author}{\bibfnamefont{F.}~\bibnamefont{Piccinini}},
  \bibinfo{author}{\bibfnamefont{A.~D.} \bibnamefont{Polosa}},
  \bibnamefont{and} \bibinfo{author}{\bibfnamefont{V.}~\bibnamefont{Riquer}},
  \bibinfo{journal}{Phys. Rev. Lett.} \textbf{\bibinfo{volume}{93}},
  \bibinfo{pages}{212002} (\bibinfo{year}{2004}), \eprint{hep-ph/0407017}.

\bibitem[{\citenamefont{'t~Hooft et~al.}(2008)\citenamefont{'t~Hooft, Isidori,
  Maiani, Polosa, and Riquer}}]{Hooft:2008we}
\bibinfo{author}{\bibfnamefont{G.}~\bibnamefont{'t~Hooft}},
  \bibinfo{author}{\bibfnamefont{G.}~\bibnamefont{Isidori}},
  \bibinfo{author}{\bibfnamefont{L.}~\bibnamefont{Maiani}},
  \bibinfo{author}{\bibfnamefont{A.~D.} \bibnamefont{Polosa}},
  \bibnamefont{and} \bibinfo{author}{\bibfnamefont{V.}~\bibnamefont{Riquer}},
  \bibinfo{journal}{Phys. Lett.} \textbf{\bibinfo{volume}{B662}},
  \bibinfo{pages}{424} (\bibinfo{year}{2008}), \eprint{0801.2288}.

\bibitem[{\citenamefont{Ishida et~al.}(1995)\citenamefont{Ishida, Ishida,
  Sawazaki, Yamada, Ishida, Kinashi, Takamatsu, and Tsuru}}]{Ishida:1995km}
\bibinfo{author}{\bibfnamefont{S.}~\bibnamefont{Ishida}},
  \bibinfo{author}{\bibfnamefont{M.~Y.} \bibnamefont{Ishida}},
  \bibinfo{author}{\bibfnamefont{H.}~\bibnamefont{Sawazaki}},
  \bibinfo{author}{\bibfnamefont{K.}~\bibnamefont{Yamada}},
  \bibinfo{author}{\bibfnamefont{T.}~\bibnamefont{Ishida}},
  \bibinfo{author}{\bibfnamefont{T.}~\bibnamefont{Kinashi}},
  \bibinfo{author}{\bibfnamefont{K.}~\bibnamefont{Takamatsu}},
  \bibnamefont{and} \bibinfo{author}{\bibfnamefont{T.}~\bibnamefont{Tsuru}}, in
  \emph{\bibinfo{booktitle}{{Meson nucleon physics and the structure of the
  nucleon. Proceedings, 6th International Symposium, Blaubeuren, Germany, July
  10-14, 1995. Vols. 1, 2}}} (\bibinfo{year}{1995}), pp.
  \bibinfo{pages}{454--456}.

\bibitem[{\citenamefont{Achasov et~al.}(1979)\citenamefont{Achasov, Devyanin,
  and Shestakov}}]{Achasov:1979xc}
\bibinfo{author}{\bibfnamefont{N.~N.} \bibnamefont{Achasov}},
  \bibinfo{author}{\bibfnamefont{S.~A.} \bibnamefont{Devyanin}},
  \bibnamefont{and} \bibinfo{author}{\bibfnamefont{G.~N.}
  \bibnamefont{Shestakov}}, \bibinfo{journal}{Phys. Lett.}
  \textbf{\bibinfo{volume}{88B}}, \bibinfo{pages}{367} (\bibinfo{year}{1979}).

\bibitem[{\citenamefont{Ablikim et~al.}(2018{\natexlab{f}})}]{Ablikim:2018pik}
\bibinfo{author}{\bibfnamefont{M.}~\bibnamefont{Ablikim}} \bibnamefont{et~al.}
  (\bibinfo{collaboration}{BESIII}), \bibinfo{journal}{Phys. Rev. Lett.}
  \textbf{\bibinfo{volume}{121}}, \bibinfo{pages}{022001}
  (\bibinfo{year}{2018}{\natexlab{f}}), \eprint{1802.00583}.

\bibitem[{\citenamefont{Dubnickova et~al.}(1996)\citenamefont{Dubnickova,
  Dubnicka, and Rekalo}}]{Dubnickova:1992ii}
\bibinfo{author}{\bibfnamefont{A.~Z.} \bibnamefont{Dubnickova}},
  \bibinfo{author}{\bibfnamefont{S.}~\bibnamefont{Dubnicka}}, \bibnamefont{and}
  \bibinfo{author}{\bibfnamefont{M.~P.} \bibnamefont{Rekalo}},
  \bibinfo{journal}{Nuovo Cim.} \textbf{\bibinfo{volume}{A109}},
  \bibinfo{pages}{241} (\bibinfo{year}{1996}).

\bibitem[{\citenamefont{F{\"a}ldt and Kupsc}(2017)}]{Faldt:2017kgy}
\bibinfo{author}{\bibfnamefont{G.}~\bibnamefont{F{\"a}ldt}} \bibnamefont{and}
  \bibinfo{author}{\bibfnamefont{A.}~\bibnamefont{Kupsc}},
  \bibinfo{journal}{Phys. Lett.} \textbf{\bibinfo{volume}{B772}},
  \bibinfo{pages}{16} (\bibinfo{year}{2017}), \eprint{1702.07288}.

\bibitem[{\citenamefont{Ablikim et~al.}(2018{\natexlab{g}})}]{Ablikim:2018zay}
\bibinfo{author}{\bibfnamefont{M.}~\bibnamefont{Ablikim}} \bibnamefont{et~al.}
  (\bibinfo{collaboration}{BESIII}) (\bibinfo{year}{2018}{\natexlab{g}}),
  \eprint{1808.08917}.

\bibitem[{\citenamefont{Ablikim et~al.}(2018{\natexlab{h}})}]{Ablikim:2017pyl}
\bibinfo{author}{\bibfnamefont{M.}~\bibnamefont{Ablikim}} \bibnamefont{et~al.}
  (\bibinfo{collaboration}{BESIII}), \bibinfo{journal}{Phys. Rev.}
  \textbf{\bibinfo{volume}{D97}}, \bibinfo{pages}{032013}
  (\bibinfo{year}{2018}{\natexlab{h}}), \eprint{1709.10236}.

\bibitem[{\citenamefont{Pacetti et~al.}(2015)\citenamefont{Pacetti,
  Baldini~Ferroli, and Tomasi-Gustafsson}}]{Pacetti:2015iqa}
\bibinfo{author}{\bibfnamefont{S.}~\bibnamefont{Pacetti}},
  \bibinfo{author}{\bibfnamefont{R.}~\bibnamefont{Baldini~Ferroli}},
  \bibnamefont{and}
  \bibinfo{author}{\bibfnamefont{E.}~\bibnamefont{Tomasi-Gustafsson}},
  \bibinfo{journal}{Phys. Rept.} \textbf{\bibinfo{volume}{550-551}},
  \bibinfo{pages}{1} (\bibinfo{year}{2015}).

\bibitem[{\citenamefont{Cronin and Overseth}(1963)}]{Cronin:1963zb}
\bibinfo{author}{\bibfnamefont{J.~W.} \bibnamefont{Cronin}} \bibnamefont{and}
  \bibinfo{author}{\bibfnamefont{O.~E.} \bibnamefont{Overseth}},
  \bibinfo{journal}{Phys. Rev.} \textbf{\bibinfo{volume}{129}},
  \bibinfo{pages}{1795} (\bibinfo{year}{1963}).

\bibitem[{\citenamefont{Overseth and Roth}(1967)}]{Overseth:1967zz}
\bibinfo{author}{\bibfnamefont{O.~E.} \bibnamefont{Overseth}} \bibnamefont{and}
  \bibinfo{author}{\bibfnamefont{R.~F.} \bibnamefont{Roth}},
  \bibinfo{journal}{Phys. Rev. Lett.} \textbf{\bibinfo{volume}{19}},
  \bibinfo{pages}{391} (\bibinfo{year}{1967}).

\bibitem[{\citenamefont{Dauber et~al.}(1969)\citenamefont{Dauber, Berge,
  Hubbard, Merrill, and Muller}}]{Dauber:1969hg}
\bibinfo{author}{\bibfnamefont{P.~M.} \bibnamefont{Dauber}},
  \bibinfo{author}{\bibfnamefont{J.~P.} \bibnamefont{Berge}},
  \bibinfo{author}{\bibfnamefont{J.~R.} \bibnamefont{Hubbard}},
  \bibinfo{author}{\bibfnamefont{D.~W.} \bibnamefont{Merrill}},
  \bibnamefont{and} \bibinfo{author}{\bibfnamefont{R.~A.}
  \bibnamefont{Muller}}, \bibinfo{journal}{Phys. Rev.}
  \textbf{\bibinfo{volume}{179}}, \bibinfo{pages}{1262} (\bibinfo{year}{1969}).

\bibitem[{\citenamefont{Cleland et~al.}(1972)\citenamefont{Cleland, Conforto,
  Eaton, Gerber, Reinharz, Gautschi, Heer, Revillard, and
  Von~Dardel}}]{Cleland:1972fa}
\bibinfo{author}{\bibfnamefont{W.~E.} \bibnamefont{Cleland}},
  \bibinfo{author}{\bibfnamefont{G.}~\bibnamefont{Conforto}},
  \bibinfo{author}{\bibfnamefont{G.~H.} \bibnamefont{Eaton}},
  \bibinfo{author}{\bibfnamefont{H.~J.} \bibnamefont{Gerber}},
  \bibinfo{author}{\bibfnamefont{M.}~\bibnamefont{Reinharz}},
  \bibinfo{author}{\bibfnamefont{A.}~\bibnamefont{Gautschi}},
  \bibinfo{author}{\bibfnamefont{E.}~\bibnamefont{Heer}},
  \bibinfo{author}{\bibfnamefont{C.}~\bibnamefont{Revillard}},
  \bibnamefont{and}
  \bibinfo{author}{\bibfnamefont{G.}~\bibnamefont{Von~Dardel}},
  \bibinfo{journal}{Nucl. Phys.} \textbf{\bibinfo{volume}{B40}},
  \bibinfo{pages}{221} (\bibinfo{year}{1972}).

\bibitem[{\citenamefont{Astbury et~al.}(1975)}]{Astbury:1975hn}
\bibinfo{author}{\bibfnamefont{P.}~\bibnamefont{Astbury}} \bibnamefont{et~al.},
  \bibinfo{journal}{Nucl. Phys.} \textbf{\bibinfo{volume}{B99}},
  \bibinfo{pages}{30} (\bibinfo{year}{1975}).

\bibitem[{\citenamefont{Barnes et~al.}(1996)}]{Barnes:1996si}
\bibinfo{author}{\bibfnamefont{P.~D.} \bibnamefont{Barnes}}
  \bibnamefont{et~al.}, \bibinfo{journal}{Phys. Rev.}
  \textbf{\bibinfo{volume}{C54}}, \bibinfo{pages}{1877} (\bibinfo{year}{1996}).

\bibitem[{\citenamefont{Donoghue et~al.}(1986)\citenamefont{Donoghue, He, and
  Pakvasa}}]{Donoghue:1986hh}
\bibinfo{author}{\bibfnamefont{J.~F.} \bibnamefont{Donoghue}},
  \bibinfo{author}{\bibfnamefont{X.-G.} \bibnamefont{He}}, \bibnamefont{and}
  \bibinfo{author}{\bibfnamefont{S.}~\bibnamefont{Pakvasa}},
  \bibinfo{journal}{Phys. Rev.} \textbf{\bibinfo{volume}{D34}},
  \bibinfo{pages}{833} (\bibinfo{year}{1986}).

\bibitem[{\citenamefont{Olsen et~al.}(1970)\citenamefont{Olsen, Pondrom,
  Handler, Limon, Smith, and Overseth}}]{Olsen:1970vb}
\bibinfo{author}{\bibfnamefont{S.}~\bibnamefont{Olsen}},
  \bibinfo{author}{\bibfnamefont{L.}~\bibnamefont{Pondrom}},
  \bibinfo{author}{\bibfnamefont{R.}~\bibnamefont{Handler}},
  \bibinfo{author}{\bibfnamefont{P.}~\bibnamefont{Limon}},
  \bibinfo{author}{\bibfnamefont{J.~A.} \bibnamefont{Smith}}, \bibnamefont{and}
  \bibinfo{author}{\bibfnamefont{O.~E.} \bibnamefont{Overseth}},
  \bibinfo{journal}{Phys. Rev. Lett.} \textbf{\bibinfo{volume}{24}},
  \bibinfo{pages}{843} (\bibinfo{year}{1970}).

\bibitem[{\citenamefont{Gonzalez and Illana}(1994)}]{Gonzalez:1994zc}
\bibinfo{author}{\bibfnamefont{E.}~\bibnamefont{Gonzalez}} \bibnamefont{and}
  \bibinfo{author}{\bibfnamefont{J.~I.} \bibnamefont{Illana}}, in
  \emph{\bibinfo{booktitle}{{Tau charm factory. Proceedings, 3rd Workshop,
  Marbella, Spain, June 1-6, 1993}}} (\bibinfo{year}{1994}), pp.
  \bibinfo{pages}{525--538}.

\bibitem[{\citenamefont{Morrissey and Ramsey-Musolf}(2012)}]{Morrissey:2012db}
\bibinfo{author}{\bibfnamefont{D.~E.} \bibnamefont{Morrissey}}
  \bibnamefont{and} \bibinfo{author}{\bibfnamefont{M.~J.}
  \bibnamefont{Ramsey-Musolf}}, \bibinfo{journal}{New J. Phys.}
  \textbf{\bibinfo{volume}{14}}, \bibinfo{pages}{125003}
  (\bibinfo{year}{2012}), \eprint{1206.2942}.

\bibitem[{\citenamefont{Aushev et~al.}(2010)}]{Aushev:2010bq}
\bibinfo{author}{\bibfnamefont{T.}~\bibnamefont{Aushev}} \bibnamefont{et~al.}
  (\bibinfo{year}{2010}), \eprint{1002.5012}.

\bibitem[{\citenamefont{Bediaga et~al.}(2012)}]{Bediaga:2012uyd}
\bibinfo{author}{\bibfnamefont{I.}~\bibnamefont{Bediaga}} \bibnamefont{et~al.}
  (\bibinfo{collaboration}{LHCb}) (\bibinfo{year}{2012}).

\bibitem[{\citenamefont{Aaij et~al.}(2018{\natexlab{b}})}]{Bediaga:2018lhg}
\bibinfo{author}{\bibfnamefont{R.}~\bibnamefont{Aaij}} \bibnamefont{et~al.}
  (\bibinfo{collaboration}{LHCb}) (\bibinfo{year}{2018}{\natexlab{b}}),
  \eprint{1808.08865}.

\bibitem[{\citenamefont{Acciarri et~al.}(2016)}]{Acciarri:2016crz}
\bibinfo{author}{\bibfnamefont{R.}~\bibnamefont{Acciarri}} \bibnamefont{et~al.}
  (\bibinfo{collaboration}{DUNE}) (\bibinfo{year}{2016}), \eprint{1601.05471}.

\bibitem[{\citenamefont{Abe et~al.}(2018)}]{Abe:2018uyc}
\bibinfo{author}{\bibfnamefont{K.}~\bibnamefont{Abe}} \bibnamefont{et~al.}
  (\bibinfo{collaboration}{Hyper-Kamiokande}) (\bibinfo{year}{2018}),
  \eprint{1805.04163}.

\bibitem[{\citenamefont{Aaij et~al.}(2019)}]{Aaij:2019kcg}
\bibinfo{author}{\bibfnamefont{R.}~\bibnamefont{Aaij}} \bibnamefont{et~al.}
  (\bibinfo{collaboration}{LHCb}) (\bibinfo{year}{2019}), \eprint{1903.08726}.

\bibitem[{\citenamefont{Khodjamirian and Petrov}(2017)}]{Khodjamirian:2017zdu}
\bibinfo{author}{\bibfnamefont{A.}~\bibnamefont{Khodjamirian}}
  \bibnamefont{and} \bibinfo{author}{\bibfnamefont{A.~A.}
  \bibnamefont{Petrov}}, \bibinfo{journal}{Phys. Lett.}
  \textbf{\bibinfo{volume}{B774}}, \bibinfo{pages}{235} (\bibinfo{year}{2017}),
  \eprint{1706.07780}.

\bibitem[{\citenamefont{Golden and Grinstein}(1989)}]{Golden:1989qx}
\bibinfo{author}{\bibfnamefont{M.}~\bibnamefont{Golden}} \bibnamefont{and}
  \bibinfo{author}{\bibfnamefont{B.}~\bibnamefont{Grinstein}},
  \bibinfo{journal}{Phys. Lett.} \textbf{\bibinfo{volume}{B222}},
  \bibinfo{pages}{501} (\bibinfo{year}{1989}).

\bibitem[{\citenamefont{Buccella et~al.}(1995)\citenamefont{Buccella,
  Lusignoli, Miele, Pugliese, and Santorelli}}]{Buccella:1994nf}
\bibinfo{author}{\bibfnamefont{F.}~\bibnamefont{Buccella}},
  \bibinfo{author}{\bibfnamefont{M.}~\bibnamefont{Lusignoli}},
  \bibinfo{author}{\bibfnamefont{G.}~\bibnamefont{Miele}},
  \bibinfo{author}{\bibfnamefont{A.}~\bibnamefont{Pugliese}}, \bibnamefont{and}
  \bibinfo{author}{\bibfnamefont{P.}~\bibnamefont{Santorelli}},
  \bibinfo{journal}{Phys. Rev.} \textbf{\bibinfo{volume}{D51}},
  \bibinfo{pages}{3478} (\bibinfo{year}{1995}), \eprint{hep-ph/9411286}.

\bibitem[{\citenamefont{Bianco et~al.}(2003)\citenamefont{Bianco, Fabbri,
  Benson, and Bigi}}]{Bianco:2003vb}
\bibinfo{author}{\bibfnamefont{S.}~\bibnamefont{Bianco}},
  \bibinfo{author}{\bibfnamefont{F.~L.} \bibnamefont{Fabbri}},
  \bibinfo{author}{\bibfnamefont{D.}~\bibnamefont{Benson}}, \bibnamefont{and}
  \bibinfo{author}{\bibfnamefont{I.}~\bibnamefont{Bigi}},
  \bibinfo{journal}{Riv. Nuovo Cim.} \textbf{\bibinfo{volume}{26N7}},
  \bibinfo{pages}{1} (\bibinfo{year}{2003}), \eprint{hep-ex/0309021}.

\bibitem[{\citenamefont{Grossman et~al.}(2007)\citenamefont{Grossman, Kagan,
  and Nir}}]{Grossman:2006jg}
\bibinfo{author}{\bibfnamefont{Y.}~\bibnamefont{Grossman}},
  \bibinfo{author}{\bibfnamefont{A.~L.} \bibnamefont{Kagan}}, \bibnamefont{and}
  \bibinfo{author}{\bibfnamefont{Y.}~\bibnamefont{Nir}},
  \bibinfo{journal}{Phys. Rev.} \textbf{\bibinfo{volume}{D75}},
  \bibinfo{pages}{036008} (\bibinfo{year}{2007}), \eprint{hep-ph/0609178}.

\bibitem[{\citenamefont{Sakharov}(1948)}]{Sakharov:1948yq}
\bibinfo{author}{\bibfnamefont{A.~D.} \bibnamefont{Sakharov}},
  \bibinfo{journal}{Zh. Eksp. Teor. Fiz.} \textbf{\bibinfo{volume}{18}},
  \bibinfo{pages}{631} (\bibinfo{year}{1948}), \bibinfo{note}{[Usp. Fiz.
  Nauk161,no.5,29(1991)]}.

\bibitem[{\citenamefont{Arbuzov and Kopylova}(2012)}]{Arbuzov:2011ff}
\bibinfo{author}{\bibfnamefont{A.~B.} \bibnamefont{Arbuzov}} \bibnamefont{and}
  \bibinfo{author}{\bibfnamefont{T.~V.} \bibnamefont{Kopylova}},
  \bibinfo{journal}{JHEP} \textbf{\bibinfo{volume}{04}}, \bibinfo{pages}{009}
  (\bibinfo{year}{2012}), \eprint{1111.4308}.

\bibitem[{\citenamefont{Lees et~al.}(2013)}]{Lees:2013ebn}
\bibinfo{author}{\bibfnamefont{J.~P.} \bibnamefont{Lees}} \bibnamefont{et~al.}
  (\bibinfo{collaboration}{BaBar}), \bibinfo{journal}{Phys. Rev.}
  \textbf{\bibinfo{volume}{D87}}, \bibinfo{pages}{092005}
  (\bibinfo{year}{2013}), \eprint{1302.0055}.

\bibitem[{\citenamefont{Akhmetshin et~al.}(2018)}]{CMD-3:2018kql}
\bibinfo{author}{\bibfnamefont{R.~R.} \bibnamefont{Akhmetshin}}
  \bibnamefont{et~al.} (\bibinfo{collaboration}{CMD-3}) (\bibinfo{year}{2018}),
  \eprint{1808.00145}.

\bibitem[{\citenamefont{Antonelli et~al.}(1998)}]{Antonelli:1998fv}
\bibinfo{author}{\bibfnamefont{A.}~\bibnamefont{Antonelli}}
  \bibnamefont{et~al.} (\bibinfo{collaboration}{FENICE}),
  \bibinfo{journal}{Nucl. Phys.} \textbf{\bibinfo{volume}{B517}},
  \bibinfo{pages}{3} (\bibinfo{year}{1998}).

\bibitem[{\citenamefont{Achasov et~al.}(2014)}]{Achasov:2014ncd}
\bibinfo{author}{\bibfnamefont{M.~N.} \bibnamefont{Achasov}}
  \bibnamefont{et~al.} (\bibinfo{collaboration}{SND}), \bibinfo{journal}{Phys.
  Rev.} \textbf{\bibinfo{volume}{D90}}, \bibinfo{pages}{112007}
  (\bibinfo{year}{2014}), \eprint{1410.3188}.

\bibitem[{\citenamefont{Bisello et~al.}(1990)}]{Bisello:1990rf}
\bibinfo{author}{\bibfnamefont{D.}~\bibnamefont{Bisello}} \bibnamefont{et~al.}
  (\bibinfo{collaboration}{DM2}), \bibinfo{journal}{Z. Phys.}
  \textbf{\bibinfo{volume}{C48}}, \bibinfo{pages}{23} (\bibinfo{year}{1990}).

\bibitem[{\citenamefont{Aubert et~al.}(2007{\natexlab{b}})}]{Aubert:2007uf}
\bibinfo{author}{\bibfnamefont{B.}~\bibnamefont{Aubert}} \bibnamefont{et~al.}
  (\bibinfo{collaboration}{BaBar}), \bibinfo{journal}{Phys. Rev.}
  \textbf{\bibinfo{volume}{D76}}, \bibinfo{pages}{092006}
  (\bibinfo{year}{2007}{\natexlab{b}}), \eprint{0709.1988}.

\bibitem[{\citenamefont{Haidenbauer and
  Mei{\ss}ner}(2016)}]{Haidenbauer:2016won}
\bibinfo{author}{\bibfnamefont{J.}~\bibnamefont{Haidenbauer}} \bibnamefont{and}
  \bibinfo{author}{\bibfnamefont{U.~G.} \bibnamefont{Mei{\ss}ner}},
  \bibinfo{journal}{Phys. Lett.} \textbf{\bibinfo{volume}{B761}},
  \bibinfo{pages}{456} (\bibinfo{year}{2016}), \eprint{1608.02766}.

\bibitem[{\citenamefont{Ablikim et~al.}(2012{\natexlab{b}})}]{BESIII:2011ab}
\bibinfo{author}{\bibfnamefont{M.}~\bibnamefont{Ablikim}} \bibnamefont{et~al.}
  (\bibinfo{collaboration}{BESIII}), \bibinfo{journal}{Phys. Rev. Lett.}
  \textbf{\bibinfo{volume}{108}}, \bibinfo{pages}{222002}
  (\bibinfo{year}{2012}{\natexlab{b}}), \eprint{1111.0398}.

\bibitem[{\citenamefont{Ablikim et~al.}(2012{\natexlab{c}})}]{Ablikim:2012ur}
\bibinfo{author}{\bibfnamefont{M.}~\bibnamefont{Ablikim}} \bibnamefont{et~al.}
  (\bibinfo{collaboration}{BESIII}), \bibinfo{journal}{Phys. Rev.}
  \textbf{\bibinfo{volume}{D86}}, \bibinfo{pages}{092009}
  (\bibinfo{year}{2012}{\natexlab{c}}), \eprint{1209.4963}.

\bibitem[{\citenamefont{Ablikim et~al.}(2019{\natexlab{k}})}]{Ablikim:2019hff}
\bibinfo{author}{\bibfnamefont{M.}~\bibnamefont{Ablikim}} \bibnamefont{et~al.}
  (\bibinfo{collaboration}{BESIII}) (\bibinfo{year}{2019}{\natexlab{k}}),
  \eprint{1912.05983}.

\bibitem[{\citenamefont{Wang et~al.}(2016)}]{Wang:2016bzv}
\bibinfo{author}{\bibfnamefont{X.}~\bibnamefont{Wang}} \bibnamefont{et~al.},
  \bibinfo{journal}{JINST} \textbf{\bibinfo{volume}{11}},
  \bibinfo{pages}{C08009} (\bibinfo{year}{2016}), \eprint{1604.02701}.

\end{thebibliography}

\end{document}